\documentclass[11pt]{article}
\usepackage{jheppub}

\usepackage{amssymb}
\usepackage{amsmath}
\usepackage{amsfonts}

\usepackage{graphicx, subfigure, placeins, float}

\usepackage{array} 
\usepackage{longtable} 
\usepackage{colortab} 
\usepackage{colortbl}
\usepackage{arydshln}

\usepackage{dsfont,bbm}
\usepackage{mathrsfs}  
\usepackage{xfrac}
\usepackage{slashed}
\usepackage{mathabx}
\usepackage{enumitem}
\usepackage{shuffle}
\usepackage{stmaryrd}
\usepackage{pifont}
\usepackage{relsize}
\usepackage{arydshln}

\makeatletter
\newcommand\xleftrightarrow[2][]{%
  \ext@arrow 9999{\longleftrightarrowfill@}{#1}{#2}}
\newcommand\longleftrightarrowfill@{%
  \arrowfill@\leftarrow\relbar\rightarrow}
\makeatother

\newcommand{\Rome}[1]{\uppercase\expandafter{\romannumeral#1}}

\newcommand{\itbf}[1]{\textbf{\textit{#1}}}

\newcolumntype{C}[1]{>{\centering\arraybackslash}m{#1}}

\makeatletter
\newcommand{\mathleft}{\@fleqntrue\@mathmargin0pt}
\newcommand{\mathcenter}{\@fleqnfalse}
\makeatother

\newcommand\wideDownarrow{\mathrel{\scalebox{1.5}[1]{$\Downarrow$}}}

\newcommand\wideLongrightarrow{\mathrel{\scalebox{2}[1]{$\Longrightarrow$}}}

\usepackage[textsize=footnotesize,textwidth=2.25cm]{todonotes}

\usepackage{extarrows}

\newcommand{\QQ}{{\bf q}}

\title{Double copy for tree-level form factors. Part II. Generalizations and special topics}

\author[a,b]{Guanda Lin,}
\emailAdd{linguandak@pku.edu.cn}
\author[a,c,d]{Gang Yang}
\emailAdd{yangg@itp.ac.cn}
\affiliation[a]{CAS Key Laboratory of Theoretical Physics, Institute of Theoretical Physics, \\Chinese Academy of Sciences, Beijing 100190, China}
\affiliation[b]{Department of Physics, University of California, Berkeley, CA 94720, U.S.A.}
\affiliation[c]{School of Fundamental Physics and Mathematical Sciences, Hangzhou Institute for Advanced Study, UCAS, Hangzhou 310024, China}
\affiliation[d]{International Centre for Theoretical Physics Asia-Pacific, Beijing/Hangzhou, China}

\abstract{
Both the Bern, Carrasco, and Johansson (BCJ) and the Kawai, Lewellen, and Tye (KLT) double-copy formalisms have been recently generalized to a class of scattering matrix elements (so-called form factors) that involve local gauge-invariant operators. 
In this paper, we continue the study of double copy for form factors.
First, we generalize the double-copy prescription to form factors of higher-length operators ${\rm tr}(\phi^m)$ with $m\geq3$.
These higher-length operators introduce new non-trivial color identities, but the double-copy prescription works perfectly well. 
The closed formulae for the CK-dual numerators are also provided.
Next, we discuss the $\vec{v}$ vectors which are central ingredients appearing in the factorization relations of both the KLT kernels and the gauge form factors. We present a general construction rule for the $\vec{v}$ vectors and discuss their universal properties.
Finally, we consider the double copy for the form factor of the ${\rm tr}(F^2)$ operator in pure Yang-Mills theory.
In this case, we propose a new prescription which involves a gauge invariant decomposition for the form factor and a mixture of different CK-dual numerators appearing in the expansion. The new prescription for the more complicated double copy has been verified up to five external gluons.
}

\begin{document}

\maketitle

\setcounter{footnote}{0}

\section{Introduction and review}\label{sec:intro}

The study of scattering amplitudes has revealed many hidden structures in gauge and gravity theories. Among these, a significant finding is the so-called double copy relations, exposing the intimate connection between gauge and gravity theories. Various double copy formalism has been developed, including the Kawai, Lewellen and Tye (KLT) relations \cite{Kawai:1985xq}, the Bern, Carrasco and Johansson (BCJ) double copy \cite{Bern:2008qj, Bern:2010ue}, as well as the Cachazo, He and Yuan (CHY) formula \cite{Cachazo:2013hca, Cachazo:2014xea}.
While significant progress has been made in understanding the double copy for amplitudes (see reviews \cite{Bern:2019prr,Bern:2022wqg,Adamo:2022dcm}), much less is understood for the double copy of other physical quantities, such as matrix elements involving gauge invariant operators. 

In our recent work \cite{Lin:2021pne, Lin:2022jrp}, the double-copy construction was generalized for the first time to the form factor observables using the BCJ and KLT formalism. 
The form factors are natural extensions of on-shell amplitudes when local operator insertions are included \cite{Maldacena:2010kp,Brandhuber:2010ad, Bork:2010wf}, defined as
\begin{equation}\label{eq:defFF}
\itbf{F}_{{\cal O},n}(1,..,n) =  \int d^{D} x \, e^{-i q \cdot x}\langle 1, .. , n |{\cal O}(x)| 0\rangle = \delta^{D} (q-\sum_{i=1}^n p_i) \langle 1, .. , n |{\cal O}(0)| 0\rangle \,.
\end{equation} 
In this context, $1,..,n$ are $n$ on-shell asymptotic states carrying momenta $p_i, i=1,..,n$, $\mathcal{O}(x)$ is the (gauge-invariant) local operator, and $q=\sum_i p_i$ is the off-shell momentum associated with the operator.

In the previous work \cite{Lin:2021pne, Lin:2022jrp}, we mainly focused on form factors of length-2 operators like $\operatorname{tr}(\phi^2)$. 
The aim of this paper is to provide a concrete generalization to more general operators such as the higher-length operators $\operatorname{tr}(\phi^m)$ and the pure YM operator ${\rm tr}(F^2)$. Other important topics such as the general structure of $\vec{v}$ vectors and certain universality properties observed in the double copy for form factors will also be discussed in detail.

Before diving into the structure of this paper, we will first summarize some of the key points from \cite{Lin:2021pne, Lin:2022jrp}, which will introduce relevant notations and set the stage for the topics in this paper. 

The guiding principle is to construct  \emph{diffeomorphism invariant} quantities via double copy. 
As mentioned in \cite{Lin:2021pne,Lin:2022jrp}, such a construction can be achieved by imposing the condition of \emph{color-kinematics(CK) duality}.  The idea of CK duality, first introduced for amplitudes in \cite{Bern:2008qj}, stipulates that every Jacobi relation satisfied by the color factors $C_i$ of certain cubic diagrams has a dual relation satisfied by the numerators $N_i$ of the same diagrams:
\begin{equation}
C_i + C_j + C_k =0 \qquad \Rightarrow \qquad N_i + N_j + N_k =0 \,.
\end{equation} 
For form factors, the inclusion of local operators provides new color relations and thus induces new numerator relations, called the \emph{operator-induced relations}
\begin{equation}
\sum_{i_{\cal O}} C_{i_{\cal O}} =0 \qquad \Rightarrow \qquad \sum_{i_{\cal O}} N_{i_{\cal O}} =0 \,.
\end{equation} 
Such relations are in general not Jacobi relations and can involve four or more color factors as we will show in this paper. 
Combining all these relations, one can solve for the CK-dual numerators for form factors, which exhibit the following properties: 
(1) because of the additional relations, the CK-dual numerators can be \emph{uniquely} determined and are manifestly gauge invariant; 
(2) these numerators contain \emph{``spurious"-type poles} in the sense that these poles do not genuinely exist in the full gauge-theory form factors. 
Given the CK-dual solutions, the double copy of the form factor can be executed as
\begin{equation}
\itbf{F} = \sum_i {C_i N_i \over \prod_\alpha D_{i,\alpha} } \qquad \Rightarrow \qquad 
{\cal G} =  \sum_i {N_i^2 \over \prod_\alpha D_{i,\alpha} } \,.
\end{equation} 
Interestingly, after performing the double copy via squaring cubic diagram numerators,\footnote{Note that the vertex associated with a high-length operator is not necessarily cubic, but for convenience, we will still use ``cubic diagrams".} 
the ``spurious"-type poles survive and become real \emph{physical poles} in gravity. 
In particular, the double copy quantities ${\cal G}$ have nice factorization behaviors on these poles.

In parallel with this cubic diagram representation, there is an alternative double-copy prescription by introducing color-ordered form factor basis and propagator matrices.  
The two representations are closely related to each other.
The CK duality makes it possible to relate the numerators basis $\vec{N}$ to the color-ordered form factor basis $\vec{\mathcal{F}}$ via a \emph{propagator matrix} $\Theta^{\mathcal{F}}$:
\begin{equation}
    \vec{\mathcal{F}}=\Theta^{\mathcal{F}} \cdot \vec{N}\,.
\end{equation}
Compared to the propagator matrices in amplitudes \cite{Vaman:2010ez}, the new feature for form factors is that the propagator matrices are invertible. One can inverse the propagator matrix and define the \emph{KLT kernel} $\mathbf{S}^{\cal F}$ as 
\begin{equation}
    {\vec N}=\mathbf{S}^{\cal F}\, \cdot \vec{\mathcal{F}}\,.
\end{equation}
What is crucial is that one can explicitly see the ``spurious"-type poles in $\mathbf{S}^{\cal F}$, and consequently in $\vec{N}$. 
Explicitly, the double-copy result $\mathcal{G}$ takes the form 
\begin{equation}\label{eq:defGn}
    \mathcal{G}=\vec{N}^{\scriptscriptstyle \rm T}\cdot \vec{\mathcal{F}}=\vec{\mathcal{F}}^{\scriptscriptstyle \rm T} \cdot \mathbf{S}^{\cal F} \cdot \vec{\mathcal{F}}\,,
\end{equation}
in which the pole structure of $\mathcal{G}$ becomes transparent|there are two types of poles: the first are those ``physical"-type poles inherited from the gauge form factor $\mathcal{F}$, and the other are the new ``spurious"-type poles from $N$.  

To investigate the factorization of $\mathcal{G}$ on its poles, some intriguing structures involving the building blocks in \eqref{eq:defGn} are in order. 
As mentioned in \cite{Lin:2021pne,Lin:2022jrp}, a special set of vectors $\vec{v}$ as rational functions of Mandelstam variables play important roles. First, the $\vec{v}$ vectors appear in the following \emph{hidden factorization} relation for gauge form factors:
\begin{equation}
\label{eq:generalfactorization}
\vec{v} \cdot \vec{\mathcal{F}}_n  \big|_{s_{\rm sp}=0}= \mathcal{F}_{m}\times \mathcal{A}_{m'}\,,
\end{equation}
where $\mathcal{F}_{m}$ and $\mathcal{A}_{m'}$ are lower-point form factors and amplitudes,
and the special kinematics of spurious pole $s_{\rm sp}=0$ is taken. 
Second, we have the \emph{matrix decomposition} relation for the KLT kernel:
\begin{equation}\label{eq:Snfact}
 \textrm{Residue of} \   \mathbf{S}^{\cal F}_n\  \text{on}\  \textrm{``spurious"-type poles} = \mathbf{V}^{\scriptscriptstyle \rm T}\cdot (\mathbf{S}_{m}^{\cal F}\otimes \mathbf{S}_{m'}^{\cal A})\cdot \mathbf{V} \,,
\end{equation}
where each row of the matrix $\mathbf{V}$ corresponds to a $\vec{v}$ vector mentioned above. 

In this paper, we present further generalizations of the above framework for a broader range of form factors and discuss some universal properties. 

In Section~\ref{sec:highlength}, we first exemplify the prescription by studying ${\rm tr}(\phi^3)$ form factors. 
Then we demonstrate that the generalization to ${\rm tr}(\phi^m)$ $(m\geq 4)$ operators is also achievable. 
One notable feature is that such higher-length form factors contain non-trivial operator-induced color relations (satisfied also by the CK-dual numerators) that involve $m\geq 4$ terms.  
We also study two kinds of multi-scalar multi-gluon form factors: 
one is for the $\operatorname{tr}(\phi^2)$ operator and the other is for the $\operatorname{tr}(\phi^m)$ operator.
We find some interesting universal structures in the CK-dual numerators for these form factors.

In Section~\ref{sec:vvec}, we focus on the $\vec{v}$ vectors and discuss their properties in various details. 
The $\vec{v}$ vectors are important for following reasons:
(1) they induce the structure of both $\mathcal{F}$ in \eqref{eq:generalfactorization} and $\mathbf{S}^{\cal F}$ in \eqref{eq:Snfact}; 
and (2) one can combine \eqref{eq:generalfactorization} and \eqref{eq:Snfact} together and confirm the double-copy factorization on the ``spurious"-type poles.\footnote{This part is one central topic in \cite{Lin:2022jrp}, where the $\vec{v}$ vectors used there was regarded as predetermined.}
Given the importance of the $\vec{v}$ vectors, a closed formula for computing $\vec{v}$ vectors will be presented.
Interestingly, it turns out that such a closed formula for $\vec{v}$ vectors has a universal structure in the sense that it applies to  different form factors. 

In Section~\ref{sec:trF2}, we generalize the double-copy approach  to the pure gluonic form factors of $\operatorname{tr}(F^2)$. These form factors have novel structures that render the generalization highly non-trivial. 
As a result, we must modify the double-copy prescription, in particular, a gauge-invariant decomposition is required, which leads to interesting new features when performing the double copy. 

We conclude in Section~\ref{sec:discussion}  by summarizing the form factor double-copy prescription and offering some outlooks for future research. 

Some further remarks or technical details are included in the appendices. 
Appendix~\ref{ap:vremarks} contains some further explanation on the construction of $\vec{v}$ vectors.  In Appendix~\ref{ap:scalartheory} we present various Lagrangians, as well as some generalizations, of the theories mentioned in this paper. Finally, we present some data of the ${\rm tr}(F^2)$ double copy in Appendix~\ref{ap:Fsqsols}.

\section{Double copy for form factors of high-length operators}
\label{sec:highlength}

In this section, we generalize the double copy prescription to form factors of high-length operators.
Specifically, we focus on the ${\rm tr}(\phi^m)$ form factors in the Yang-Mills scalar (YMS) theory, with $m$ scalars and any number of gluons. 
We start with considering the ${\rm tr}(\phi^3)$ form factor in Section~\ref{ssec:FFL3} which is similar to the ${\rm tr}(\phi^2)$ case. 
Next we consider the general ${\rm tr}(\phi^m)$ $(m\geq 4)$ form factors in Section~\ref{ssec:FFL4}, 
and some new features such as the operator-induced relation like \eqref{eq:phi4color} with multiple terms will be discussed.
Finally, in Section~\ref{ssec:universalnum}, we consider the expressions of the CK-dual numerators and the closed formula will be given.

\subsection{Form factor of ${\rm tr}(\phi^3)$}\label{ssec:FFL3}

We first consider the form factors of the length-3 operator ${\rm tr}(\phi^3)$:
\begin{equation}\label{eq:phi3FCFF}
\itbf{F}_{{\rm tr}(\phi^3), n}(1^{\phi},2^{\phi},3^{\phi}, 4^g, \ldots, n^g) =  \int d^{D} x \, e^{-i q \cdot x}\langle \phi(p_1) \, \phi(p_2) \, \phi(p_3) g(p_4) \ldots g(p_n) |{\rm tr}(\phi^3)(x)| 0\rangle \,.
\end{equation} 

The minimal form factor has three external scalars
\begin{equation}\label{eq:phi3FCFF3pt}
\itbf{F}_{{\rm tr}(\phi^3),3}(1^{\phi},2^{\phi},3^{\phi}) = (2\pi)^4 \delta^{(4)}(q - \sum_{i=1}^3 p_i)\ d^{a_1 a_2 a_3} \,,
\end{equation} 
where the fully symmetric color factor is
\begin{equation}\label{eq:d123}
d^{a_1 a_2 a_3} =  {\rm tr}(T^{a_1} T^{a_2} T^{a_3}) +  {\rm tr}(T^{a_1} T^{a_3} T^{a_2})  \,.
\end{equation} 

The double copy  of the minimal form factor is trivial, reading $\mathcal{G}_3^{[\phi^3]}=1$ (omitting the delta function of momentum conservation). 

\begin{figure}[t]
    \centering
 \includegraphics[height=0.13\linewidth]{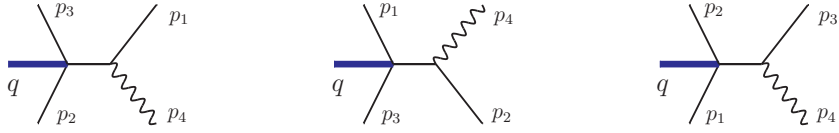}
    \caption{Feynman diagrams for the four-point form factor of ${\rm tr}(\phi^3)$. In the second diagram, $p_2$ and $p_4$ are intentionally twisted to get a desired sign of its color factor.}
    \label{fig:F4treeLength3}
\end{figure}

\subsubsection{The four-point case}

Let us go through the double copy procedure described in Section~\ref{sec:intro} for the first non-trivial example $\itbf{F}_{{\rm tr}(\phi^3),4}$.
Since the gluons can couple to each of the three scalars, there are  three diagrams to consider, as shown in Figure~\ref{fig:F4treeLength3}.

\paragraph{CK duality.}
The form factor can be expanded in terms of  cubic diagrams as 
\begin{equation}\label{eq:4ptL3}
	\itbf{F} _{{\rm tr}(\phi^3), 4}(1^{\phi},2^{\phi},3^{\phi},4^{g})=\frac{C_1 N_1}{s_{14}}+\frac{C_2 N_2}{s_{24}}+\frac{C_3 N_3}{s_{34}}\,,
\end{equation}
where the color factors are
\begin{equation}
\label{eq:4ptL3color}
C_1 = d^{a_2 a_3 b} f^{b a_1 a_4} \,, \ 
C_2 = - d^{a_3 a_1 b} f^{b a_2 a_4} \,, \ 
C_3 = d^{a_1 a_2 b} f^{b a_3 a_4}  \,,
\end{equation}
and they satisfy
\begin{equation}\label{eq:4ptL3colorJacobilike}
C_1 + C_3 = C_2  \,.
\end{equation}
Note that this is similar to the color Jacobi relation. 
By imposing the CK duality, one requires that
\begin{equation}
N_1^{\rm \scriptscriptstyle CK}  + N_3^{\rm \scriptscriptstyle CK} = N_2^{\rm \scriptscriptstyle CK}  \,.
\end{equation}
To solve the CK-dual numerators, we can extract the color-ordered form factors $\mathcal{F}_{{\rm tr}(\phi^3), 4}$ from \eqref{eq:4ptL3} and obtain 
\begin{equation}\label{eq:4ptL3Neq}
\begin{pmatrix}
\mathcal{F}_{{\rm tr}(\phi^3), 4} (1^{\phi},4^{g}, 3^{\phi},2^{\phi}) \\
\mathcal{F}_{{\rm tr}(\phi^3), 4} (1^{\phi},3^{\phi},4^{g},2^{\phi})
\end{pmatrix} 
=
\begin{pmatrix}
 {1\over s_{14}} +  {1\over s_{34}}&  - {1\over s_{34}} \\
- {1\over s_{34}} &   {1\over s_{24}} + {1\over s_{34}}
\end{pmatrix} 
\cdot
\begin{pmatrix}
N_1^{\rm \scriptscriptstyle CK} \\
 N_2^{\rm \scriptscriptstyle CK}
\end{pmatrix} ,
\end{equation}
where the color-ordered basis has two elements, and the matrix on the RHS is the propagator matrix $\Theta^{\mathcal{F}}$: 
\begin{equation}
\label{eq:4ptL3PMatrix}
\Theta^{\mathcal{F}_{{\rm tr}(\phi^3)}}_{4} = 
\begin{pmatrix}
 {1\over s_{14}} +  {1\over s_{34}}&  - {1\over s_{34}} \\
- {1\over s_{34}} &   {1\over s_{24}} + {1\over s_{34}}
\end{pmatrix}, 
\quad \text{ with }
\det\left(\Theta^{\mathcal{F}_{{\rm tr}(\phi^3)}}_{4}\right)=-{s_{123} - q^2 \over s_{14} s_{24} s_{34} }\,.
\end{equation}
One can see that the propagator matrix is a full-ranked 2 by 2 matrix.
Thus the master numerators can be uniquely determined from \eqref{eq:4ptL3Neq} as
\begin{align}\label{eq:xxx}
\begin{pmatrix}
N_1^{\rm \scriptscriptstyle CK} \\
N_2^{\rm \scriptscriptstyle CK}
\end{pmatrix}
= \mathbf{S}^{\mathcal{F}_{{\rm tr}(\phi^3)}}_{4} \cdot 
\begin{pmatrix}
\mathcal{F}_{{\rm tr}(\phi^3), 4} (1^{\phi},4^{g}, 3^{\phi},2^{\phi}) \\
\mathcal{F}_{{\rm tr}(\phi^3), 4} (1^{\phi},3^{\phi},4^{g},2^{\phi})
\end{pmatrix},
\end{align}
where the four-point KLT kernel for the ${\rm tr}(\phi^3)$ form factors is 
\begin{equation}
   \mathbf{S}^{\mathcal{F}_{{\rm tr}(\phi^3)}}_{4} = {-1\over s_{123} - q^2}
\begin{pmatrix}
s_{24} (s_{14} + s_{34})  &  s_{14} s_{24} \\
s_{14} s_{24} &  s_{14} (s_{24} + s_{34}) \,.
\end{pmatrix}
\end{equation}

The color-ordered four-point form factor results can be obtained as
\begin{equation}
\label{eq:F4L3colorordered}
\mathcal{F}_{{\rm tr}(\phi^3),4} (1^{\phi},4^{g}, 3^{\phi},2^{\phi})  = \frac{\varepsilon_4 \cdot p_1}{s_{14}}-\frac{\varepsilon_4 \cdot p_3}{s_{34}} \,, \quad \mathcal{F}_{{\rm tr}(\phi^3),4}(1^{\phi},3^{\phi},4^{g},2^{\phi}) = \frac{\varepsilon_4 \cdot p_3}{s_{34}} - \frac{\varepsilon_4 \cdot p_2}{s_{24}} \,,
\end{equation} 
and the numerators can be given in the manifestly gauge invariant expressions:
\begin{equation}\label{eq:4ptL3numsol}
    N_{1}^{\rm \scriptscriptstyle CK}=-\frac{2\, {\rm f}_{4}^{\mu\nu}p_{1,\mu}p_{(2+3),\nu}}{s_{123}-q^2}\,, \qquad 
    N_{2}^{\rm \scriptscriptstyle CK}=-\frac{2\, {\rm f}_{4}^{\mu\nu}p_{(1+2),\mu}p_{3,\nu}}{s_{123}-q^2}\,,
\end{equation}
where ${\rm f}_i^{\mu\nu}\equiv p_i^{\mu}\varepsilon_i^{\nu}{-}p_i^{\nu}\varepsilon_i^{\mu}$ is the linearized field strength. 
We can see that a spurious-type pole $s_{123}{-}q^2$ in the CK-dual numerators (which also appears in the numerator of $\det\big(\Theta^{\mathcal{F}_{{\rm tr}(\phi^3)}}_{4}\big)$ in \eqref{eq:4ptL3PMatrix}).

\vskip 5pt

\paragraph{Double copy.}
Now we perform the double copy. By squaring the numerators, the double copy of the form factor is 
\begin{equation}\label{eq:G4L3}
\mathcal{G}_4^{[\phi^3]}= \frac{(N_{1}^{\rm \scriptscriptstyle CK})^2}{s_{14}}+\frac{(N_{2}^{\rm \scriptscriptstyle CK})^2}{s_{24}}+\frac{(N_{3}^{\rm \scriptscriptstyle CK})^2}{s_{34}} \,.
\end{equation}
Equivalently, it can written as
\begin{equation}\label{eq:G4L3-2}
	\mathcal{G}_4^{[\phi^3]}
	 = \left(N_{1}^{\rm \scriptscriptstyle CK},\ N_{2}^{\rm \scriptscriptstyle CK} \right)\cdot \Theta^{\mathcal{F}_{{\rm tr}(\phi^3)}}_{4} \cdot  \begin{pmatrix}
N_1^{\rm \scriptscriptstyle CK} \\
N_2^{\rm \scriptscriptstyle CK}
\end{pmatrix} =  
	\left(\mathcal{F}_1,\ \mathcal{F}_2\right)\cdot \mathbf{S}_4^{\mathcal{F}_{{\rm tr}(\phi^3)}}\cdot \begin{pmatrix}
\mathcal{F}_1 \\ 
\mathcal{F}_2
\end{pmatrix} \\
 \,, 
\end{equation}
where $\mathcal{F}_{1,2}$ are the two color-ordered form factors in \eqref{eq:F4L3colorordered} and $\mathbf{S}^{\mathcal{F}_{{\rm tr}(\phi^3)}}_4 = \big(\Theta^{\mathcal{F}_{{\rm tr}(\phi^3)}}_{4} \big)^{-1}$ is the KLT kernel explicitly given on the RHS of \eqref{eq:xxx}.

\begin{figure}[t]
    \centering
 \includegraphics[height=0.14\linewidth]{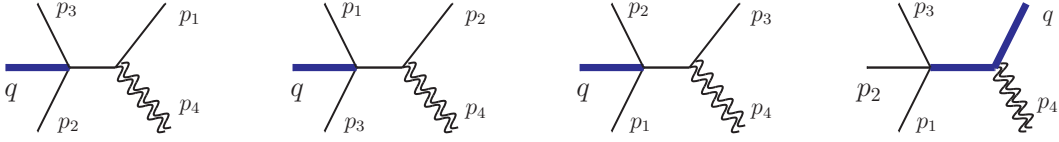}
    \caption{Feynman diagrams for the double copy of the four-point form factor of ${\rm tr}(\phi^3)$.}
    \label{fig:G4treeLength3}
\end{figure}

Furthermore, one can check that the above double-copy quantity is indeed a well-defined gravitational observable.
First, the gauge invariance of the numerators immediately tells the diffeomorphism invariance of $\mathcal{G}_4^{[\phi^3]}$ under the transformation $\varepsilon_{4}^{\mu\nu} \rightarrow \varepsilon_{4}^{\mu\nu} + p_4^{(\mu} \xi^{\nu)}$.
Second, after double copy, the pole $s_{123}{-}q^2$ becomes a real pole in $\mathcal{G}_4^{[\phi^3]}$.
Consider the factorization of ${\cal G}_4^{[\phi^3]}$  w.r.t this new pole, one has the nice factorization properties
\begin{equation}\label{eq:4ptL3GRAfactorize}
\begin{aligned}
\mathrm{Res}_{s_{123}=q^2} \big[\mathcal{G}_4^{[\phi^3]} \big] 
& = \Big[ s_{24}\mathcal{F}_{{\rm tr}(\phi^3),4} (1,3,4,2) + (s_{24}+s_{34}) \mathcal{F}_{{\rm tr}(\phi^3),4} (1,4,3,2) \Big]^2_{s_{123}=q^2}  \\
 & = \mathcal{G}_3^{(\phi^3)} (1^\phi,2^\phi,3^\phi) \  \mathcal{M}_{3}(\QQ_3^{S}, -q^{S}, 4^h)  \,.
\end{aligned}
\end{equation}
This corresponds to a new diagram with a massive scalar propagator $s_{123}{-}q^2$, which is given in  the last diagram in Figure~\ref{fig:G4treeLength3}.
One can check that, summing up the Feynman diagrams in Figure~\ref{fig:G4treeLength3} coincides with \eqref{eq:G4L3}. 

For completeness, we also give the decomposition of the KLT kernel $\mathbf{S}^{\mathcal{F}_{{\rm tr}(\phi^3)}}$. 
By taking the residue of $\mathbf{S}^{\mathcal{F}_{{\rm tr}(\phi^3)}}$ on the $s_{123}{-}q^2$ pole, we see it factorize to be 
\begin{equation}\label{eq:phi34ptrelations1}
\begin{aligned}
&    \text{Res}_{s_{123}=q^2}[\mathbf{S}^{\mathcal{F}_{{\rm tr}(\phi^3)}}_{4}]=\left(s_{24},s_{24}+s_{34}\right)^{\rm \scriptscriptstyle T}\cdot \mathbf{S}^{\mathcal{F}_{{\rm tr}(\phi^3)}}_{3}\cdot \left(s_{24},s_{24}+s_{34}\right)\,,
\end{aligned}
\end{equation}
note that $\mathbf{S}^{\mathcal{F}_{{\rm tr}(\phi^3)}}_{3}$ is just 1. This can be checked by a direct calculation. 
The vector $(s_{24},s_{24}{+}s_{34})$ is actually a simple  $\vec{v}$ vector in this case. 

Moreover, we have the hidden factorization relation satisfied by the color-ordered form factor as 
\begin{equation}\label{eq:phi34ptrelations}
\begin{aligned}
&    s_{24}\mathcal{F}_{{\rm tr}(\phi^3),4} (1,3,4,2) + (s_{24}+ s_{34}) \mathcal{F}_{{\rm tr}(\phi^3),4} (1,4,3,2)\big|_{s_{123}=q^2}=\mathcal{F}_{{\rm tr}(\phi^3),3}(1,3,2)\mathcal{A}_3(\QQ_3,4,-q)\,, \\
\end{aligned}
\end{equation}
in which the same $\vec{v}$ vector is involved.

\subsubsection{Higher-point cases}

Now we consider the $n$-point form factor $\itbf{F}_{{\rm tr}(\phi^3), n}$. 

\paragraph{Propagator matrix.}
To begin with, we define the propagator matrix. 
As discussed in \cite{Lin:2022jrp}, 
the propagator matrix elements can be interpreted as form factors in the bi-adjoint scalar theory. This picture also applies to higher-length form factors.
Concretely, we need a bi-adjoint scalar theory with two different types of scalars $\{\phi,\Phi\}$, having the Lagrangian (this is also equation (4.11) in \cite{Lin:2022jrp})
\begin{equation}\label{eq:ymsL}
\begin{aligned}
    \mathcal{L}^{\phi^3}=&\frac{1}{2} \operatorname{tr}_{\mathrm{C}}\left(D_{\mu} \phi^{I} D^{\mu} \phi^{I}\right)
    +\frac{1}{2} \operatorname{tr}_{\mathrm{C}}\left(D_{\mu} \Phi^{I} D^{\mu} \Phi^{I}\right)
    -\frac{\lambda_3}{3 !} \tilde{f}^{I J K} f^{abc} \phi^{I, a} \phi^{J, b} \Phi^{K, c} \\
    &-\frac{\lambda_1}{3 !} \tilde{f}^{I J K} f^{abc} \phi^{I, a} \phi^{J, b} \phi^{K, c}-\frac{\lambda_2}{3 !} \tilde{f}^{I J K} f^{abc} \Phi^{I, a} \Phi^{J, b} \Phi^{K, c}
    \,, 
\end{aligned}
\end{equation}
where we use $\{I, J, K\}$ to denote the flavor (FL) index and $\{a, b, c\}$ to denote the color (C) index. 
Roughly speaking, the little $\phi$ is the same as the scalars in the YMS form factors \eqref{eq:phi3FCFF}, while the capital $\Phi$ plays the role of gluons in the YMS form factors.

To obtain the propagator matrix, it is also necessary to properly define a gauge invariant operator in the bi-adjoint scalar theory \cite{Du:2011js, Bjerrum-Bohr:2012kaa, Cachazo:2013iea}.
The operator is introduced as 
$\mathcal{O}_{\phi^3}=(1/2!)^2 d^{abc} \tilde{d}^{IJK} \phi^{I,a}\phi^{J,b}\phi^{K,c}$, 
and the reason for selecting this operator is that the color and flavour factors must be identical, and the color part matches the color factor of the minimal form factor \eqref{eq:phi3FCFF3pt}. 
Given all these definitions, the propagator matrix can be defined as
\begin{align}\label{eq:deftheta}
    \Theta_n^{{\cal F}_{{\rm tr}(\phi^3)}}[\alpha &|\beta]
    =\int d^{D}x\  e^{\mathrm{i}q\cdot x} \langle 1^\phi, 2^\phi, 3^\phi, 4^{\Phi} \ldots n^\Phi | \mathcal{O}_{\phi^3}(x) | 0 \rangle \big|_{\operatorname{tr}_{\rm C}(\alpha) \operatorname{tr}_{\rm FL}(\beta)}\,,
\end{align}
where $\alpha$ refers to an ordering of color indices (C) and $\beta$ to an ordering of flavor (FL) indices. 

As in the previous four-point case, the propagator matrix is full-ranked, with the following numerator/denominator of the determinants
\begin{equation}\label{eq:detphi3}
\begin{aligned}
	\text{numerator}=&{\prod_{r=0}^{n-4}\prod_{\substack{\{i_1,\ldots,i_r\}\\
	\ \ \subset \{4,\ldots,n\}}}}\left(s_{123i_1\ldots i_r}-q^2\right)^{r!\times (n-4-r)! }\,,\\
	\text{denominator}=&\prod_{j=1}^{3}\Bigg(\ {\prod_{r=1}^{n-3}\prod_{\substack{\{i_1,\ldots,i_r\}\\
	\ \ \subset \{4,\ldots,n\}}}}(s_{j i_1 \cdots i_r})^{(r-1)! \times (n-2-r)!}\Bigg)\,.
\end{aligned}
\end{equation}
The zeros of the numerator provide the possible spurious poles, which are $s_{123\cdots}{-}q^2$ with $\cdots$ representing certain gluon momenta. 

The inverse of the propagator matrix is the KLT kernel $\mathbf{S}^{{\cal F}_{{\rm tr}(\phi^3)}}$, which is an $(n-2)!$ by $(n-2)!$ matrix with only simple poles $s_{123\cdots}{-}q^2$. 
The residues of such a propagator matrix on these poles satisfy the matrix decomposition relation like \eqref{eq:phi34ptrelations1}, see also \eqref{eq:generalmatdec} below.

\paragraph{Choice of basis.}
Next we choose a proper color basis and define the corresponding (basis) color-ordered form factors and (basis) CK numerators.
We define the basis by introducing the generalized DDM color basis associated with the following trivalent graph (with the blue dot associated with the $d^{abc}$ color factor from the $\operatorname{tr}(\phi^3)$ operator)
\begin{equation}\label{eq:3scalara}
\Gamma[\{i\};\{j\}]:=
\begin{aligned}
    \includegraphics[width=0.36\linewidth]{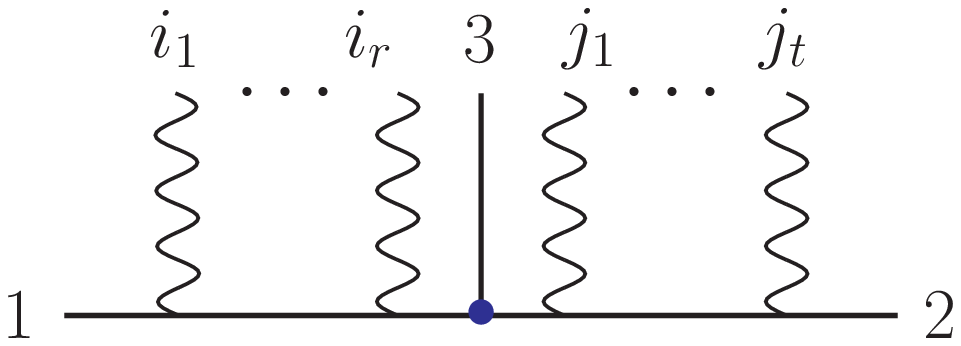}
\end{aligned} ,
\end{equation}
where $\{i_1,\ldots,i_r\}$ and  $\{j_1,\ldots,j_t\}$ (here $r{+}t=n{-}3$) are complementary subsets of the gluon set $\{4,\ldots,n\}$. 
The orderings in both subsets can be arbitrary.\footnote{For example, when $n=5$, $\{\{i_1,\ldots,i_r\},\{j_1,\ldots,j_t\}\}$ can be $\{\emptyset,\{4,5\}\}$, $\{\emptyset,\{5,4\}\}$, $\{\{4,5\},\emptyset\}$, $\{\{5,4\},\emptyset\}$, $\{\{4\},\{5\}\}$ and $\{\{5\},\{4\}\}$.} 
Note that we take two scalars $\{1,2\}$ to be special such that no gluon is attached to the leg of scalar $3$.
The specificity of such a basis choice is that all the gluons are directly connected to the scalar ``skeleton", \emph{i.e.}~no gluon self-interaction vertices appear in these basis cubic diagrams, which will also be valid for form factors of higher-length operators. The reason why the color factors of the diagrams in \eqref{eq:3scalara} form a basis should be clear:
due to the color Jacobi relation, there exists a color basis containing no gluon self-interaction vertex; likewise, given the color relations similar to \eqref{eq:4ptL3color}, we can remove all the gluons on one of the scalar lines.

\paragraph{CK-dual numerators and double copy.}
We further consider the numerators. The CK-dual numerators of the half ladder diagram \eqref{eq:3scalara} can be identified as\footnote{As commented in the ${\rm tr}(\phi^2)$ discussion, we point out that \eqref{eq:phi3nums} is a special case for simple operators like $\phi^{2}$ and $\phi^{3}$. We will see the generalization in Section~\ref{ssec:FFL4}. }
\begin{equation}\label{eq:phi3nums}
    N\bigg(\begin{aligned}
    \includegraphics[width=0.32\linewidth]{figure/3scalars2.eps}
\end{aligned} \bigg)=N[1,i_1,\ldots,i_r,3,j_1,\ldots,j_t,2]\,, 
\end{equation}
where the numerators are related to the color-ordered form factors as 
\begin{equation}\label{eq:theta3FandN}
    {\mathcal{F}_{{\rm tr}(\phi^3)}(1,\beta,2)}=\sum_{
\beta'\in S_{n-2}}\Theta^{{\mathcal{F}_{{\rm tr}(\phi^3)}}}[\beta|\beta'] {N}[1,\beta',2]\,.
\end{equation}
Examining the expressions, we observe that the numerators \eqref{eq:phi3nums} have only $s_{123\cdots}{-}q^2$ spurious-type poles, and can be spelt out via a Hopf-algebra-based  closed formula, see Section~\ref{ssec:universalnum}.

Finally, we explicitly present the double copy as 
\begin{equation}
    \mathcal{G}_n^{[\phi^3]}=\sum_{\beta_{1,2}\in S_{n-2}} N[\beta_1] \Theta^{{\mathcal{F}_{{\rm tr}(\phi^3)}}}[\beta_1|\beta_2]  N[\beta_2]=\sum_{\beta\in S_{n-2}} N[\beta] \mathcal{F}_{{\rm tr}(\phi^3)}[\beta]\,,
\end{equation}
and assert that ${\cal G}_n$  has the desired pole structures and the corresponding factorization properties, which are corroborated by the hidden factorization relations and matrix decompositions discussed in Section~\ref{ssec:universalv}.

\subsection{Form factor of ${\rm tr}(\phi^m)$}\label{ssec:FFL4}

Similar discussions can be generalized form factors of higher-length operators ${\rm tr}(\phi^m)$. 
Here a main new feature is that, unlike the color relation \eqref{eq:4ptL3colorJacobilike} in the length-3 case that is similar to the standard Jacobi relations, we will have some more general color relations.
In the following, we will mainly focus on the length-4 operator ${\rm tr}(\phi^4)$ which can be discussed  in explicit expressions and at the same time can capture the salient features for the higher-length operators.

The minimal form factor has four external scalars
\begin{equation}
\itbf{F}_4^{(\phi^4)}(1^{\phi},2^{\phi},3^{\phi},4^{\phi}) = (2\pi)^4 \delta^{(4)}(q - \sum_{i=1}^4 p_i) d^{a_1 a_2 a_3 a_4}
\,,
\end{equation} 
where the color factor $d$ is fully symmetric as
\begin{equation}
d^{a_1 a_2 a_3 a_4} :=  \sum_{\sigma \in S_4/Z_4} {\rm tr}(T^{a_{\sigma(1)}} T^{a_{\sigma(2)}} T^{a_{\sigma(3)}} T^{a_{\sigma(4)}})\,.
\end{equation}
We also trivially have the double copy as $\mathcal{G}_4^{[\phi^4]}=1$.

\begin{figure}[t]
    \centering
 \includegraphics[height=0.13\linewidth]{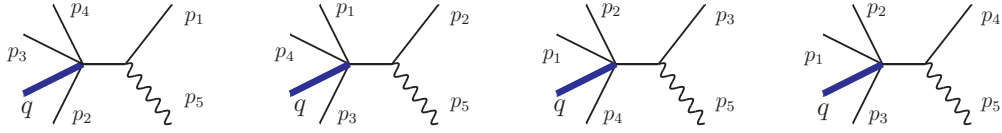}
    \caption{Feynman diagrams for the five-point form factor of ${\rm tr}(\phi^4)$.}
    \label{fig:F5treeLength4}
\end{figure}

\paragraph{The five-point case.}
To make the discussion less trivial, we add gluons to the form factors. 
The five-point form factor can be expanded in terms of four cubic diagrams shown in Figure~\ref{fig:F5treeLength4} as 
\begin{equation}\label{eq:5ptL4}
	\itbf{F} _{5}^{(\phi^4)}(1^{\phi},2^{\phi},3^{\phi},4^{\phi},5^{g})=\frac{C_1 N_1}{s_{15}}+\frac{C_2 N_2}{s_{25}}+\frac{C_3 N_3}{s_{35}}+\frac{C_4 N_4}{s_{45}}\,,
\end{equation}
with the color factors
\begin{equation}
\label{eq:5ptL4color}
C_1 = d^{a_2 a_3 a_4 b} f^{b a_1 a_5} \,, \ 
C_2 = d^{a_1 a_3 a_4 b} f^{b a_2 a_5} \,, \ 
C_3 = d^{a_1 a_2 a_4 b} f^{b a_3 a_5}  \,, \
C_4 = d^{a_1 a_2 a_3 b} f^{b a_4 a_5}  \,,
\end{equation}
and they satisfy
\begin{equation}\label{eq:phi4color}
C_1 + C_2 + C_3 + C_4 = 0 \,.
\end{equation}
Note that this is different from the usual Jacobi relation.
Remarkably, as we will see below, the CK duality and double copy still work nicely with these more general relations.

Imposing the CK duality, one requires that
\begin{equation}\label{eq:phi4num}
N_1^{\rm \scriptscriptstyle CK} + N_2^{\rm \scriptscriptstyle CK} + N_3^{\rm \scriptscriptstyle CK} + N_4^{\rm \scriptscriptstyle CK} = 0  \,.
\end{equation}
The equation \eqref{eq:phi4num} is saying that there are three independent numerators. 
One can take three of the numerators, for example $N_{1,2,4}$, as the basis numerators.
Alternatively, it turns out to be useful to choose a different basis as 
\begin{equation}\label{eq:phi4orderednum}
    N_a=N_1^{\scriptscriptstyle \rm CK}, \qquad N_b=N_2^{\scriptscriptstyle \rm CK}+N_1^{\scriptscriptstyle \rm CK},\qquad N_c=-N_4^{\scriptscriptstyle \rm CK}\,.
\end{equation}
And $N_3^{\scriptscriptstyle \rm CK}=-N_b+N_c$ from \eqref{eq:phi4num}. 
As we will see shortly, this special choice of numerators has the following advantages:  (i) they are related to color-ordered form factors via a \emph{symmetric} propagator matrix, and (ii) they have a ``closer" relationship to the ${\rm tr}(\phi^2)$ numerators.

Applying the color decomposition, one has
\begin{equation}\label{eq:5ptL4Neq}
\begin{pmatrix}
\mathcal{F}_{{\rm tr}(\phi^4),5}(1,5,3,4,2) \\
\mathcal{F}_{{\rm tr}(\phi^4),5}(1,3,5,4,2) \\
\mathcal{F}_{{\rm tr}(\phi^4),5}(1,3,4,5,2) 
\end{pmatrix} 
=
\begin{pmatrix}
{1\over s_{15}} +  {1\over s_{35}}&  -{1\over s_{35}}& 0 \\
-{1\over s_{35}} &  {1\over s_{35}}+ {1\over s_{45}} &  -{1\over s_{45}} \\
0 & - {1\over s_{45}}&  {1\over s_{25}} + {1\over s_{45}}
\end{pmatrix} 
\cdot
\begin{pmatrix}
N_a \\
N_b \\
N_c
\end{pmatrix} ,
\end{equation}
in which the propagator matrix, as well as its determinant, is
\begin{equation}
\label{eq:5ptL4PMatrix}
\Theta^{\mathcal{F}_{{\rm tr}(\phi^4)}}_{5}= 
\begin{pmatrix}
{1\over s_{15}} +  {1\over s_{35}}&  -{1\over s_{35}}& 0 \\
-{1\over s_{35}} &  {1\over s_{35}}+ {1\over s_{45}} &  -{1\over s_{45}} \\
0 & - {1\over s_{45}}&  {1\over s_{25}} + {1\over s_{45}}
\end{pmatrix} 
\,, \qquad
\det(\Theta^{\mathcal{F}_{{\rm tr}(\phi^4)}}_{5}) = -{ s_{1234} - q^2 \over s_{15} s_{25} s_{35} s_{45} } \,.
\end{equation}

As for the numerators, we can see again that using \eqref{eq:5ptL4Neq}, they are {uniquely} determined by the requirement of CK duality and are also manifestly gauge invariant%
\footnote{
Here we remind the reader that we use the $\tau$ factors to express the results concisely: $\tau_{ij}\equiv 2 p_i\cdot p_j$ and $\tau_{i,(j+..+k)}\equiv \tau_{ij}+..+\tau_{ik}$.}
\begin{align}
\begin{pmatrix}
N_a \\
N_b \\
N_c
\end{pmatrix}
= \frac{-1}{s_{1234}-q^2}\begin{pmatrix}
s_{15} \tau_{5,(2+3+4)} &  s_{15} \tau_{5,(2+4)} &  s_{15} s_{25} \\
s_{25} \tau_{5,(2+4)} & \tau_{5,(1+3)}  \tau_{5,(2+4)}  & s_{25} \tau_{5,(1+3)}  \\
s_{15} s_{25} & s_{25} \tau_{5,(1+3)} &  s_{25} \tau_{5,(1+3+4)}
\end{pmatrix} 
\begin{pmatrix}
\mathcal{F}_{4}^{(\phi^3)} (1,2,3,4,5) \\
\mathcal{F}_{4}^{(\phi^3)} (1,5,2,3,4) \\
\mathcal{F}_{4}^{(\phi^3)} (1,2,5,3,4)
\end{pmatrix} \,,
\end{align}
resulting in
\begin{equation}\label{eq:phi4newnums}
    N_a=-\frac{2\,{\rm f}_5^{\mu\nu}p_{1,\mu}p_{(2+3+4),\nu}}{s_{1234}-q^2}, \ \  N_b=-\frac{2\, {\rm f}_5^{\mu\nu}p_{(1+3),\mu}p_{(2+4),\nu}}{s_{1234}-q^2}, \ \  N_c=-\frac{2\,{\rm f}_5^{\mu\nu}p_{(1+3+4),\mu}p_{2,\nu}}{s_{1234}-q^2}\,.
\end{equation}
One can observe that these numerators are similar to \eqref{eq:4ptL3numsol}, which is not a coincidence and will be discussed in Section~\ref{ssec:universalnum}. 

\begin{figure}[t]
    \centering
 \includegraphics[height=0.11\linewidth]{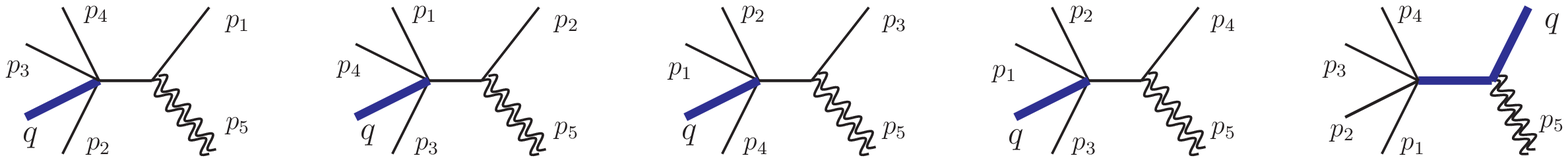}
    \caption{Feynman diagrams for the double copy of the five-point form factor of ${\rm tr}(\phi^4)$. {\color{red}}}
    \label{fig:G5treeLength4}
\end{figure}

We are now ready to make the double copy and obtain
\begin{equation}\label{eq:G5L4}
	\mathcal{G}_5^{[\phi^4]}= \frac{(N_{1}^{\rm \scriptscriptstyle CK})^2}{s_{15}}+\frac{(N_{2}^{\rm \scriptscriptstyle CK})^2}{s_{25}}+\frac{(N_{3}^{\rm \scriptscriptstyle CK})^2}{s_{35}} +\frac{(N_{4}^{\rm \scriptscriptstyle CK})^2}{s_{45}}  \,.
\end{equation}
The gauge invariance of the numerators immediately implies the diffeomorphism invariance of $\mathcal{G}_5^{[\phi^4]}$.
Moreover, on the new pole $s_{1234}{-}q^2$, ${\cal G}_5^{[\phi^4]}$ factorizes as
\begin{align}\label{eq:4ptL4GRAfactorize}
	\mathrm{Res}_{s_{1234}=q^2} \big[\mathcal{G}_5^{[\phi^4]} \big]
	& = \mathcal{G}_4^{[\phi^4]} (1^\phi,2^\phi,3^\phi,4^\phi) \  \mathcal{M}_{3}(\QQ_4^{S}, -q^{S}, 5^h)  \,,
\end{align}
which corresponds to a new diagram in Figure~\ref{fig:G5treeLength4}(the last one).

\paragraph{The $n$-point generalization.}

The generalization with more external legs for the $\phi^4$ form factor is straightforward, and below we summarize the main features.
\begin{enumerate}
[topsep=3pt,itemsep=1ex,partopsep=1ex,parsep=1ex]
    \item 
    To begin with, the propagator matrix is defined similarly to \eqref{eq:deftheta} as 
\begin{align}\label{eq:deftheta4}
    \Theta_n^{{\cal F}_{{\rm tr}(\phi^4)}}[\alpha &|\beta]
    =\int d^{D}x\  e^{\mathrm{i}q\cdot x} \langle 1^\phi, \ldots, 4^{\phi}, 5^{\Phi},\ldots n^\Phi | \mathcal{O}_{\phi^4}(x) | 0 \rangle \big|_{\operatorname{tr}_{\rm C}(\alpha) \operatorname{tr}_{\rm FL}(\beta)}\,,
\end{align}
with the operator $$\mathcal{O}_{\phi^4}=(1/3!)^2 d^{1234} 
{\tilde{d}}^{1234} \prod_{k=1}^{4}\phi^{a_k I_k}\,.$$
with $\tilde{d}^{1234}=\sum_{\sigma \in S_4/\mathbb{Z}_4} {\rm tr}_{\rm FL}(T^{I_{\sigma(1)}}T^{I_{\sigma(2)}}T^{I_{\sigma(3)}}T^{I_{\sigma(4)}})$. 
Specifically, the $\alpha/\beta$ involved here belong to $S_{n-2}/S_2$, which permute $\{3^{\phi},4^{\phi},5^{g},\ldots,n^{g}\}$ but leave the relative ordering of $\{3^{\phi},4^{\phi}\}$ invariant. 
Using this definition, one can check that the five-point propagator matrix is indeed the one in \eqref{eq:5ptL4PMatrix}.

    \item Next we come to color basis and ordered form factor basis.
The color factors satisfy two types of relations, the normal Jacobi relations and the operator-induced relations like \eqref{eq:phi4color}. 
We can specify one set of color bases, which are color factors of a special subset of cubic diagrams. These diagrams can be chosen as
(see also the explanation below \eqref{eq:3scalara})
\begin{equation}\label{eq:4scalara}
\Gamma[\{i\};\{j\};\{k\}]:=
    \begin{aligned}
        \includegraphics[width=0.4\linewidth]{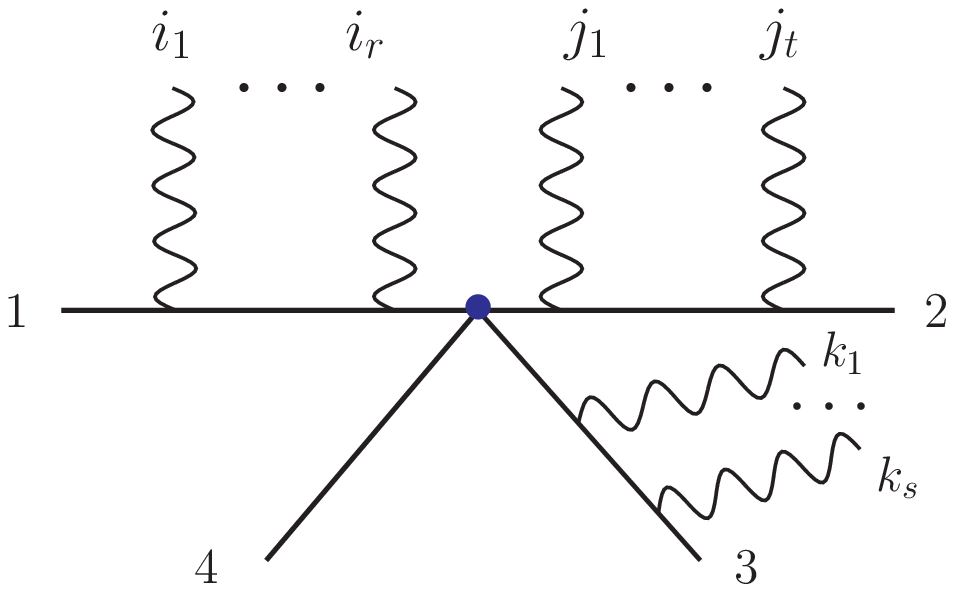}
    \end{aligned}
\end{equation}
where $r{+}s{+}t=n{-}4=n_g$, which is the total number of external gluons. 
The number of elements in the basis is $(n{-}2)!/2=(n_{g}{+}2)!/2!$, which is consistent with the size of the propagator matrix above.  

The basis of color-ordered form factors can be taken to be $\mathcal{F}_{{\rm tr}(\phi^4)}(1,\alpha,2)$ with $\alpha \in S_{n-2}/S_2$, the same as stated above. For $n=5$, we need $\mathcal{F}_{{\rm tr}(\phi^4)}(1,3,4,5,2)$, $\mathcal{F}_{{\rm tr}(\phi^4)}(1,3,5,4,2)$, and $\mathcal{F}_{{\rm tr}(\phi^4)}(1,5,3,4,2)$, exactly those in \eqref{eq:5ptL4Neq}. 

    \item Now we discuss the numerators. 
    In the CK-dual representation, we have for each trivalent diagram in \eqref{eq:4scalara} a numerator $N(\Gamma)$, which will form a basis of CK-dual numerators.
On the other hand, we will also look for an alternative set of numerators $N[\beta]$ satisfying 
\begin{equation}\label{eq:theta4FandN}
    \mathcal{F}_{{\rm tr}(\phi^4)}(1,\alpha,2)=\sum_{\beta \in S_{n-2}/S_2} \Theta_n^{{\cal F}_{{\rm tr}(\phi^4)}}[\alpha |\beta] \, N[\beta]\,,
\end{equation}
where the propagator matrix $\Theta_n^{{\cal F}_{{\rm tr}(\phi^4)}}$ is defined in \eqref{eq:deftheta4}.
Like the remarks given around \eqref{eq:phi4orderednum}, the advantages of introducing such numerators are: (i) they are related to the color-ordered form factors by the propagator matrix in \eqref{eq:deftheta4}, and (ii) their expressions can be given in a compact closed formula, 
which will be discussed in the next subsection. 
We will discuss more on these two sets of numerators (namely, $N[\Gamma]$ and $N[\beta]$) below.

    \item 
As a result, given $N[\beta]$, the length-4 double copy is formally defined  as
\begin{equation}
    \mathcal{G}_n^{[\phi^4]}=\sum_{\beta_{1,2}\in S_{n-2}/S_2} N[\beta_1] \Theta_n^{{\mathcal{F}_{{\rm tr}(\phi^4)}}}[\beta_1|\beta_2]  N[\beta_2]=\sum_{\beta\in S_{n-2}/S_2} N[\beta] \mathcal{F}_{{\rm tr}(\phi^4)}[\beta]\,.
\end{equation}
    
\end{enumerate}

\paragraph{Brief comment on the ${\rm tr}(\phi^m)$ form factors.} 

The generalization to the higher-length operators is straightforward, and we only briefly comment on a few main points.
The first point is that the operator itself has a fully symmetric color structure so that the minimal form factor is proportional to the color  factor $d^{a_1\cdots a_m}=\sum_{\sigma\in S_m/\mathbb{Z}_m}{\rm tr}\left(T^{a_{\sigma(1)}}\cdots T^{a_{\sigma(m)}}\right)$.
Moreover, the high-length operator  ${\rm tr}(\phi^m)$ induces an $m$-term color relation generalizing \eqref{eq:phi4color}
(and an $m$-term numerator relation generalizing \eqref{eq:phi4num}):
\begin{equation}\label{eq:phimcolor}
C_1 + C_2 + \ldots + C_m = 0 \,,
\end{equation}
with $C_1 = d^{a_2 a_3 .. a_{m} b} f^{b a_1 a_{m+1}}$ and $C_i$ by permutations.
Imposing these relations, as well as the dual Jacobi relations, will uniquely determine all the CK-dual numerators, and the subsequent double copy is straightforward. 
Importantly, the ``spurious"-type pole in the ${\rm tr}(\phi^m)$ case can only be $(\QQ_m{+}\text{gluon momenta})^2-q^2$. 

We will not go into details, since one can follow the above steps and consider the cubic diagram basis, the propagator matrix, the color-ordered form factor basis, and so on. Here we only mention that the number of elements in these bases are $(n-2)!/(m-2)!$, and the cubic diagram basis is similar to \eqref{eq:4scalara} having one scalar leg untouched and gluons distributed in all possible ways, and ordered form factors can be chosen as $\mathcal{F}(1,\alpha,2)$ with $\alpha\in S_{n-2}/S_{m-2}$ maintaining the relative position of $m$-scalars. 

\paragraph{Remark on the two kinds of numerators.} 
In the discussion above, we meet two different kinds of numerators: (1) The CK-dual numerators $N(\Gamma)$ defined for cubic diagrams $\Gamma$, which are obtained by imposing all the dual Jacobi relations and the high-length-operator-induced relations like \eqref{eq:phi4color}; and (2) the new $N[\beta]$ in \eqref{eq:theta4FandN}, which are, as we will explain, numerators defined according color orderings $\beta$.
We refer to the former as the \emph{cubic-diagram numerator} and the latter as the \emph{color-ordered numerator}.

The definition of  the color-ordered numerators deserves further clarification.
The basic idea is that we can naturally define a color factor $C[\beta]$ associated to an ordering $\beta$, which is nothing but the color trace. 
We can express the color traces in terms of cubic color factors $C(\Gamma)$, which are color factors of cubic diagrams. 
Within this relation, we directly replace both kinds of color factors with numerators $N[\beta]$ and $N[\Gamma]$. 
Then there comes a relation between two kinds of numerators. 

Concretely, we find special linear combinations of the (cubic) color basis $C(\Gamma)$, where $\Gamma$ are of the cubic diagrams in \eqref{eq:4scalara}, with linear coefficients $c_{\beta \Gamma}$, to give $C[\beta]$ as 
\begin{equation}\label{eq:orderedC}
   C[\beta]\equiv\sum_{\Gamma \in \eqref{eq:4scalara}} c_{\beta \Gamma} \, C(\Gamma)\,.
\end{equation}
The equations needed to constrain these coefficients $c_{\beta \Gamma}$ are  
\begin{equation}\label{eq:orderedcolorcondition}
    C[\beta]\big|_{{\rm tr}(1,\beta',2)}= \delta_{\beta \beta'}\, \text{ for } \beta' \in S_{n-2}/S_2\,.
\end{equation}
where $C|_{{\rm tr}}$ means for a color factor $C$ taking the coefficient of the corresponding trace in the subscript. 
Let us see how \eqref{eq:orderedcolorcondition} works as constraints on $c_{\beta \Gamma}$. 
Expanding the cubic color factor $C(\Gamma)$ will give a linear combination of all the color traces, and we are interested in the traces like ${\rm tr}(1,\beta',2)$. 
Among all these traces, only the coefficient of ${\rm tr}(1,\beta,2)$ is 1 for a specific $\beta$. 
This means $c_{\beta \Gamma}$ in \eqref{eq:orderedC} have to be special.
To see why $c_{\beta\Gamma}$ can be completely determined, we refer to the following counting argument:  \eqref{eq:orderedcolorcondition}  can be translated into $(n{-}2)!/2$ linear equations, and uniquely fix all the numbers $c_{\beta \Gamma}$ (recall that the number of $\beta$ is also $(n{-}2)!/2$). 

Then the CK-duality comes into play. 
Starting from \eqref{eq:orderedC}, we translate $C(\Gamma)$ to $N(\Gamma)$, the cubic-diagram numerators, and more importantly $C[\beta]$ to $N[\beta]$, which is the color-ordered numerator that we would like to define here:
\begin{equation}\label{eq:orderednum}
    N[\beta]= \sum_{\Gamma \in \eqref{eq:4scalara}} c_{\beta \Gamma } N(\Gamma) \,,
\end{equation}
where $N(\Gamma)$ is the CK-dual numerator of the cubic diagram $\Gamma$. 
Working with this new definition, it is easy to check that the $N_{a,b,c}$ used in \eqref{eq:phi4newnums} are simply
\begin{equation}
N_a =  N[1,3,4,5,2] \,,\qquad N_b = N[1,3,5,4,2] \,,\qquad N_c = N[1,5,3,4,2] \,.
\end{equation}

We give a comment about this definition \eqref{eq:orderednum}. 
Here we use the ${\rm tr}(\phi^4)$ form factor to illustrate the definition. 
When applying the definition to the ${\rm tr}(\phi^2)$ and ${\rm tr}(\phi^3)$ form factors, 
we reproduce the $N[\beta]$ that we used\footnote{The previous definitions of $N[\beta]$ for the ${\rm tr}(\phi^2)$ and ${\rm tr}(\phi^3)$ form factors are \eqref{eq:phi3nums} in this paper and (4.4) in the previous paper \cite{Lin:2022jrp}.} previously for the these operators. 
For instance, when looking back at the ${\rm tr}(\phi^3)$ numerators defined in \eqref{eq:phi3nums}, we find that we are secretly using a equation like \eqref{eq:orderednum}, 
requiring that for the ordering $\beta_0=\{1,i_1,\ldots,i_r,3,j_1,\ldots,j_t,2\}$, 
only the numerator of the half-ladder diagram $\Gamma_0= \hskip -5pt \begin{aligned}
    \includegraphics[width=0.24\linewidth]{figure/3scalars2.eps}
\end{aligned}$ 
contributes. In other words, only $c_{\beta_0\Gamma_0}$ is 1 and all other $c_{\beta_0\Gamma}$ vanish. 
This is why we end up with $N(\Gamma_0)=N[\beta_0]$ in \eqref{eq:phi3nums}, where we have a cubic-diagram numerator on one side but have a color-ordered numerator on the other side. 

We would like to introduce the \emph{commutator} of color-ordered numerators, which is understandable in the context of color-kinematics duality. 
The commutator can be defined for color factors. 
Color commutators can be expressed as 
\begin{equation}
    T^{a_i}T^{a_j}-T^{a_j}T^{a_i}=f^{ijk}T^{a_k} \,,
\end{equation}
and one can embed it into color traces as follows.
Consider two orderings only differ by swapping an adjacent pair of labels $\beta_1=\{1,\ldots,i,j,\ldots,2\}$ and $\beta_2=\{1,\ldots,j,i,\ldots,2\}$, and we define a commutator $[,]$ 
\begin{equation}
    {\rm tr}(1,\ldots,[i,j],\ldots,2)\equiv {\rm tr}(1,\ldots,i,j,\ldots,2)-{\rm tr}(1,\ldots,j,i,\ldots,2)=f^{ijk}{\rm tr}(1,\ldots,k,\ldots,2)\,.
\end{equation}
Furthermore, we take the numerator of $C[\beta_{1,2}]$ and define its commutator as 
\begin{align}
    C[1,\ldots,&[i,j],\ldots,2]\equiv C[\beta_1]-C[\beta_2]\\
    &=\big({\rm tr}(1,\ldots,i,j,\ldots,2)\pm \text{other traces}\big)-\big({\rm tr}(1,\ldots,j,i,\ldots,2)\pm \text{other traces}\big)\nonumber \\
    &=\big({\rm tr}(1,\ldots,[i,j],\ldots,2)\pm \text{other traces' commutators}\big)\nonumber \,.
\end{align}
For numerators, the CK-duality tells us that 
\begin{equation}\label{eq:numcomms}
    N[1,\ldots,[i,j],\ldots,2]\equiv N[\beta_1]-N[\beta_2]\,,
\end{equation}
which serves as the definition of commutators for color-ordered numerators. 

Finally, we give two remarks:

 $\bullet$ These two kinds of numerators can be transformed into each other. 
One relation we know is \eqref{eq:orderednum}, and the inverse relation has actually commutators \eqref{eq:numcomms} involved.  
This is a reasonable expectation because commutators of trace color factors create cubic color factors, and the color-kinematics duality is saying that the same commutators of color-ordered numerators should give cubic-diagram numerators.\footnote{In the Hopf algebra constructions in \cite{Brandhuber:2021bsf,Chen:2022nei} the commutators of ``pre-numerators" give cubic diagram numerators. Our definition of color-ordered numerators is essentially equivalent to the pre-numerators defined there. See further discussions below.}

 $\bullet$ These two kinds of numerators are preferred in different scenarios: the dual relations like Jacobi relations and operator-induced relations are naturally expressed in terms of the cubic-diagram numerators $N(\Gamma)$, while the color-ordered numerators $N[\beta]$ are closer related to the color-ordered form factors using equations like \eqref{eq:theta3FandN} and \eqref{eq:theta4FandN}.

\subsection{The universal master numerators}\label{ssec:universalnum}

In this subsection, we discuss the expressions of the numerators.
We will see that the numerators take some universal structures, which allow us to obtain compact closed formulas.
For simplicity, we discuss only scalar+gluon form factors in this subsection.

The story begins with observing that all the CK-dual numerators of form factors with only one gluon are taking the same form, see \eqref{eq:4ptL3numsol} and \eqref{eq:G5L4} in this paper and (3.12) and (6.13) in the previous paper \cite{Lin:2022jrp}. 
Basically, if, in a cubic diagram as below, the gluon separates the scalars into two groups denoted as $\phi_{L,1}\cdots \phi_{L,m}$  and $\phi_{R,1}\cdots \phi_{R,m^{\prime}}$ 
\begin{equation}
   \Gamma= \begin{aligned}
        \includegraphics[width=0.5\linewidth]{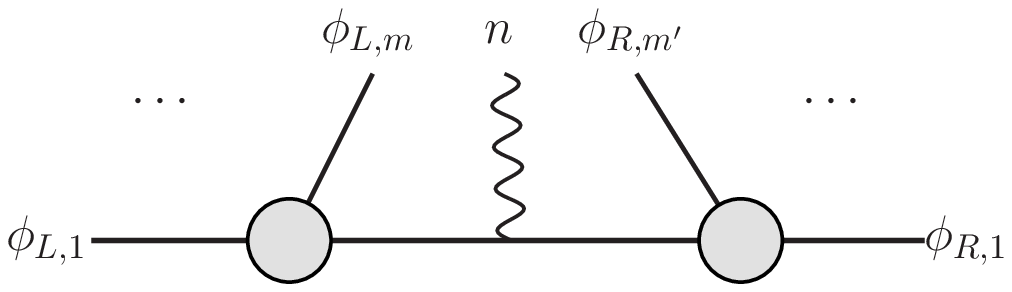}
    \end{aligned}\,,
\end{equation}
then the corresponding master numerator is roughly 
\begin{equation}
    N(\Gamma)= -\frac{2\, {\rm f}_{n}^{\mu\nu}\big(\sum_{i}p_{\phi_{L,i}}\big)_{\mu}\big(\sum_{j}p_{\phi_{R,j}}\big)_{\nu}}{\big(\sum_{i}p_{\phi_{L,i}}+\sum_{j}p_{\phi_{R,j}}\big)^2-q^2} \times \text{ possible flavor factors}\,,
\end{equation}
where we recall the definition of (linearized field strength)
\begin{equation}
{\rm f}_i^{\mu\nu}\equiv p_i^{\mu}\varepsilon_i^{\nu}-p_i^{\nu}\varepsilon_i^{\mu} \,.
\end{equation}
Such an expression is manifestly gauge invariant and has the expected spurious pole $(\sum p_\phi)^2$ $-q^2$. Moreover, it satisfies the corresponding dual Jacobi or beyond-Jacobi relations. 

The above formula can be generalized to cases with multiple external gluons and it will also present a property of universality,
meaning that: for a large class of  form factors, their numerators have a universal kinematical part as a function of momenta and polarizations, together with a flavor factor which can be trivially spelled out. 
Below we show this by establishing a map from the  numerators of the ${\rm tr}(\phi^2)$ $m$-scalar form factors to the numerators of the ${\rm tr}(\phi^m)$ $m$-scalar form factors. 
The reason why we consider these two theories is that the former is the theory that has been thoroughly discussed in the previous paper \cite{Lin:2022jrp}, while the latter has the simplest flavor structure, which is just the identity. 
Other form factors will be mentioned in the Appendix~\ref{ap:scalartheory}.

\subsubsection{Review of the ${\rm tr}(\phi^2)$ numerators}

Below we give a review of the ${\rm tr}(\phi^2)$ numerators, which was originally given in \cite{Lin:2022jrp}. Here we will focus more on the multi-scalar cases, 
and the theory is YMS theory with also scalar self-interaction $-\frac{\lambda_1}{3 !} \tilde{f}^{I J K} f^{abc} \phi^{I, a} \phi^{J, b} \phi^{K, c}$ in the Lagrangian. 
Note that the scalars carry flavor indices, and thus the numerators will contain a flavor factor.

Let us consider the color-ordered numerator $N[1^{\phi},\alpha(3^{\phi},..,m^{\phi},(m{+}1)^{g},..,n^{g}),2^{\phi}]$. 
The permutation $\alpha$ mixes the positions of gluons and scalars, and also changes the relative positions among the scalars or gluons.
One denotes the scalar ordering as $\mathbf{s}=\{1^{\phi},i_1,i_2,..,i_{m-2},2^{\phi}\}$ and the gluon ordering as $\mathbf{g}=\{j_1,j_2,..,j_{n-m}\}$, which are two ordered subsets of the combined ordering $\{1,\alpha,2\}$. 

\paragraph{The ${\rm tr}(\phi^2)$ numerator formula.} The numerator is a sum of contributions from all the ordered partitions, reading 
\begin{equation}\label{eq:BCJnumFgen}
	N[1^{\phi},\alpha(3^{\phi},..,m^{\phi},(m{+}1)^{g},..,n^{g}),2^{\phi}]={\rm tr}_{\rm FL}(t^{I_1}t^{I_{i_1}}\cdots t^{I_{i_{m-2}}}t^{I_{2}})\sum_{r=1}^{|\mathbf{g}|}\sum_{\tau\in \mathbf{P}_{\mathbf{g}}^{(r)} } n^{(1\alpha 2)}_{(\tau_1),\ldots,(\tau_r)}\,,
\end{equation}
where $\mathbf{P}_{\mathbf{g}}^{(r)}$ is the ordered partition as the collection of all possible ways dividing the gluon set $\mathbf{g}$ into $r$ (ordered) subsets, $|\mathbf{g}|$ is the length of the gluon set, $\tau$ is one particular way dividing $\mathbf{g}$ into $r$ subsets, and these $r$ subsets are $(\tau_1),\ldots,(\tau_r)$, and the flavor trace factor ${\rm tr}_{\rm FL}$ is defined according to the scalar ordering $\mathbf{s}=\{1^{\phi},i_1,i_2,\ldots,i_{m-2},2^{\phi}\}$. 

Next, we write down the concrete expression for $n^{(1\alpha 2)}_{(\tau_1),\ldots,(\tau_r)}$. 
The so-called ``musical diagrams" are needed here, of which the definition was given in \cite{Lin:2022jrp,Chen:2022nei}. 
The diagram, uniquely defined by the scalar ordering  $\mathbf{s}$, the partition $\tau$ of the gluon ordering $\mathbf{g}$ and combined ordering $\{1,\alpha,2\}$, can be drawn with the following two steps. 
(1) we put the scalars as well as the partitions of gluons $\tau_1$ to $\tau_{r}$ on different levels with $\tau_{i+1}$ above $\tau_i$ and $\tau_1$ above  $\mathbf{s}$.
(2) projecting the elements on all the levels onto the bottom line has to be exactly $\{1,\alpha,2\}$.
We use again the following example: the total ordering is 
$\{1^{\phi},5^{g},6^{g},4^{\phi},8^{g},3^{\phi},9^{g},7^{g},2^{\phi}\}$,
and $\mathbf{s}=\{1^{\phi},4^{\phi},3^{\phi},2^{\phi}\}$ while $\mathbf{g}=\{5^{g},6^{g},8^{g},9^{g},7^{g}\}$. We split $\mathbf{g}$ into two parts, $\tau_1=\{6^{g},9^{g},7^{g}\}$ and $\tau_2=\{5^{g},8^{g}\}$. Then the musical diagram is
\begin{equation}\label{eq:musical}
\begin{aligned}
    \includegraphics[width=0.55\linewidth]{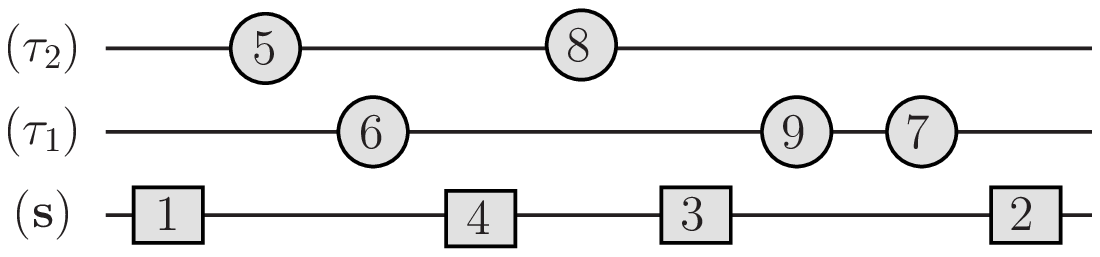}
\end{aligned}\,.
\end{equation}

Given the musical diagram, we can write down the following expression for $n^{(1\alpha 2)}_{(\tau_1),\ldots,(\tau_r)}$ as 
 \begin{align}\label{eq:num1alpha2}
 n^{(1\alpha 2)}_{(\tau_1),\ldots,(\tau_r)}={(-2)^r\prod\limits_{i=1}^r \Big(p_{\widetilde{\Xi}_L(\tau_i)}\cdot {\rm f}_{(\tau_i)}\cdot p_{\widetilde{\Xi}_R(\tau_i)}\Big) \over  (p_{\mathbf{s}}^2-q^2) (p_{\mathbf{s}\tau_1}^2-q^2)\cdots (p_{\mathbf{s}\tau_1\cdots \tau_{r-1}}^2-q^2)} \,,
\end{align}
where $\widetilde{\Xi}_{L}(\tau_i)$ is the collection of indices in the musical diagram,  lower-left to the first element of $\tau_{i}$,  and $\widetilde{\Xi}_{R}(\tau_i)$ is the collection of indices lower-right to the last element of $\tau_{i}$, including the scalars; also $\mathrm{f}_{(\tau_i)}$ is a contraction of field strength with two open indices $\rho,\sigma$, reading 
$$\mathrm{f}_{\tau_{i,1},\nu_1}^{\rho}\mathrm{f}_{\tau_{i,2},\nu_2}^{\nu_1}\cdots \mathrm{f}_{\tau_{i,r},\sigma}^{\mu_{r-1}} \,,$$
where $\tau_{i,k}$ is the $k$-th element of $\tau_i$ and $r$ is the length of $\tau_i$.
Taking the example in  \eqref{eq:musical}, we have 
\begin{equation}
	p_{\widetilde\Xi_{L}(\tau_1)}=p_1, \quad p_{\widetilde\Xi_{R}(\tau_1)}=p_2, \quad p_{\widetilde\Xi_{L}(\tau_2)}=p_1,\quad p_{\widetilde\Xi_{R}(\tau_2)}=p_2{+}p_{3}{+}p_{7}{+}p_9=p_{2379},
\end{equation}
so that
\begin{equation}
 \begin{aligned}
 n^{(156483972)}_{(679),(58)}&={(-2)^2 \Big(p_{1}\cdot {\rm f}_{697}\cdot p_{2}\Big) \Big(p_{1}\cdot {\rm f}_{58}\cdot p_{2379}\Big)  \over  (p_{1234}^2-q^2) (p_{1234679}^2-q^2)}\\
 &={(-2)^2 \Big(p_{1}\cdot {\rm f}_{6}\cdot{\rm f}_{9}\cdot {\rm f}_{7} \cdot p_{2}\Big) \Big(p_{1}\cdot {\rm f}_{5}\cdot {\rm f}_{8}\cdot (p_2+p_{3}+p_{7}+p_9)\Big)  \over  (p_{1234}^2-q^2) (p_{1234679}^2-q^2)}\,.
\end{aligned}
\end{equation}

\vskip 6pt

The above discussion gives us a clear presentation of the color-ordered numerators $N[1,\beta,2]$.
Now we discuss the cubic-diagram numerators $N[\Gamma]$. 
In particular, we focus on the numerators for the DDM basis diagrams.
Their expressions then take a simple form as
\begin{equation}\label{eq:phi2numc-1}
	 N\bigg( \hskip -3pt \begin{aligned}
        \includegraphics[width=0.22\linewidth]{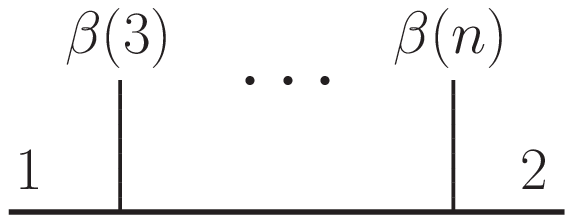}
        \end{aligned}\hskip -4pt\bigg)=\tilde{f}^{1i_1\mathrm{x}_1}\tilde{f}^{\mathrm{x}_1i_2\mathrm{x}_2}\cdots \tilde{f}^{\mathrm{x}_{m-3}i_{m-2}2}\sum_{r=1}^{|\mathbf{g}|}\sum_{\tau\in \mathbf{P}_{\mathbf{g}}^{(r)} } n^{(1\beta 2)}_{(\tau_1),\ldots,(\tau_r)} \,,
\end{equation}
in which the $n^{(1\beta 2)}_{(\tau_1),\ldots,(\tau_r)}$ are given in \eqref{eq:num1alpha2}. 
We also would like to remind the reader that $\tilde{f}$ is the flavor group structure constant, and $i_{k}$ are elements in the scalar ordering $\mathbf{s}$. 
The important point is that \eqref{eq:BCJnumFgen} and \eqref{eq:phi2numc-1} only differ by their flavor factor. 
Specifically, we define $\widetilde{N}$ to represent the kinematic part, so that 
\begin{equation}\label{eq:phi2numc}
\begin{aligned}
     N\bigg( \hskip -3pt \begin{aligned}
        \includegraphics[width=0.22\linewidth]{figure/mscalars.eps}
        \end{aligned}\hskip -4pt\bigg)&=\tilde{f}^{1i_1\mathrm{x}_1}\tilde{f}^{\mathrm{x}_1i_2\mathrm{x}_2}\cdots \tilde{f}^{\mathrm{x}_{m-3}i_{m-2}2} \widetilde{N}[1,\beta,2]\,,\\
       N[1^{\phi},\beta(3^{\phi},..,m^{\phi},(m{+}1)^{g},..,n^{g}),2^{\phi}]&={\rm tr}_{\rm FL}(t^{I_1}t^{I_{i_1}}\cdots t^{I_{i_{m-2}}}t^{I_{2}}) \widetilde{N}[1,\beta,2]\,,
\end{aligned}
\end{equation}
or 
\begin{equation}\label{eq:phi2numd}
    N\bigg( \hskip -3pt \begin{aligned}
        \includegraphics[width=0.22\linewidth]{figure/mscalars.eps}
        \end{aligned}\hskip -4pt\bigg)_{\text{flavor factor}\rightarrow 1}=N[1^{\phi},\beta(3^{\phi},..,m^{\phi},(m{+}1)^{g},..,n^{g}),2^{\phi}]_{\text{flavor factor}\rightarrow 1}\,.
\end{equation}

\vskip 6pt

We finally comment on the relation between the two kinds of numerators ($N[1,\beta,2]$ and $N[\Gamma]$), which was originally proposed in \cite{Brandhuber:2021bsf,Chen:2022nei}. 
In a word, the cubic-diagram numerators are (nested) commutators of the color-ordered numerators.
For example, we have 
\begin{equation}
    N\bigg( \hskip -3pt \begin{aligned}
        \includegraphics[width=0.22\linewidth]{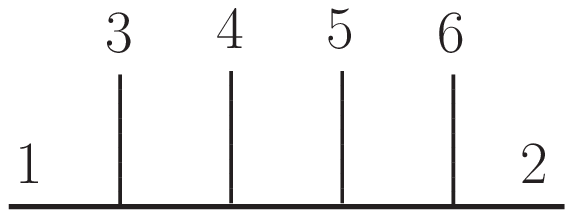}
        \end{aligned}\hskip -4pt\bigg)=N\big[{\color{blue}[[[[}1,3{\color{blue}]},4{\color{blue}]},5{\color{blue}]},6{\color{blue}]},2\big]\,,
\end{equation}
where the  small blue square brackets indicate commutators as in \eqref{eq:numcomms}. 
Such a nested commutator relation holds for any numerators of DDM basis diagrams, and one can prove the nested commutators give indeed \eqref{eq:phi2numc}, see \cite{Chen:2022nei}. 

The above discussions have two-fold importance to this paper:
\begin{enumerate}[topsep=3pt,itemsep=-1ex,partopsep=1ex,parsep=1ex]

    \item They are the reason why the map from the ${\rm tr}(\phi^2)$ numerators to the ${\rm tr}(\phi^m)$ numerators are reasonable.

    \item They provide an understanding of the relation between the numerators defined for orderings and for cubic diagrams.\footnote{The statement in this paper is a closely related  statement from the fundamental understanding given in \cite{Chen:2022nei}. We will not focus on the kinematic algebra as the authors did in \cite{Brandhuber:2021bsf,Chen:2022nei}, so we use this alternative understanding instead. }
\end{enumerate}
This ends our reviews of ${\rm tr}(\phi^2)$ numerators, and next we discuss their relations to the ${\rm tr}(\phi^m)$ numerators.

\subsubsection{Form factors of ${\rm tr}(\phi^3)$ with three scalars}

We first discuss the form factors with three external scalars, and we want to show that the ${\rm tr}(\phi^3)$ numerator can be simply obtained from the ${\rm tr}(\phi^2)$ ones via
\begin{equation}\label{eq:phi2tophi3}
    N^{\mathcal{F}_{{\rm tr}(\phi^3)}}=N^{\mathcal{F}_{{\rm tr}(\phi^2)}}|_{\text{flavor factor}\rightarrow 1}\,.
\end{equation}
We stress that here the numerators are with three external scalar particles plus an arbitrary number of gluons. Also, since the ${\rm tr}(\phi^3)$ numerators considered here have a trivial overall flavor factor, we simply set it to be one.

The ${\rm tr}(\phi^2)$ numerators are given as a special case of \eqref{eq:BCJnumFgen}:
\begin{align}\label{eq:FFT2}
    &N^{\mathcal{F}_{{\rm tr}(\phi^2)}}[1^{\phi},4^{g},\ldots, k^{g},3^{\phi},(k{+}1)^{g},\ldots,n^{g},2^{\phi}]={\rm tr}_{\rm FL}(t^{I_1}t^{I_{3}}t^{I_{2}})  \widetilde{N}[1,4,\ldots, k,3,(k{+}1),\ldots,n,2]\,,\nonumber \\
    &\widetilde{N}[1,4,\ldots, k,3,(k{+}1),\ldots,n,2]=\sum_{r=1}^{n-3}
	\sum_{\tau\in \mathbf{P}_{\mathbf{g}}^{(r)}}
	{(-2)^r\prod\limits_{i=1}^r \Big(p_{\Phi_{L}\Xi_L(i)}\cdot {\rm f}_{(\tau_i)}\cdot p_{\Phi_{R}\Xi_R(i)}\Big)\over  (p_{123}^2{-}q^2) (p_{123\tau_1}^2{-}q^2)\cdots (p_{123\tau_1\cdots \tau_{r-1}}^2{-}q^2)} \,,
\end{align}
where $\mathbf{g}$ is the gluon set $\{4^g,\ldots,n^g\}$. 
A word about notation: comparing to \eqref{eq:num1alpha2}, here we denote $p_{\widetilde{\Xi}_L(\tau_i)}$ as $p_{\Phi_L\Xi_L(i)}$. Roughly speaking, $\Xi_{L,R}$ is deleting the scalars from $\widetilde{\Xi}_{L,R}$.
$\Xi_{L,R}(i)$ denote two special subsets of  $\Xi(i){\equiv}\{\tau_1,\tau_2,\ldots, \tau_{i-1}\}$: $\Xi_{L}(i)$ contain the elements in $\Xi(i)$ which are smaller than the first element in $\tau_{i}$;  $\Xi_{R}(i)$ contain the elements in $\Xi(i)$ which are bigger than the last element in $\tau_{i}$. 
The rules of assigning proper scalars to $\Phi_L$ and $\Phi_R$ are: 
\begin{align}
& \textrm{if  min}(\tau_i)<k, \  \Phi_{L}=\{1^{\phi}\}, \qquad \textrm{otherwise } \Phi_{L}=\{1^{\phi},3^{\phi}\};  \nonumber\\
& \textrm{if max}(\tau_i)> k, \  \Phi_{R}=\{2^{\phi}\}, \qquad \textrm{otherwise } \Phi_{R}=\{2^{\phi},3^{\phi}\}. \nonumber
\end{align}

Then we claim that exact the same formula \eqref{eq:FFT2} can be used to describe the ${\rm tr}(\phi^3)$ numerators as in \eqref{eq:phi2tophi3}:
\begin{equation}\label{eq:phi2eqphi3}
    N^{\mathcal{F}_{{\rm tr}(\phi^3)}}[1^{\phi},4^{g},\ldots, k^{g},3^{\phi},(k{+}1)^{g},\ldots,n^{g},2^{\phi}]=\widetilde{N}[1,4,\ldots, k,3,(k{+}1),\ldots,n,2]\,.
\end{equation}
This claim is not strange because these two numerators should have the same structure: first, they should both have the same spurious poles after double copy, which are $s_{123\cdots}{-}q^2$; second, they should be both manifestly gauge invariant; most importantly, they have to satisfy dual Jacobi relations.\footnote{Technically, the dual Jacobi relations are suitably expressed for the cubic-diagram numerators. But since the two kinds of numerators are expressionally equivalent, it does not really matter to distinguish them.}
The last requirement is highly non-trivial. 
One can check that the particular form \eqref{eq:FFT2} indeed satisfies all the desired properties (see \cite{Lin:2022jrp}).

So far we consider the color-ordered numerators as in \eqref{eq:FFT2} and \eqref{eq:phi2eqphi3}. In this case, it is also straightforward to work out the numerators $N[\Gamma]$ as associated with cubic diagrams as
\begin{equation}\label{eq:phi2eqphi32}
\begin{aligned}
        N^{\mathcal{F}_{{\rm tr}(\phi^2)}}\bigg( \hskip -3pt \begin{aligned}
        \includegraphics[width=0.28\linewidth]{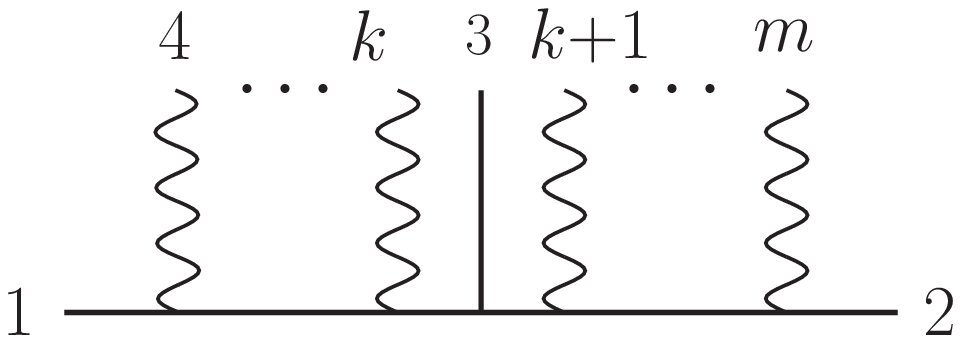}
        \end{aligned}\hskip -6pt\bigg)&=\tilde{f}^{123}\widetilde{N}[1,4,\ldots,k,3,k{+}1,\ldots,m,2] \,,\\
        N^{\mathcal{F}_{{\rm tr}(\phi^3)}}\bigg( \hskip -3pt \begin{aligned}
        \includegraphics[width=0.28\linewidth]{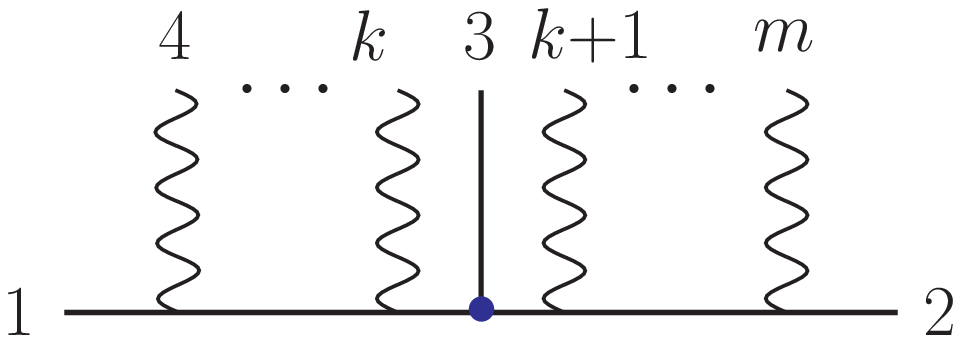}
        \end{aligned}\hskip -6pt\bigg)&=\widetilde{N}[1,4,\ldots,k,3,k{+}1,\ldots,m,2] \,.
\end{aligned}
\end{equation}
Here the first line of \eqref{eq:phi2eqphi32} is true because we have used the relation \eqref{eq:phi2numd}, saying that ignoring the flavor factor, for the ${\rm tr}(\phi^2)$ form factors, the color-ordered numerator and the cubic-diagram numerator (of the DDM basis of course) are the same.\footnote{There is another case worth considering: We take the theory to be bi-adjoint $\phi^3$ coupled to Yang-Mills, and the operator to be either  $\mathcal{O}_{\phi^3}$ before in \eqref{eq:deftheta} or the three-scalar interaction vertex (with off-shell momentum $q$). Then the flavor factor can be either $\tilde{d}^{123}$ or $\tilde{f}^{123}$, see the Appendix~\ref{ap:scalartheory} for details.}
As for the second line, it comes from straightforward calculations. 

From \eqref{eq:phi2eqphi32}, \eqref{eq:FFT2} and \eqref{eq:phi2eqphi3}, one can see clearly for both the two kinds of numerators, the simple relation \eqref{eq:phi2tophi3} holds. 
We comment that the three-scalar case is particularly simple so that there is no need to distinguish the two types of numerators (note that \eqref{eq:phi2eqphi3} and the second line of \eqref{eq:phi2eqphi32} are the same). 
More scalars will bring more intriguing structures, as will be clarified in the next subsection.

\subsubsection{Form factors of ${\rm tr}(\phi^m)$ with four or more scalars}
For form factors with four scalars, again we want to confirm that 
\begin{equation}\label{eq:phi2tophi4}
    N^{\mathcal{F}_{{\rm tr}(\phi^4)}}=N^{\mathcal{F}_{{\rm tr}(\phi^2)}}|_{\text{flavor factor}\rightarrow 1} \,, \quad \text{the case with four external scalars}\,. 
\end{equation}
To do this, we take the strategy that assuming \eqref{eq:phi2tophi4} is true, we can get the (conjectured) ${\rm tr}(\phi^4)$ numerators from the ${\rm tr}(\phi^2)$ ones; then we show that these conjectured numerators satisfy all the desired relations and they are indeed the correct numerators that we are looking for.

\paragraph{Color-ordered numerators.}

It is easier to focus first on the color-ordered numerators $N[1,\beta,2]$. 
By specializing \eqref{eq:BCJnumFgen} to the four-point case, we have 
\begin{equation}\label{eq:4scalarnum0}
\begin{aligned}
    N^{{\rm tr}(\phi^2)}[1^{\phi},5^{g},&\ldots,k_1^{g},3^{\phi}, (k_1{+}1)^{g},\ldots,k_2^{g},4^{\phi}, (k_2{+}1)^{g},\ldots,n^{g},2^{\phi}]\\
    &=  {\rm tr}_{\rm FL}(T^{I_1}T^{I_3}T^{I_4}T^{I_2})
    \widetilde{N}[1,5,\ldots,k_1,3, (k_1{+}1),\ldots,k_2,4, (k_2{+}1),\ldots,n,2] \,, \\
     \widetilde{N}[1,5,\ldots,k_1,& 3, (k_1{+}1),\ldots,k_2,4, (k_2{+}1),\ldots,n,2] \\
     &=
    \sum_{r=1}^{n-4}\sum_{\mathbf{P}_{\tau}^{(r)}} {(-2)^r\prod\limits_{i=1}^r \Big(p_{\Phi_{L}\Xi_L(i)}\cdot {\rm f}_{(\tau_i)}\cdot p_{\Phi_{R}\Xi_R(i)}\Big)\over  (p_{1234}^2{-}q^2) (p_{1234\tau_1}^2{-}q^2)\cdots (p_{1234\tau_1\cdots \tau_{r-1}}^2{-}q^2)}\,.
\end{aligned}
\end{equation}
Here $\mathbf{s}=\{1,3,4,2\}$ is the ordered scalar set and 
the rules for the $\Phi$ functions are 
\begin{equation}
\begin{aligned}
    & \Phi_L=p_1 \text{ if } \text{min}(\tau_i)\leq k_1, \quad \qquad   \Phi_R=p_{24} \text{ if } m<\text{min}(\tau_i)\leq r, \\
    &\Phi_L=p_{134} \text{ if } k_2<\text{min}(\tau_i), \ \quad \quad \Phi_R=p_{234} \text{ if } \text{max}(\tau_i)\leq k_1, \\
    &\Phi_L=p_{13} \text{ if } k_1<\text{max}(\tau_i)\leq k_2, \ \Phi_R=p_4 \text{ if } r<\text{max}(\tau_i). 
\end{aligned}
\end{equation}
Then, assuming \eqref{eq:phi2tophi4}, the ${\rm tr}(\phi^4)$ numerator should be given by the  ${\rm tr}(\phi^2)$ ones via
\begin{equation}\label{eq:4scalarnumx}
       N^{{\rm tr}(\phi^4)}[1,\ldots,2]=N^{{\rm tr}(\phi^2)}[1,\ldots,2]\big|_{\text{flavor factor}\rightarrow 1}\,.
\end{equation}
One can check the above equation gives the right numerators in the sense that they satisfy the relation \eqref{eq:theta4FandN} involving also propagator matrix and ordered form factors. 

\paragraph{Cubic-diagram numerators.}

When it comes to the cubic-diagram numerators $N(\Gamma)$, things are a bit more complicated. The diagrams for ${\rm tr}(\phi^2)$ and ${\rm tr}(\phi^4)$ are different, because of the four-point vertex associated with the operator for the latter. 
Diagrammatically, we need to shrink propagators to create a quadruple vertex, and we expect the following relation to hold 
\begin{equation}\label{eq:phi2tophi4cubic}
    N^{\mathcal{F}_{{\rm tr}(\phi^4)}}
    \Bigg( \hskip -3pt \begin{aligned}
        \includegraphics[width=0.17\linewidth]{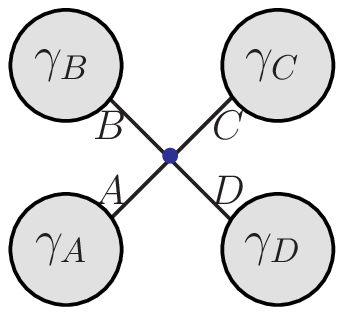}
        \end{aligned}\hskip -4pt\Bigg)
    = 
    N^{\mathcal{F}_{{\rm tr}(\phi^2)}}
    \Bigg( \hskip -3pt \begin{aligned}
        \includegraphics[width=0.17\linewidth]{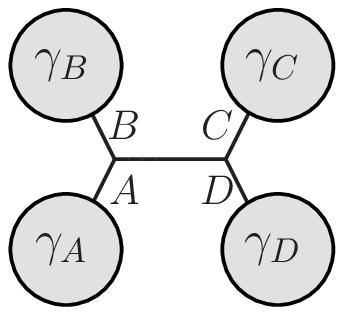}
        \end{aligned}\hskip -4pt\Bigg)_{\text{flavor factor}\rightarrow 1}\,,
\end{equation}
where $A,B,C,D$ are scalar lines and $\gamma_{A,B,C,D}$ denote the remaining subdiagrams. 

\vskip 6pt

One question naturally arises from \eqref{eq:phi2tophi4cubic}: there are three different ways of shrinking a numerator to get  the LHS diagram; will they lead to the same answer?
The answer is positive. 
We have the following relation 
\begin{equation}\label{eq:stu}
\begin{aligned}
    N^{\mathcal{F}_{{\rm tr}(\phi^2)}}
    \Bigg( \hskip -3pt \begin{aligned}
        \includegraphics[width=0.17\linewidth]{figure/tree4CK1.eps}
        \end{aligned}\hskip -3pt\Bigg)_{\text{flavor factor}\rightarrow 1}
        = 
        N^{\mathcal{F}_{{\rm tr}(\phi^2)}}
    \Bigg( \hskip -3pt \begin{aligned}
        \includegraphics[width=0.17\linewidth]{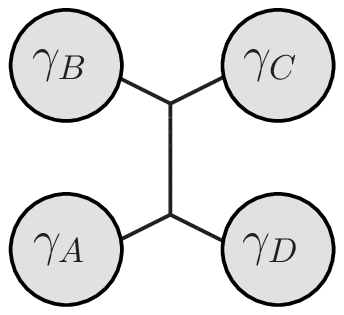}
        \end{aligned}\hskip -3pt\Bigg)_{\text{flavor factor}\rightarrow 1}\\
        =N^{\mathcal{F}_{{\rm tr}(\phi^2)}}
    \Bigg( \hskip -3pt \begin{aligned}
        \includegraphics[width=0.17\linewidth]{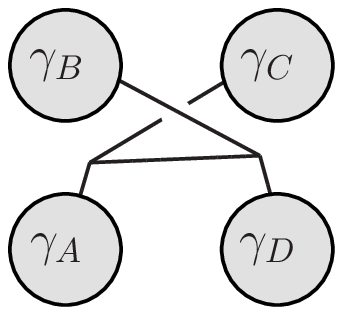}
        \end{aligned}\hskip -3pt\Bigg)_{\text{flavor factor}\rightarrow 1}\,, 
\end{aligned}
\end{equation}
which is not surprising after realizing the following fact: 
if we have 
\begin{equation}
    N_s-N_u=N_t,\qquad N_s\big|_{\text{flavor factor}}-N_u\big|_{\text{flavor factor}}=N_t\big|_{\text{flavor factor}}\,,
\end{equation}
then it is reasonable to expect (which is just \eqref{eq:stu})
\begin{equation}
    N_s\big|_{\text{flavor factor}\rightarrow 1}=N_u\big|_{\text{flavor factor}\rightarrow 1}=N_t\big|_{\text{flavor factor}\rightarrow 1}\,,
\end{equation}
where 
\begin{equation}
\begin{aligned}
    N_s&=N^{\mathcal{F}_{{\rm tr}(\phi^2)}}
    \Bigg( \hskip -3pt \begin{aligned}
        \includegraphics[width=0.17\linewidth]{figure/tree4CK1.eps}
        \end{aligned}\hskip -3pt\Bigg)\,,\quad 
    N_s\big|_{\text{flavor factor}}=\tilde{f}^{AB\text{x}}\tilde{f}^{\text{x}CD}\,,\\
    N_t&=N^{\mathcal{F}_{{\rm tr}(\phi^2)}}
    \Bigg( \hskip -3pt \begin{aligned}
        \includegraphics[width=0.17\linewidth]{figure/treeCK2.eps}
        \end{aligned}\hskip -3pt\Bigg)\,,\quad  
    N_t\big|_{\text{flavor factor}}=\tilde{f}^{DA\text{x}}\tilde{f}^{\text{x}BC}\,,\\
    N_u&=N^{\mathcal{F}_{{\rm tr}(\phi^2)}}
    \Bigg( \hskip -3pt \begin{aligned}
        \includegraphics[width=0.17\linewidth]{figure/treeCK4.eps}
        \end{aligned}\hskip -3pt\Bigg)\,,\quad 
    N_u\big|_{\text{flavor factor}}=\tilde{f}^{AC\text{x}}\tilde{f}^{\text{x}BD}\,.\\
\end{aligned}       
\end{equation}

To make the above argument less abstract, we consider the following example. 
If there is no gluons between 3,4, then $\Phi_{L}(i)$ is $p_1$ or $p_1{+}(p_3{+}p_4)$; $\Phi_{R}(i)$ is $p_2$ or $p_2{+}(p_3{+}p_4)$.
Importantly, $p_{3,4}$ are always combined together. 
This means, following from checking the concrete expressions,
\begin{equation}
    \tilde{N}[1,\ldots,3,4,\ldots,2]=\tilde{N}[1,\ldots,4,3,\ldots,2]\equiv \tilde{N}_0\,,
\end{equation}
where we define $\tilde{N}_0$ just to simplify notations, and 
\begin{equation}
\begin{aligned}
   N^{\mathcal{F}_{{\rm tr}(\phi^2)}} \bigg(\hskip -3pt \begin{aligned}
        \includegraphics[width=0.2\linewidth]{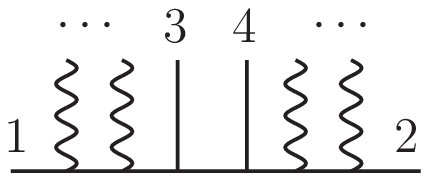}
    \end{aligned}\hskip -5pt\bigg)&=\tilde{f}^{13\text{x}}\tilde{f}^{\text{x}42}\widetilde{N}_0,\ 
    N^{\mathcal{F}_{{\rm tr}(\phi^2)}} \bigg(\hskip -3pt \begin{aligned}
        \includegraphics[width=0.2\linewidth]{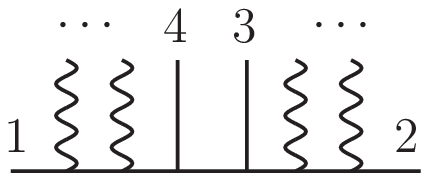}
    \end{aligned}\hskip -5pt\bigg)=\tilde{f}^{14\text{x}}\tilde{f}^{\text{x}32}\widetilde{N}_0\,.
\end{aligned}
\end{equation}
Moreover, one can also derive 
\begin{equation}
     N^{\mathcal{F}_{{\rm tr}(\phi^2)}} \bigg(\hskip -4pt \begin{aligned}
        \includegraphics[width=0.2\linewidth]{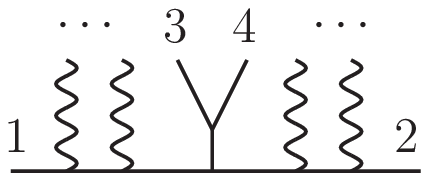}
    \end{aligned}\hskip -8pt\bigg)=\tilde{f}^{21\text{x}}\tilde{f}^{\text{x}34}\widetilde{N}_0
\end{equation}
from the dual Jacobi relations satisfied by these three numerators. Thus, we can see
\begin{equation}\label{eq:4scalarsnumstu}
\begin{aligned}
    N^{\mathcal{F}_{{\rm tr}(\phi^2)}} \bigg(\hskip -3pt \begin{aligned}
        \includegraphics[width=0.2\linewidth]{figure/4scalarsbsub6.eps}
    \end{aligned}\hskip -5pt\bigg)_{\text{flavor factor}\rightarrow1}
    &=N^{\mathcal{F}_{{\rm tr}(\phi^2)}} \bigg(\hskip -3pt \begin{aligned}
        \includegraphics[width=0.2\linewidth]{figure/4scalarsbsub7.eps}
    \end{aligned}\hskip -5pt\bigg)_{\text{flavor factor}\rightarrow1}\\
    &=N^{\mathcal{F}_{{\rm tr}(\phi^2)}} \bigg(\hskip -3pt \begin{aligned}
        \includegraphics[width=0.2\linewidth]{figure/4scalarsbsub8.eps}
    \end{aligned}\hskip -5pt\bigg)_{\text{flavor factor}\rightarrow1}
\end{aligned}
\end{equation}
which is an example of \eqref{eq:stu}.
All the three terms in \eqref{eq:4scalarsnumstu} give $N^{\mathcal{F}_{{\rm tr}(\phi^4)}} \bigg(\hskip -3pt \begin{aligned}
\includegraphics[width=0.2\linewidth]{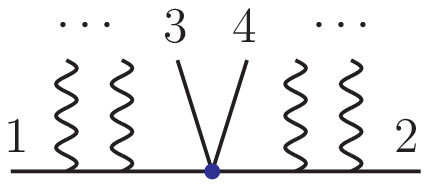}
    \end{aligned}\hskip -5pt\bigg)$.
For general cases in which each of $\gamma_{A,B,C,D}$ in \eqref{eq:stu} may contain some gluons, one can show that \eqref{eq:stu} is true by explicit calculations.
Now we can conclude that \eqref{eq:phi2tophi4cubic} is a consistent map with no ambiguity. 

\vskip 6pt

Then we verify that the ${\rm tr}(\phi^4)$ numerator given in \eqref{eq:phi2tophi4cubic} satisfies the dual relations.
If \eqref{eq:phi2tophi4cubic} is indeed true, then it should automatically be consistent with the Jacobi relations and the operator-induced relations. 
To check these relations, working with concrete examples and expressions is one way out, but we can do better. 
We will show that, given the dual Jaocbi-relations satisfied by the ${\rm tr}(\phi^2)$ numerators on the RHS of \eqref{eq:phi2tophi4}, one can prove that the numerators on the LHS of \eqref{eq:phi2tophi4} indeed satisfy all the required relations.

Let us start by considering the following two Jacobi relations satisfied by the ${\rm tr}(\phi^2)$ numerators 
\begin{equation}
\begin{aligned}
     &N^{\mathcal{F}_{{\rm tr}(\phi^2)}}\Bigg(\hskip -3pt \begin{aligned}
     \includegraphics[width=0.17\linewidth]{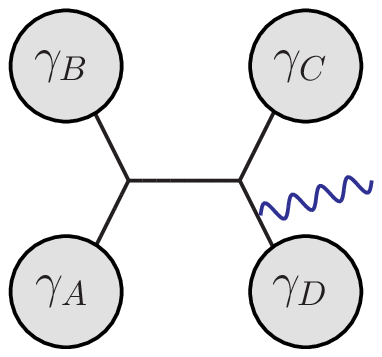}
     \end{aligned}\hskip -3pt \Bigg)-
     N^{\mathcal{F}_{{\rm tr}(\phi^2)}}\Bigg(\hskip -3pt\begin{aligned}
     \includegraphics[width=0.17\linewidth]{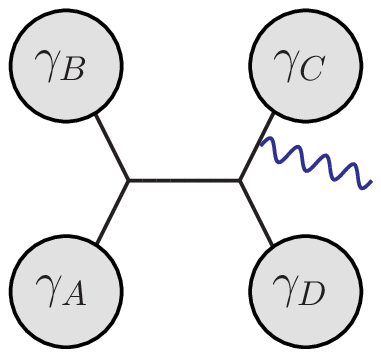}
     \end{aligned}\hskip -3pt\Bigg)=
     N^{\mathcal{F}_{{\rm tr}(\phi^2)}}\bigg(\begin{aligned}
     \includegraphics[width=0.17\linewidth]{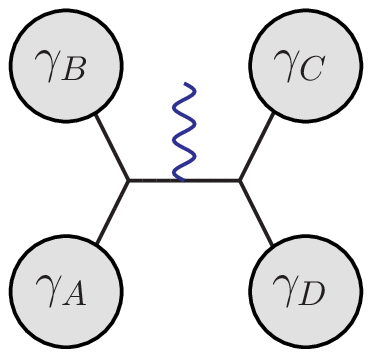}
     \end{aligned}\hskip -3pt \Bigg), \\
     & N^{\mathcal{F}_{{\rm tr}(\phi^2)}}\Bigg(\hskip -3pt\begin{aligned}
     \includegraphics[width=0.17\linewidth]{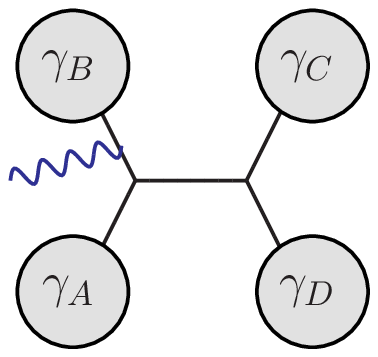}
     \end{aligned}\hskip -3pt \Bigg)-
     N^{\mathcal{F}_{{\rm tr}(\phi^2)}}\Bigg(\hskip -3pt\begin{aligned}
     \includegraphics[width=0.17\linewidth]{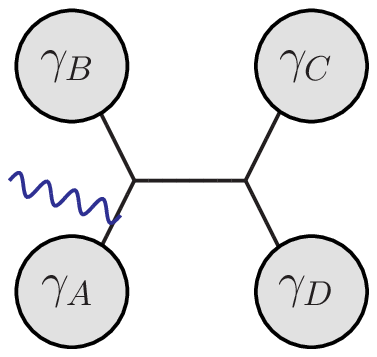}
     \end{aligned}\hskip -3pt \Bigg)=
     N^{\mathcal{F}_{{\rm tr}(\phi^2)}}\Bigg(\hskip -3pt\begin{aligned}
     \includegraphics[width=0.17\linewidth]{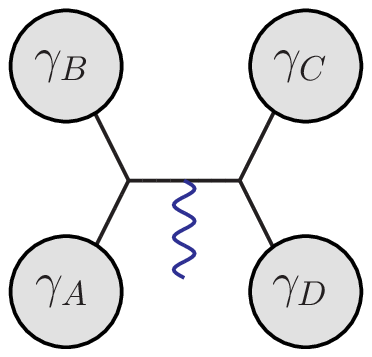}
     \end{aligned}\hskip -3pt \Bigg).
\end{aligned}
\end{equation}
After summing up the above two equations, what we get is a four-term identity 
\begin{equation}\label{eq:phi24scalarCK}
\begin{aligned}
     &N^{\mathcal{F}_{{\rm tr}(\phi^2)}}\Bigg(\hskip -3pt\begin{aligned}
     \includegraphics[width=0.17\linewidth]{figure/4scalarsjacobi2.eps}
     \end{aligned}\hskip -3pt \Bigg)-
     N^{\mathcal{F}_{{\rm tr}(\phi^2)}}\Bigg(\hskip -3pt\begin{aligned}
     \includegraphics[width=0.17\linewidth]{figure/4scalarsjacobi.eps}
     \end{aligned}\hskip -3pt \Bigg)=\\
     &\hskip 3cm 
     N^{\mathcal{F}_{{\rm tr}(\phi^2)}}\Bigg(\hskip -3pt\begin{aligned}
     \includegraphics[width=0.17\linewidth]{figure/4scalarsjacobi3.eps}
     \end{aligned}\hskip -3pt \Bigg)-
     N^{\mathcal{F}_{{\rm tr}(\phi^2)}}\Bigg(\hskip -3pt\begin{aligned}
     \includegraphics[width=0.17\linewidth]{figure/4scalarsjacobi3b.eps}
     \end{aligned}\hskip -3pt \Bigg)\,.
\end{aligned}
\end{equation}

Such a four-term identity for ${\rm tr}(\phi^2)$ numerators is meaningful because they correspond to the operator-induced relation for high-length operators such as in \eqref{eq:phi4num}.
To show this, we take the map from ${\rm tr}(\phi^2)$ numerators to the ${\rm tr}(\phi^4)$ numerators. 
Note that the flavor structure of all the numerators involved is the same so that  one can cancel the same flavor factor on both sides of \eqref{eq:phi24scalarCK}. This gives us
\begin{equation}\label{eq:phi24scalarCK2}
\begin{aligned}
     &N^{\mathcal{F}_{{\rm tr}(\phi^4)}}\Bigg(\hskip -3pt\begin{aligned}
     \includegraphics[width=0.17\linewidth]{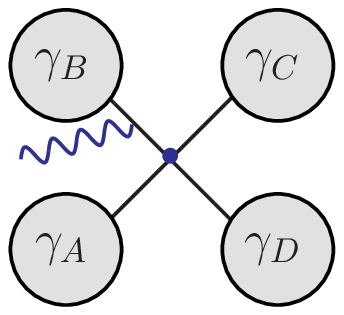}
     \end{aligned}\hskip -3pt \Bigg)-
     N^{\mathcal{F}_{{\rm tr}(\phi^4)}}\Bigg(\hskip -3pt\begin{aligned}
     \includegraphics[width=0.17\linewidth]{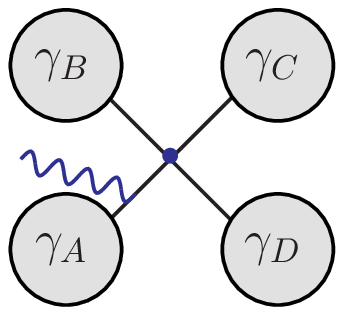}
     \end{aligned}\bigg)=\\
     &\hskip 3cm 
     N^{\mathcal{F}_{{\rm tr}(\phi^4)}}\Bigg(\hskip -3pt\begin{aligned}
     \includegraphics[width=0.17\linewidth]{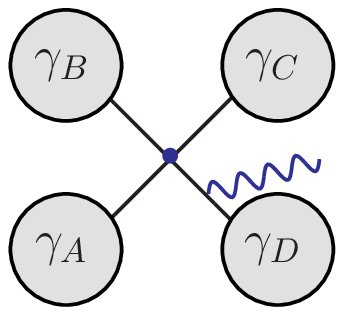}
     \end{aligned}\hskip -3pt \Bigg)-
     N^{\mathcal{F}_{{\rm tr}(\phi^4)}}\Bigg(\hskip -3pt\begin{aligned}
     \includegraphics[width=0.17\linewidth]{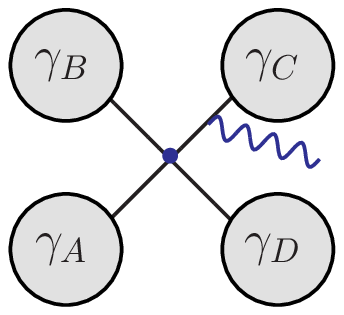}
     \end{aligned}\hskip -3pt \Bigg),
\end{aligned}
\end{equation}
which is exactly the four-term equation describing the ${\rm tr}(\phi^4)$ operator-induced numerators relation.\footnote{In \eqref{eq:phi24scalarCK2} and its derivation, we have used a fact that flipping a gluon against a subdiagram gives a minus sign.} 

We also mention that the ordinary Jacobi relations of the ${\rm tr}(\phi^4)$ numerators trivially stem from the Jacobi relations of the ${\rm tr}(\phi^4)$ numerators, and we will omit the details here. 

To summarize, we have shown that it is convincing that  the map \eqref{eq:phi2tophi4} is valid, because it is consistently defined, gives the right relation to amplitudes, and the ${\rm tr}(\phi^4)$ numerators inherit all the desired structures and relations from  the ${\rm tr}(\phi^2)$ ones.

\paragraph{Further discussion on the relation between two types of numerators.}

 The final point that we want to make is to understand the relation between the two kinds of numerators as an application of the map \eqref{eq:phi2tophi4}. We want to establish such a ``commutative diagram" as
\begin{equation}\label{eq:comdiagram}
\begin{aligned}
     N^{\mathcal{F}_{\rm tr}(\phi^2)}[1,\beta,2]\qquad  & \xrightarrow{\text{flavor factor}\rightarrow 1 }& N^{\mathcal{F}_{\rm tr}(\phi^4)} [1,\beta,2]\qquad \\
    \Downarrow \ {[,]} \qquad\qquad   & & {\color{red}{\wideDownarrow}} \ {\color{red}[,]} \qquad \qquad   \\
    N^{\mathcal{F}_{{\rm tr}(\phi^2)}}
    \Bigg( \hskip -3pt \begin{aligned}
        \includegraphics[width=0.17\linewidth]{figure/tree4CK1.eps}
        \end{aligned}\hskip -3pt\Bigg) & \xrightarrow{\text{flavor factor}\rightarrow 1 }  &   N^{\mathcal{F}_{{\rm tr}(\phi^4)}}
    \Bigg( \hskip -3pt \begin{aligned}
        \includegraphics[width=0.17\linewidth]{figure/tree4CK1.eps}
        \end{aligned}\hskip -3pt\Bigg)
\end{aligned}
\end{equation}
In this diagram, the top line contains the color-ordered numerators, while the bottom line has the cubic-diagram numerators.
The horizontal arrows have been discussed above, and the down arrows represent \textbf{commutators}. 

We want to use \eqref{eq:comdiagram} to understand better the red down-arrow, which is the relation between the two kinds of ${\rm tr}(\phi^4)$ numerators.  
We already discussed the other three arrows in \eqref{eq:comdiagram}, and the last red one will become self-evident after getting the rest three clarified. 

The left down-arrow corresponds to the action of performing commutators for ${\rm tr}(\phi^2)$ numerators: as commented in \eqref{eq:phi2numc} and \eqref{eq:phi2numd}, $N^{\mathcal{F}_{\rm tr}(\phi^2)}[1,\beta,2]$ and $N^{\mathcal{F}_{\rm tr}(\phi^2)}\bigg( \hskip -3pt \begin{aligned}
\includegraphics[width=0.2\linewidth]{figure/mscalars.eps}
\end{aligned}\hskip -4pt\bigg)$ are basically the same, and the latter serve as master numerators, whose commutators give the CK-dual numerator of any cubic diagram
\begin{equation}
    N^{\mathcal{F}_{\rm tr}(\phi^2)}\bigg( \hskip -3pt \begin{aligned}
\includegraphics[width=0.22\linewidth]{figure/mscalars.eps}
\end{aligned}\hskip -4pt\bigg) \overset{\text{commutators}}{\wideLongrightarrow}N^{\mathcal{F}_{{\rm tr}(\phi^2)}}
    \Bigg( \hskip -3pt \begin{aligned}
        \includegraphics[width=0.17\linewidth]{figure/tree4CK1.eps}
        \end{aligned}\hskip -3pt\Bigg)\,.
\end{equation}
Meanwhile, the two horizontal arrows correspond to the map taking the flavor factor to 1. Note that the map is linear. 
Therefore, combining all these three arrows gives an understanding of the last red arrow: the ${\rm tr}(\phi^4)$ cubic-diagram numerator, which comes from removing the color factors of the ${\rm tr}(\phi^2)$ cubic-diagram numerator, can be expressed as commutators of ${\rm tr}(\phi^4)$ color-ordered numerators, of which the pre-image (as ${\rm tr}(\phi^2)$ color-ordered numerators) can also take the same commutators to get the aforementioned ${\rm tr}(\phi^2)$ cubic-diagram numerator. 

We  use a simple two-gluon example to illustrate the commutative diagram \eqref{eq:comdiagram}.
We start from the upper-left corner of \eqref{eq:comdiagram}. 
Since we will consider commutators, we need at least two ${\rm tr}(\phi^2)$ numerators, which are 
\begin{align}
\label{eq:NFphi2order1}
    N^{{\cal F}_{{\rm tr}(\phi^2)}} & [1,5,3,4,6,2]={\rm tr}_{\rm FL}(T^{I_1}T^{I_3}T^{I_4}T^{I_2})\times\\
    &\bigg[ \frac{-2 p_{1}\cdot {\rm f}_{56} \cdot p_{2}}{s_{1234}-q^2}+\frac{4 (p_{1}\cdot {\rm f}_{5} \cdot p_{234})(p_{1345}\cdot {\rm f}_{6} \cdot p_{2}) }{(s_{1234}-q^2)(s_{12345}-q^2)}-\frac{4 (p_{134}\cdot {\rm f}_{6} \cdot p_{2})(p_{1}\cdot {\rm f}_{5} \cdot p_{2346}) }{(s_{1234}-q^2)(s_{12346}-q^2)} \bigg]\,,\nonumber \\
\label{eq:NFphi2order2}
    N^{{\cal F}_{{\rm tr}(\phi^2)}} & [1,5,3,6,4,2]={\rm tr}_{\rm FL}(T^{I_1}T^{I_3}T^{I_4}T^{I_2})\times \\
    &\bigg[ \frac{-2 p_{1}\cdot {\rm f}_{56} \cdot p_{24}}{s_{1234}-q^2}+\frac{4 (p_{1}\cdot {\rm f}_{5} \cdot p_{234})(p_{135}\cdot {\rm f}_{6} \cdot p_{234}) }{(s_{1234}-q^2)(s_{12345}-q^2)}+\frac{4 (p_{1}\cdot {\rm f}_{6} \cdot p_{24})(p_{1}\cdot {\rm f}_{5} \cdot p_{2346}) }{(s_{1234}-q^2)(s_{12346}-q^2)} \bigg]\,.\nonumber
\end{align}
And apparently we can obtain $N^{{\cal F}_{{\rm tr}(\phi^2)}}\bigg(\hskip -3pt \begin{aligned}
\includegraphics[width=0.2\linewidth]{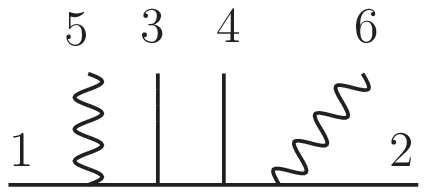}
\end{aligned}\hskip -4pt \bigg)$ and $N^{{\cal F}_{{\rm tr}(\phi^2)}}\bigg(\hskip -3pt \begin{aligned}
\includegraphics[width=0.2\linewidth]{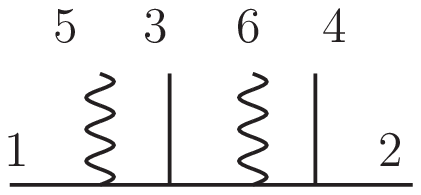}
\end{aligned}\hskip -4pt \bigg)$ by replacing ${\rm tr}_{\rm FL}(T^{I_1}T^{I_3}T^{I_4}T^{I_2})$ with $\tilde{f}^{13\text{x}}\tilde{f}^{\text{x}42}$ in \eqref{eq:NFphi2order1} and \eqref{eq:NFphi2order2}, respectively. 

We then follow the first right-arrow in \eqref{eq:comdiagram}. The ${\rm tr}(\phi^4)$ numerators can be directly read out as 
\begin{equation}
\begin{aligned}
      N^{{\cal F}_{{\rm tr}(\phi^4)}}[1,5,3,4,6,2]& = N^{{\cal F}_{{\rm tr}(\phi^2)}}[1,5,3,4,6,2]\big|_{\text{flavor factor}\rightarrow 1}\\
      &= N^{{\cal F}_{{\rm tr}(\phi^2)}}\bigg(\hskip -3pt \begin{aligned}
\includegraphics[width=0.2\linewidth]{figure/4s2g1b.eps}
\end{aligned}\hskip -4pt \bigg)_{\text{flavor factor}\rightarrow 1}\,,\\
    N^{{\cal F}_{{\rm tr}(\phi^4)}}[1,5,3,6,4,2]& = N^{{\cal F}_{{\rm tr}(\phi^2)}}[1,5,3,6,4,2]\big|_{\text{flavor factor}\rightarrow 1}\\
      &= N^{{\cal F}_{{\rm tr}(\phi^2)}}\bigg(\hskip -3pt \begin{aligned}
\includegraphics[width=0.2\linewidth]{figure/4s2g2b.eps}
\end{aligned}\hskip -4pt \bigg)_{\text{flavor factor}\rightarrow 1}\,.
\end{aligned}
\end{equation}
Next, we consider the black down-arrow in \eqref{eq:comdiagram}. 
Taking the commutator of the above two ${\rm tr}(\phi^2)$ numerators is nothing but the dual Jacobi relation
\begin{equation}\label{eq:commsegs}
    N^{{\cal F}_{{\rm tr}(\phi^2)}}\bigg(\hskip -3pt \begin{aligned}
\includegraphics[width=0.2\linewidth]{figure/4s2g2b.eps}
\end{aligned}\hskip -4pt \bigg) {-}N^{{\cal F}_{{\rm tr} (\phi^2)}}\bigg(\hskip -3pt \begin{aligned}
\includegraphics[width=0.2\linewidth]{figure/4s2g1b.eps}
\end{aligned}\hskip -4pt \bigg)
{=} N^{{\cal F}_{{\rm tr}(\phi^2)}}\bigg(\hskip -3pt \begin{aligned}
\includegraphics[width=0.2\linewidth]{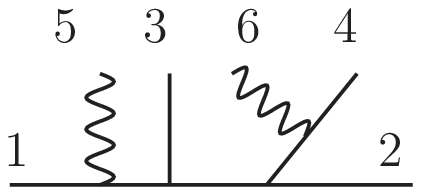}
\end{aligned}\hskip -4pt \bigg)\,,
\end{equation}
where the numerator on the RHS is 
\begin{align}\label{eq:Nphi2cubic3}
    N^{{\cal F}_{{\rm tr}(\phi^2)}}&\bigg(\hskip -3pt \begin{aligned}
\includegraphics[width=0.2\linewidth]{figure/4s2g2c.eps}
\end{aligned}\hskip -4pt \bigg)=\tilde{f}^{13\text{x}}\tilde{f}^{\text{x}42}\times\\
&\bigg[ \frac{2 p_{1}\cdot {\rm f}_{56} \cdot p_{4}}{s_{1234}-q^2}-\frac{4 (p_{1}\cdot {\rm f}_{5} \cdot p_{234})(p_{1235}\cdot {\rm f}_{6} \cdot p_{4}) }{(s_{1234}-q^2)(s_{12345}-q^2)}-\frac{4 (p_{134}\cdot {\rm f}_{6} \cdot p_{4})(p_{1}\cdot {\rm f}_{5} \cdot p_{2346}) }{(s_{1234}-q^2)(s_{12346}-q^2)} \bigg]
\,. \nonumber 
\end{align}
Note that the flavor factor does not change. 

At last, we take away the flavor factors from both sides of \eqref{eq:commsegs}, which exemplified the red down-arrow in \eqref{eq:comdiagram}, reading
\begin{equation}\label{eq:commsegs1}
    N^{\mathcal{F}_{{\rm tr}(\phi^4)}}[1,5,3,4,6,2]- N^{\mathcal{F}_{{\rm tr}(\phi^4)}}[1,5,3,6,4,2]=N^{{\cal F}_{{\rm tr}(\phi^4)}}\bigg(\hskip -3pt \begin{aligned}
\includegraphics[width=0.2\linewidth]{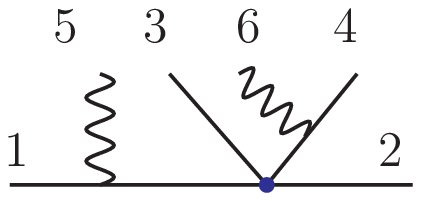}
\end{aligned}\hskip -4pt \bigg)\,,
\end{equation}
where on the RHS we have used \eqref{eq:phi2tophi4cubic} to change \eqref{eq:Nphi2cubic3} to be a ${\rm tr}(\phi^4)$ numerator. 

Independent of the derivation of \eqref{eq:commsegs1} from the commutative diagram \eqref{eq:comdiagram}, one can actually confirm that \eqref{eq:commsegs1} is true solely from the map \eqref{eq:phi2tophi4}.
The expressions for the LHS can be found using \eqref{eq:4scalarnumx}, while on the RHS one recognizes that it is the example used to illustrate \eqref{eq:phi2tophi4} discussed around \eqref{eq:4scalarsnumstu}. 
This check provides independent evidence of the assertion that \textsl{commutators of the color-ordered numerators are the cubic-diagram numerators}. 

For more general situations, we give the following remarks:
\begin{enumerate}[topsep=3pt,itemsep=-1ex,partopsep=1ex,parsep=1ex]
    \item In general, the commutators are not as simple as \eqref{eq:commsegs1}. 
Usually, they are the so-called nested commutators. 
We have learned from \eqref{eq:commsegs1} that moving one gluon requires one commutator ``$[,]$", and it is easy to expect that for a diagram like in \eqref{eq:4scalara}, 
an $s$-fold ($s$ is the number of gluons on the scalar line $3$) nested commutators $[\ldots[[,],]\ldots,]$ of $N[1,\beta,2]$ is required. 
Just to specify the nested commutator, we use the following example
\begin{equation}
\begin{aligned}
    N[1,3,[4,[[5,6],7]],8,2]=&N[1,3,4,5,6,7,8,2]-N[1,3,4,6,5,7,8,2]-\\
    &N[1,3,4,7,5,6,8,2]+N[1,3,4,7,6,5,8,2]+\\
    &N[1,3,5,6,7,4,8,2]-N[1,3,6,5,7,4,8,2]-\\
    &N[1,3,7,5,6,4,8,2]+N[1,3,7,6,5,4,8,2]
\end{aligned}
\end{equation}

    \item One should notice that $N[1,\beta,2]$ with $\beta\in S_{n-2}/S_2$ form a numerator basis, while the numerators for the cubic diagrams in \eqref{eq:4scalara} form another basis with the same number of elements. 
Then the nested commutators can be interpreted as a basis change, just as the inverse operation of \eqref{eq:orderednum} which is employed to define $N[1,\beta,2]$. 

    \item As in Section~\ref{ssec:universalv}, 
the understanding of the ${\rm tr}(\phi^4)$ numerators 
makes it straightforward to generalize to the ${\rm tr}(\phi^m)$ ($m>4$) case. For the $m>4$ case, the new operator-induced numerator relations are $m$-term relations, like the $4$-term relation \eqref{eq:phi24scalarCK2}. They can also be generated by manipulating the  ${\rm tr}(\phi^2)$ numerator identities and then taking the map. We will omit the details of the generalization here.

\end{enumerate}

Before ending this section, we comment on the relation between the construction illustrated in this subsection and the Hopf Algebra construction in \cite{Chen:2022nei}. 
We have learned above that taking commutators of color-ordered numerators give cubic diagram numerators. 
This is reminiscent of the Hopf Algebra construction in \cite{Chen:2022nei} where  commutators of ``pre-numerators" gives the  CK-dual numerator for a given cubic diagram. 
The commutators taken in both cases are the same. 
The color-ordered numerators and pre-numerators in \cite{Chen:2022nei} both play important roles in both constructions. 
The pre-numerators there were pre-determined from algebra, but here we give a more physical interpretation from \eqref{eq:orderednum}, stating that these numerators are naturally defined for orderings with the help  of CK duality.\footnote{An interesting comment is about the counting of color-ordered numerators. In \eqref{eq:orderednum}, we confine ourselves to the orderings $1,\beta,2$ with $\beta\in S_{n-2}/S_2$. The reader may wonder if it is possible to change $c_{\beta \Gamma}$ there and define more than $(n-2)!/2$ such numerators. The answer is no, which is due to the fact that the linear relations from \eqref{eq:orderedcolorcondition} have no solutions. In \cite{Chen:2022nei}, only the $N[1,\beta,2]$ numerators are non-vanishing and all the $N[1,\kappa,2,\rho]$ numerators vanish ($\rho\neq\emptyset$). This fact reflects the consistency of the two definitions of color-ordered numerators (or pre-numerators).}

\section{Further discussion on the $\vec{v}$ vectors}\label{sec:vvec}

As discussed in \cite{Lin:2022jrp}, the $\vec{v}$ vectors play a crucial role of  connecting the hidden factorization relation and the matrix decomposition relation. 
In this section, we provide a more extensive and advanced study of the $\vec{v}$ vectors.
We first give a brief review of the $\vec{v}$ vectors and how they appear in the hidden factorization relation and the matrix decomposition relation in Section~\ref{ssec:vreview}. 
In Section~\ref{ssec:closedv}, we pack the expressions for the $\vec{v}$ vectors into an all-multiplicity closed formula.
In Section~\ref{ssec:universalv}, we show further that the $\vec{v}$ vectors are universal in the sense that the same expressions can lead to the hidden factorization and the matrix decomposition  for different form factors. 
We leave some further remarks on the properties of $\vec{v}$ vectors and details of the closed formula in Appendix~\ref{ap:vremarks}.
The reader is especially encouraged to read Section~3, Section~6 and Appendix~A of \cite{Lin:2022jrp} to have some simple examples in mind before reading this section.

\subsection{The $\vec{v}$ vectors revisited}\label{ssec:vreview}
Although an elementary example has been given in Section~\ref{ssec:FFL3},  some more examples are required to initiate the thorough discussion of the $\vec{v}$ vectors in this section. 
Consider the following special case of the hidden factorization relation where the involved $\vec{v}$ vector is particularly simple
\begin{align}\label{eq:nptbcjN}
	&\sum_{\gamma \in \{3,4,..,(n{-}1)\}\shuffle \{n\}} v[1,\gamma,2] \mathcal{F}_n(1,\gamma,2)  \big|_{\QQ_{n-1}^2=q^2} \\
	&\qquad = \bigg[ \tau_{n 2}\mathcal{F}_{n}(1,3,\ldots,n,2)+\sum_{i=3}^{n-1}\tau_{n,(2+i + \cdots + (n-1))}\mathcal{F}_{n}{\small (1,3\ldots,i-1, n,i\ldots,n-1,2)} \bigg] \bigg|_{\QQ_{n-1}^2=q^2}  \nonumber \\
	&\qquad =\mathcal{F}_{n-1}(1,3,\ldots,n-1,2)\times \mathcal{A}_3(\QQ_{n-1},n,-q) \nonumber 
\end{align}
where  $\QQ_{n-1}=\sum_{i=1}^{n-1}p_i$, $\tau_{\rm ab}=2p_{\rm a}\cdot p_{\rm b}$, and the components of the $\vec{v}$ vector are 
\begin{equation}
\begin{aligned}\label{eq:nptv1}
    &v[\alpha]=\tau_{n,(2+i + \cdots + (n-1))} \quad  \textbf{ if }  \alpha=\{ 1,3\ldots,i{-}1, n,i\ldots,n{-}1,2\}\,.
\end{aligned}
\end{equation}
The above hidden factorization relation  \eqref{eq:nptbcjN} for form factors can be compared with the following BCJ relations of amplitudes \cite{Bern:2008qj}
\begin{align}\label{eq:BCJampexample}
	& \tau_{n 2}\mathcal{A}_{n}(1,3,\ldots,n,2)+\sum_{i=3}^{n-1}\tau_{n,(2+i + \cdots + (n-1))}\mathcal{A}_{n}{\small (1,3\ldots,i-1, n,i\ldots,n-1,2)} =0 \,.
\end{align}
We can see that the same $\vec{v}$ vector 
\begin{equation}
\vec{v} = \{ \tau_{n 2}, \, \tau_{n, (2+3+\ldots+(n-1))}, \, \tau_{n, (2+4+\ldots+(n-1))}, \,\ldots, \, \tau_{n, (2+(n-1))} \}
\end{equation}
appears in both \eqref{eq:nptbcjN} and \eqref{eq:BCJampexample}.
In this sense, the factorization relation \eqref{eq:nptbcjN} can be understood as the generalization of the BCJ relations for form factors.

\vskip 5pt

With a glimpse of what the $\vec{v}$ vectors are, now we review their two important properties.
The first important aspect of the $\vec{v}$ vectors is again the hidden factorization relation, but now we present it in a more general form compared to \eqref{eq:nptbcjN}. 

Specifically, for a given spurious pole $\QQ_m^2{-}q^2$ 
(with $\QQ_m$ always means $\sum_{i=1}^{m}p_i$), the hidden factorization relation takes the following form\footnote{We use the notation $v[1,\beta,2]$ in \eqref{eq:fact1} which is same as the one in \eqref{eq:nptbcjN}, but reader should keep in mind they are different in general|the $\vec{v}$ in \eqref{eq:nptbcjN} is just a simple special case.}
\begin{equation}\label{eq:fact1}
	\sum_{\beta\in S_{n-2}} v[1,\beta,2] \mathcal{F}_n(1,\beta,2)|_{\QQ_m^2=q^2}=\mathcal{F}_{m}(1,3,..,m,2)\times \mathcal{A}_{m'}(\QQ_m,(m{+}1),..,n,-q)\,.
\end{equation}
We may understand that the $\vec{v}$ vectors take higher-point form factors as an input and give lower-point form factors and amplitudes as an output. 
We explain the notion of \eqref{eq:fact1} in more detail as follows 
\begin{enumerate}[topsep=3pt,itemsep=-1ex,partopsep=1ex,parsep=1ex]
    \item[a.] $\beta$ is a permutation in $S_{n-2}$ acting on $\{3,4,\ldots,n\}$, and we use this ordering to label the vector element. 
    \item[b.]  $\mathcal{F}(1,\beta,2)$ is the $n$-point color-ordered form factor of ${\rm tr}(\phi^2)$. The ordering $\{1,\beta,2\}$ is the short of $\{1,\beta(3,4,\ldots,n),2\}$. 
    These ordered form factors form a basis.
    \item[c.] $v[1,\beta,2]$ is the vector element of the $\vec{v}$ vector. 
    \item[d.]  The form factor on the RHS is of the same type as the one on the LHS, but has a smaller number of gluons. 
    \item[e.]  The $m'=(n{-}m{+}2)$ point amplitude $\mathcal{A}_{m'}$ on the RHS has two massive external scalars (with mass square $m^2=q^2$) coupled to gluons. We will see that this amplitude factor is universal for different types of form factors. 
\end{enumerate}
We point out that \eqref{eq:fact1} is the ``normal" version of the hidden factorization relation in the sense that we have chosen a normal order of gluons as $\{3,4,..,n\}$. We can permute the gluons in both the form factor and the amplitude and the factorization relation takes the form as
\begin{equation}\label{eq:hiddenfactorization2}
    \sum_{\beta\in S_{n-2}} v_{(\bar{\kappa},\bar{\rho})}[1,\beta,2] \mathcal{F}_n(1,\beta,2)|_{\QQ_m^2=q^2}=\mathcal{F}_{m}(1,\bar{\kappa},2)\times \mathcal{A}_{m'}(\QQ_m,(m{+}1),\bar{\rho},-q)\,,
\end{equation}
where $\bar{\kappa}$ and $\bar{\rho}$ are permutations acting on $\{3,..,m\}$ and $\{(m{+}2),..,n\}$ respectively.
We denote the new $\vec{v}$ vector as $\vec{v}_{(\bar{\kappa},\bar{\rho})}$ which is related to $\vec{v}\equiv\vec{v}_{(\mathbf{1},\mathbf{1})}$ via proper permutations.
The explicit relation between $\vec{v}_{(\bar{\kappa},\bar{\rho})}$ and $\vec{v}_{(\mathbf{1},\mathbf{1})}$ will be discussed later.  

\vskip 5pt

Next, we move on to the matrix decomposition, which is the other important aspect of the $\vec{v}$ vectors. Again, the $\vec{v}$ vectors are connecting higher- and lower-point objects, and this time the object is the KLT kernel. 
However, the roles of input and output are reversed: the factorized kernel (a tensor product of lower-point kernels) is the input while the higher-point kernel (its residue if we want to be rigorous) is the output. The equation is 
\begin{align}\label{eq:nptSFdecomposition}
	&
	\sum_{\substack{\bar{\kappa}_{1,2}\in S_{m-2} \\ \bar{\rho}_{1,2}\in S_{m^{\prime}-3} } } { v}_{(\bar\kappa_1,\bar\rho_1)}[\beta_1]\left(\mathbf{S}^{\mathcal{F}}_{m}[\bar{\kappa}_1|\bar{\kappa}_2] \mathbf{S}^{\mathcal{A}}_{m^{\prime}}[\bar{\rho}_1|\bar{\rho}_2]\right) { v}_{(\bar\kappa_2,\bar\rho_2)}[\beta_2]
	={\rm Res}_{\QQ_{m}^2=q^2}\left[\mathbf{S}^{\mathcal{F}}_{n}[\beta_1|\beta_2]\right]\,, 
\end{align}
where $v[\beta]$ is just the shortened notation for the $v[1,\beta,2]$ above, $\mathbf{S}^{\cal F}$ is the form factor KLT kernel, and $\mathbf{S}^{\cal A}$ is the amplitude KLT kernel. 
Note that here we have used symmetric representations of the kernels. 
The expressions and properties of the kernels can be found in \cite{Lin:2022jrp}, and here our focus is the $\vec{v}$ vectors, which act in a bi-linear way on the kernels. 
Importantly, we want to emphasize that the two $\mathbf{S}^{\cal F}$ on both sides are the kernels used to double copy the ${\rm tr}(\phi^2)$ form factors with two external scalars|these form factors only differ by the number of gluons. 

\vskip 5pt

Before ending this review section, we suggest the readers to check the explicit expressions given in Appendix~A in \cite{Lin:2022jrp} to have a better sense of the $\vec{v}$ vectors, where all the expressions up to six points were given. 
Some properties can be observed from those expressions (and  can also be deduced from the above two aspects), and we summarize some of them below
\begin{enumerate}[topsep=3pt,itemsep=-1ex,partopsep=1ex,parsep=1ex]
    \item The $\vec{v}$ vectors have mass dimension two, which is the same as a $\tau_{ij}$ or $s_{ij}$. This simple property will be important in determining the general form of $\vec{v}$ vectors.
    \item  The $\vec{v}$ vectors in general contain poles. For the factorization related to the spurious pole $\QQ_m^2{-}q^2$, the $\vec{v}$ vectors have the following pole structure:\footnote{Even without the explicit expression, this is also understandable due to the Feynman propagator ${s_{12\cdots m \cdots (m+i) }-q^2}$ ($0{<}i{<}n{-}m$) in the amplitudes on the RHS of \eqref{eq:fact1}} 
    \begin{equation}\label{eq:generalvdeno}
	\vec{v}\ \propto \  \prod_{k=m+1}^{n-1} \frac{1}{s_{12\cdots k}-q^2} \,,
    \end{equation}
    which is a degree $(n{-}m{-}1)$ polynomial. 
    By dimension analysis, the $\vec{v}$ vector must have a numerator factor of a degree-$(n{-}m)$ polynomial,
    and the major task is to study this degree $(n{-}m)$ polynomial.
    \item From all the known explicit results, we find that the $(n{-}m)$ polynomial factorize into $(n{-}m)$ linear factors, each of which is a linear combination of some $\tau_{\text{ab}}$s. Note that the number $(n-m)$ is also the number of gluons in the amplitude in the hidden factorization relation \eqref{eq:hiddenfactorization2}.
    
    \end{enumerate}

\subsection{The closed formula}\label{ssec:closedv}

In this subsection, we present the closed formula for the $\vec{v}$ vectors. 
Although we do not have a general proof yet, highly non-trivial checks up to eight points have been done which strongly support the following compact rules for the $\vec{v}$ vectors. 
We point out that 
this subsection is relatively technical. For understanding the remaining part of the paper, knowing the existence of such a formula would be sufficient and this subsection may be skipped in a first reading.

The expressions of $\vec{v}$ vectors depend on the factorization channel.
We have seen the case associated with the spurious pole ${\QQ_{n-1}^2-q^2}$ in \eqref{eq:nptbcjN}, where the $\vec{v}$ vector has a very simple form as \eqref{eq:nptv1}. 
The expressions in other channels are generally more complicated. 
To clearly show the pattern, it is helpful to directly work out the most ``complicated" case, where the $\vec{v}$ vector appears in the following hidden factorization relation on the spurious pole $\QQ_2^2{-}q^2$ 
\begin{equation}
    \vec{v}_n\cdot \vec{\mathcal{F}}_{n}|_{\QQ_2^2=q^2}=\sum_{\beta \in S_{n-2}} {v}_{n}[\beta] {\mathcal{F}}_{n}[\beta]|_{\QQ_2^2=q^2}=\mathcal{F}_2(1^{\phi},2^{\phi})\times \mathcal{A}_n(\QQ_2,3,\ldots,n,-q)\,.
\end{equation}
We will explain the generalization for other spurious poles afterwards. As mentioned before, it is sufficient to work on the $\vec{v}_{(\mathbf{1},\mathbf{1})}$ vectors, and we will omit the subscript $(\mathbf{1},\mathbf{1})$ for simplicity. We will also discuss how to determine $\vec{v}_{(\bar{\kappa},\bar{\rho})}$ from $\vec{v}_{(\mathbf{1},\mathbf{1})}$ later.

\paragraph{The explicit rules for the  $\QQ_2^2=q^2$ pole.}

The closed formula for the $\vec{v}$ vectors, defined for the $\QQ_2^2=q^2$ pole, is
\begin{equation}\label{eq:schematicv}
\vec{v}_n = \{v_n[\beta] \} \,, \qquad
v_n[\beta]=(-1)^{\varpi(\beta)}\frac{\prod_{i=3}^{n}l_i[\beta]}{\prod_{k=3}^{n-1}(s_{12\cdots k}-q^2)}\,,
\end{equation}
where $l_{i}[\beta]$ is a linear function of certain $\tau_{\text{ab}}$ assigned for the gluon $i$, and $\varpi$ is a function controlling the sign. 
They can be determined via the following steps:
\begin{enumerate}[topsep=3pt,itemsep=1ex,partopsep=1ex,parsep=1ex]
\item[\textbf{1.}] We first pick out a subset of gluons $\{j_{1},\ldots,j_{r}\}$ such that each element $j_a$  satisfies:  $3<j_a<n$, and either $\{j_a{-}1,j_a,j_a{+}1\}$ or $\{j_a{+}1,j_a,j_a{-}1\}$ is a subordering\footnote{ We say $A$ is a subordering, or an ordered subset, of $B$, if $A\subset B$ and the order of elements in $A$ is exactly the same as their order in $B$.} of the ordering $\beta$. These are the first set of gluons. 
    
    \item[\textbf{2.}] For each of the above gluons $j_{a}$, we define 
    \begin{equation*}
    \begin{aligned}
        l_{j_{a}}[\beta]=\tau_{j_{a},\Xi_{L}(j_a;\beta)}+(s_{12\cdots (j_a-1)}{-}q^2),\  \textbf{if}\ \{j_a{-}1,j_a,j_a{+}1\} \text{ is a subordering of }\beta; \\
        l_{j_{a}}[\beta]=\tau_{j_{a},\Xi_{R}(j_a;\beta)}+(s_{12\cdots (j_a-1)}{-}q^2),\  \textbf{if}\ \{j_a{+}1,j_a,j_a{-}1\} \text{ is a subordering of }\beta.
    \end{aligned}
    \end{equation*}
     Here we use the notation $\Xi_{L}(j;\alpha)$ (or $\Xi_{R}(j;\alpha)$) to represent the sum over momenta with index $k<j$ and on the left (or right) side of the element $j$ in the ordering $\alpha$. More precisely, for a ordering $\alpha$ and an element $j\in\alpha$, we can simply delete the particles $j{+}1,\ldots,n$ in $\alpha$ (leaving $1,2,\ldots,j$ unchanged) and  get a subordering $\Xi(j,\alpha)$ of $\alpha$ such that 
     \begin{equation}
         \Xi(j,\alpha)=\{\underbrace{1,\ldots}_{\Xi_{L}(j;\alpha)},j,\underbrace{\ldots,2}_{\Xi_{R}(j;\alpha)}\}\,.
     \end{equation}
Note that $|\Xi(j,\alpha)|=j$.  For example, 
\begin{equation}
\Xi(4,\{1,3,5,4,6,2\})=\{1,3,4,2\},\ \, 
\tau_{4,\Xi_{L}(4;\{1,3,5,4,6,2\})}{=}\tau_{4,(1{+}3)} , \ \,  \tau_{4,\Xi_{R}(4;\{1,3,5,4,6,2\})}{=}\tau_{42} .
\end{equation} 
See also Appendix~C of \cite{Lin:2022jrp} for more details.

    \item[\textbf{3.}] The rest of the gluons belong to the second set. For them, we define an ordered list by deleting gluons in the first set and $\{1,2\}$: 
    $$\overline{\beta}=\beta\backslash (\{j_{1},\ldots,j_{r}\}\cup \{1,2\}) \,,$$ 
    say $\overline{\beta}=\{{\rm x}_1,\ldots,{\rm x}_{n-2-r}\}$, which is a subordering in $\beta$.

    Now $l_{\mathrm{x}_{b}}[\beta]$ is either $\tau_{\mathrm{x}_{b},\Xi_{L}(\mathrm{x}_{b};\beta)}$ or $\tau_{\mathrm{x}_{b},\Xi_{R}(\mathrm{x}_{b};\beta)}$.
    To tell which one, we sort the list $\overline{\beta}$ and get $\overline{\beta}'=\{{\rm x}_{k_1},\ldots,{\rm x}_{k_{n-r-2}} \}$ with ${\rm x}_{k_c}<{\rm x}_{k_{c+1}}$. Suppose in the new ordering, $\mathrm{x}_{1}$ is in the $s$-th position, that is $\mathrm{x}_{1}=\mathrm{x}_{k_s}$. We have that 
    \begin{equation}\label{eq:firstcond}
        l_{\mathrm{x}_{k_s}}[\beta]=l_{\mathrm{x}_1}[\beta]=\tau_{\mathrm{x}_1,\Xi_{\red L}(\mathrm{x}_1;\beta)}\,.
    \end{equation}
    
    Then we ask that the $\Xi_L$ and $\Xi_R$ appear in an alternating way according to the ordering $\overline{\beta}'$. Specifically, we have the alternating list 
    \begin{equation*}
    \hskip -8pt
    \begin{aligned}
    &.., \ l_{\mathrm{x}_{k_{s-1}}}[\beta]=\tau_{\mathrm{x}_{k_{s-1}},\Xi_{\red R}(\mathrm{x}_{k_{s-1}};\beta)},\ l_{\mathrm{x}_1}[\beta]=\tau_{\mathrm{x}_1,\Xi_{\red L}(\mathrm{x}_1;\beta)},\ l_{\mathrm{x}_{k_{s+1}}}[\beta]=\tau_{\mathrm{x}_{k_{s+1}},\Xi_{\red R}(\mathrm{x}_{k_{s+1}};\beta)}, \ ..
    \end{aligned}
    \end{equation*}
    or say
    \begin{equation}\label{eq:alterlist}
    \begin{aligned}
        l_{\mathrm{x}_{t}}[\beta]=\tau_{\mathrm{x}_{k_{t}},\Xi_{\red L}(\mathrm{x}_{k_{t}};\beta)}, \quad \textbf{if}\ (t-s)=0 (\mathrm{mod}\ 2)\,,\\
        l_{\mathrm{x}_{t}}[\beta]=\tau_{\mathrm{x}_{k_{t}},\Xi_{\red R}(\mathrm{x}_{k_{t}};\beta)}, \quad \textbf{if}\ (t-s)=1 (\mathrm{mod}\ 2)\,.
    \end{aligned}
    \end{equation}
    We would like to stress the difference of expressions between $l_{\mathrm{x}_{t}}$ for the second set and $l_{j_{a}}$ for the first set: in the first set $l_{j_{a}}$, there is an extra term $(s_{12\cdots (j_a-1)}{-}q^2)$. 
    
    \item[\textbf{4.}] Finally, we determine the overall sign for $v[\beta]$. The result is 
    \begin{equation}\label{eq:vectorSign}
        \varpi(\beta)=\big[1+\sum_{\mathrm{x}_{b}\in\overline{\beta}} (x_{b} + n-1) + n (\iota+1) \big](\text{mod }2) \,,
    \end{equation}
    where the quantity $\iota$ from $\overline{\beta}$ as 
    $$\iota=s(\text{mod }2)$$
    where $s$ means, as above, the first element in $\overline{\beta}$ is the $s$-th element in the sorted $\overline{\beta}'$.
\end{enumerate}

\vskip 5pt

We briefly recapitulate the underlying idea of the above rules. 
This idea is to separate the ordering $\beta$ into three parts: the scalars $\{1,2\}$ having no corresponding  $l[\beta]$ contributions, the gluon subset $\{j_a\}$ satisfying some relative-position conditions, and the rest gluons forming the second set $\overline{\beta}$.
Suppose initially all the gluons have the canonical $\{3,4,\ldots,n\}$ ordering, and then we act on the permutation $\beta$. 
Some of the gluons satisfy the subordering criteria, which form the first set $\{j_1,\ldots,j_r\}$ above. Their $l_{j_{a}}[\beta]$s are easier to handle. On the other hand, for the rest of gluons in $\overline{\beta}$, their positions do not satisfy the subordering criteria  in $\beta$, and it takes efforts to write down their $l[\beta]$ factor and the overall sign. 

\vskip 6pt

To make these rules more user-friendly, we give also the following explicit examples:
\begin{enumerate}[topsep=3pt,itemsep=3ex,partopsep=1ex,parsep=1ex]
    \item[\small \textcircled{1}] 
    To begin with, we consider a simple five-point example $v[\beta_1]$ with $\beta_1=\{1,5,4,3,2\}$.

    The first type of gluons has only one element $\{4\}$, where $\{5,4,3\}$ is a subordering of $\beta_1$. 
    And we can write down $l_{4}=\tau_{4,\Xi_{R}(4,\beta_1)}+(s_{123}-q^2)$, with $\tau_{4,\Xi_{R}(4,\beta_1)}=\tau_{4,(1+3)}$. 
    
    For the remaining gluons, we have $\overline{\beta}_1=\{5,3\}$. Here we see that $5$ is the first element and 3 is the second. We then sort $\overline{\beta}_1$ and get $\overline{\beta}'_1=\{3,5\}$, in which 5 is the second and 3 is the first.
    And we directly give $l_{5}=\tau_{5,\Xi_{L}(5;\beta_1)}=\tau_{51}$. 
    In $\overline{\beta}_1'$, 3 appears one site ahead of 5, so that if we use $\Xi_L$ for 5, we need $\Xi_R$ for 3, namely, $l_{3}=\tau_{3,\Xi_{R}(3;\beta_1)}=\tau_{32}$. 

    The last step is to determine the overall sign. The $\iota$ for $\overline{\beta}_1$ is 0 because the first element of $\overline{\beta}_1$ is at the second place in $\overline{\beta}_1'$. 
    Using \eqref{eq:vectorSign}, we have 
    $$\varpi(\beta_1)=[1+((3+4)+(5+4))+5 (0+1)] (\text{mod }2)=0.$$

    In the end, we get 
    \begin{equation*}
        v[\beta_1]=\frac{(-1)^{\varpi(\beta_1)}\prod_{j=3}^{5} l_{j}[\beta_1]}{(s_{123}-q^2)(s_{1234}-q^2)}=\frac{(-1)^0 \tau_{32} \tau_{51} }{s_{1234}-q^2}\Big(\frac{\tau_{4,(1+3)}}{s_{123}-q^2}+1\Big) \,.
    \end{equation*}

    \item[\small \textcircled{2}] 
    Next we consider a seven-point example $v[\beta_2]$ with $\beta_2{=}\{1,4,5,3,7,6,2\}$.

    The first type of gluons has only one element $\{5\}$, and $\{4,5,6\}$ is an ordered subset of $\beta_2$. This means $$l_5[\beta_2]= \tau_{5, \Xi_{L}(5;\beta_2)}+(s_{1234}-q^2){=}\tau_{5,(1+4)}+(s_{1234}-q^2).$$
       
    Deleting $\{1,2,5\}$ in $\beta_2$, we obtain the subordering $\overline{\beta}_2=\{4,3,7,6\}$ for the remaining gluons. And it becomes $\overline{\beta}_2'=\{3,4,6,7\}$ after sorting. The first element in $\overline{\beta}_2$ is 4 and it is the second one in $\overline{\beta}_2'$. We use $\Xi_L$ in $l_4$, and $\Xi_R$ in $l_3$ and $l_6$, and  $\Xi_L$ in $l_7$. This is to make an alternating list. 
    
 Finally we determine the sign. 
    The first element in $\overline{\beta}_2$ is the second in $\overline{\beta}'_2$, making $\iota$ to be 0. A straightforward calculation according to \eqref{eq:vectorSign} gives $\varpi(\beta_2)=0$.
    
    Thus we have the final expression
    \begin{equation*}
    \begin{aligned}
        v[\beta_2]=&\frac{(-1)^0\tau_{3,\Xi_{R}(3;\beta_2)}\tau_{4,\Xi_{L}(4;\beta_2)}\tau_{6,\Xi_{R}(6;\beta_2)}\tau_{7,\Xi_{L}(7;\beta_2)}}{(s_{123}-q^2)(s_{12345}-q^2)(s_{123456}-q^2)}\Big(\frac{\tau_{5,\Xi_{L}(5;\beta_2)}}{s_{1234}-q^2}+1\Big)\\
        =&\frac{\tau_{32}\tau_{41}\tau_{62}\tau_{7,(1+3+4+5)}}{(s_{123}-q^2)(s_{12345}-q^2)(s_{123456}-q^2)}\Big(\frac{\tau_{5,(1+4)}}{s_{1234}-q^2}+1\Big) .
    \end{aligned}
    \end{equation*}
    \item[\small \textcircled{3}] 
    Since the above examples are both odd-point ones, we conclude with an eight-point example: $v[\beta_3]$ with $\beta_3=\{1,3,4,7,8,6,5,2\}$, in which some simplifications can be introduced. 
    
    Again, we first pick out the gluons with their adjacent suborderings in $\beta_3$. We have $j_{1}=4$ with $\{3,4,5\}$ a subordering and $j_{2}=6$ with $\{7,6,5\}$ is a subordering.
    They give factors 
    \begin{equation*}
    \begin{aligned}
        l_4[\beta_3]=& \tau_{4, \Xi_{L}(4;\beta_3)}+(s_{123}-q^2)=\tau_{4, (1+3)}+(s_{123}-q^2),\\
        l_6[\beta_3]=&\tau_{6,\Xi_{R}(6;\beta_3)}+(s_{12345}-q^2)=\tau_{6,(2+5)}+(s_{12345}-q^2). 
    \end{aligned}
    \end{equation*}
    
    The subordering removing $\{1,2,j_1,j_2\}$ is $\overline{\beta}_3=\{3,7,8,5\}$ and $\overline{\beta}'_3=\{3,5,7,8\}$ sorted. The first element in $\overline{\beta}_3$ is also the first one in $\overline{\beta}'_3$. This helps us to fix $l_{3,5,7,8}$ to be $\tau_{3,\Xi_{L}(3;\beta_3)}$, $\tau_{5,\Xi_{R}(5;\beta_3)}$, $\tau_{7,\Xi_{L}(7;\beta_3)}$ and $\tau_{8,\Xi_{R}(8;\beta_3)}$ respectively, based on the alternating list requirement. 
    
    As for the sign, because $n=8$ is an even number, we do not bother to write down $\iota$ and $\varpi(\beta_3)$ is basically just summing up $\overline{\beta}_3$ (and a 1), this makes $\varpi(\beta_3)=0$. 
    
    We have the full expression
    \begin{equation*}
    \hskip -15pt 
    \begin{aligned}
        v[\beta_3]=&\frac{(-1)^0\tau_{3,\Xi_{L}(3;\beta_2)}\tau_{5,\Xi_{R}(5;\beta_2)}\tau_{7,\Xi_{L}(7;\beta_2)}\tau_{8,\Xi_{R}(8;\beta_2)}}{(s_{123}-q^2)(s_{1234}-q^2)(s_{123456}-q^2)(s_{1234567}-q^2)}\Big(\frac{\tau_{4,\Xi_{L}(4;\beta_2)}}{s_{123}-q^2}+1\Big)\Big(\frac{\tau_{6,\Xi_{R}(6;\beta_2)}}{s_{12345}-q^2}+1\Big)\\
        =&\frac{\tau_{31}\tau_{52}\tau_{7,(1+3+4)}\tau_{8,(2+5+6)}}{(s_{123}-q^2)(s_{1234}-q^2)(s_{123456}-q^2)(s_{1234567}-q^2)}\Big(\frac{\tau_{4,(1+3)}}{s_{123}-q^2}+1\Big)\Big(\frac{\tau_{6,(2+5)}}{s_{12345}-q^2}+1\Big).
    \end{aligned}
    \end{equation*}
    
\end{enumerate}

\paragraph{Generalizations to other poles.}

The generalization of the $\vec{v}$ vectors for other spurious poles like $s_{12\cdots m}{-}q^2$ is straightforward. 
Only some minor modifications are  required for the rules given above. 

First, we need a new supplementary rule in the beginning:\footnote{Again, we stress that we are dealing with the standard ${v}_{(\mathbf{1},\mathbf{1})}[\beta]$. When we have ${v}_{(\bar{\kappa},\mathbf{1})}[\beta]$, it is non-zero if $\bar{\kappa}\{3,\ldots,r\}$ is a subordering of $\beta$.  }

\noindent \  $\bullet$ Step \textbf{0.} $v[\beta]$ is non-zero only if $\{1,3,\ldots,m,2\}$ is a subordering of $\beta$.

Next, we need the following small changes to the previous rules:

\noindent \  $\bullet$  In the step \textbf{1}, for $j_a \in \{j_1,\ldots,j_r\}$, instead of requiring $3<j_a<n$, now we ask for that $(m{+}1)<j_a<n$. 

\noindent \  $\bullet$  The step \textbf{2} stays the same. 

\noindent \  $\bullet$  
In the step \textbf{3}, we modify the subordering $\overline{\beta}$ as $\overline{\beta}=\beta\backslash (\{j_{1},\ldots,j_{r}\}\cup \{1,2,\ldots, m\})$.\footnote{ In other words, the non-contributing part that needs to be dropped out in $\beta$ becomes $\{1,2,..,m\}$. }

\noindent \  $\bullet$  In the step \textbf{4}, we need to modify $\varpi(\beta)$ slightly as 
\begin{equation*}
    \begin{aligned}
        \varpi(\beta)&=\big[1+\sum_{\mathrm{x}_{b}\in\overline{\beta}} (x_{b} + n-1) + (n-m)(\iota+1) \big](\text{mod }2)\,.
    \end{aligned}
    \end{equation*}

We illustrate this briefly with the example $v[1,5,3,7,6,4,2]$ for the pole $s_{1234}{-}q^2$. 
First, this $v$ element is not zero because $\{1,3,4,2\}$ is indeed a subordering of $\{1,5,3,7,6,4,2\}$. 
Second, there are no first-type gluons, and for the gluons of the second type, we have $\overline{\beta}=\{5,7,6\}$ and $\overline{\beta'}=\{5,6,7\}$. Following the rules, 
the factors $l_{5,6,7}[\beta]$ are $\tau_{51}$, $\tau_{6,(2+4)}$, and $\tau_{7,(1+3+5)}$, respectively. As for the overall sign, we have $\iota=1$ and  
\begin{equation}
\varpi(\beta)=[1{+}((5{+}6){+}(6{+}6){+}(7{+}6)){+}3(1{+}1)](\text{mod }2)=1 \,,
\end{equation} 
thus there is an overall minus sign. In conclusion, we have that
\begin{equation}
    v[1,5,3,7,6,4,2]=-\frac{\tau_{51}\tau_{6,(2+4)}\tau_{7,(1+3+5)}}{(s_{12345}-q^2)(s_{123456}-q^2)}\,,
\end{equation}
associated with the factorization relation for the spurious pole $s_{1234}-q^2$.

\paragraph{Permutational Covariance.}

Now we consider the hidden factorization relation \eqref{eq:hiddenfactorization2} associated with general color-ordered form factors.
To relate an arbitrary $\vec{v}_{(\bar{\kappa},\bar{\rho})}$ to the standard $\vec{v}_{(\mathbf{1},\mathbf{1})}$, we need the property of the $\vec{v}$ vectors referred to as the permutation covariance.

We start by performing the following permutation on both sides of the hidden factorization relation \eqref{eq:hiddenfactorization2}
\begin{align}
    \left(\widehat{\bar{\kappa}}\right)^{-1} \left(\widehat{\bar{\rho}}\right)^{-1}
    \sum_{\beta\in S_{n-2}}{v}_{(\bar{\kappa},\bar{\rho})}[\beta] \mathcal{F}_n(1,\beta,2)&=\left(\widehat{\bar{\kappa}}\right)^{-1} \left(\widehat{\bar{\rho}}\right)^{-1} \mathcal{F}_{m}(1,\bar{\kappa},2)\mathcal{A}(\QQ_2,(m{+}1),\bar{\rho},-q)\nonumber \\
    &\ \Downarrow \nonumber\\
    \sum_{\beta\in S_{n-2}}\Big(\left(\widehat{\bar{\kappa}}\right)^{-1} \left(\widehat{\bar{\rho}}\right)^{-1}{v}_{(\bar{\kappa},\bar{\rho})}[\beta]\Big) \mathcal{F}_n(1,\bar{\kappa}^{-1}\bar{\rho}^{-1}\beta,2) &= \mathcal{F}_{m}(1,3,\ldots,m,2)\mathcal{A}(\QQ_2,(m{+}1),\ldots,n,-q) \nonumber
\end{align}
where $\left(\widehat{\bar{\kappa}}\right)^{-1}, \left(\widehat{\bar{\rho}}\right)^{-1}$ are sub-permutations acting on $\{3,..,m\}$ and $\{(m{+})2,..,n\}$.

In comparison, we also have 
\begin{align}
    \sum_{\beta\in S_{n-2}} v_{(\mathbf{1},\mathbf{1})}[\beta] \mathcal{F}_{n} (1,\beta,2) & = \mathcal{F}_{m}(1,3,\ldots,m,2)\mathcal{A}(\QQ_2,(m{+}1),\ldots,n,-q) \nonumber \\
    &\ \Downarrow \nonumber\\
    \sum_{\beta\in S_{n-2}} v_{(\mathbf{1},\mathbf{1})}[\bar{\kappa}^{-1}\bar{\rho}^{-1}\beta] \mathcal{F}_{n} (1,\bar{\kappa}^{-1}\bar{\rho}^{-1}\beta,2) & = \mathcal{F}_{m}(1,3,\ldots,m,2)\mathcal{A}(\QQ_2,(m{+}1),\ldots,n,-q) \nonumber \,.
\end{align}

The consistency of the above two equations gives
\begin{equation}\label{eq:permcov}
    v_{(\mathbf{1},\mathbf{1})}[\bar{\kappa}^{-1}\bar{\rho}^{-1}\beta]=\left(\widehat{\bar{\kappa}}\right)^{-1} \left(\widehat{\bar{\rho}}\right)^{-1}{v}_{(\bar{\kappa},\bar{\rho})}[\beta] \ \Rightarrow\ \widehat{\bar{\kappa}}\  \widehat{\bar{\rho}}\  v_{(\mathbf{1},\mathbf{1})}[\bar{\kappa}^{-1}\bar{\rho}^{-1}\beta]={v}_{(\bar{\kappa},\bar{\rho})}[\beta] \,.
\end{equation}
This is the permutational covariance of the $\vec{v}$ vectors, meaning that it is sufficient to just determine the standard $\vec{v}_{(\mathbf{1},\mathbf{1})}$ and write down other $\vec{v}$'s by applying permutations.

Below we use some concrete examples to illustrate the relation \eqref{eq:permcov}. 
The first non-trivial example is the  five-point $\vec{v}$ vectors for the $s_{12}{-}q^2$ spurious pole. The two $\vec{v}$ vectors are $v_{5,(\mathbf{1},\mathbf{1})}$ and $v_{5,(\mathbf{1},\sigma_2)}=v_{5,(\mathbf{1},(4,5))}$, with the following concrete expressions 
\begin{align}
    &v_{5,(\mathbf{1},\mathbf{1})}[1,3,4,5,2]=-\frac{\tau_{31}\tau_{52}}{s_{1234}-q^2}\Big(\frac{\tau_{4,(1+3)}}{s_{123}-q^2}+1\Big), \  v_{5,(\mathbf{1},\mathbf{1})}[1,3,5,4,2]= -\frac{\tau_{31}\tau_{42}\tau_{5,(1+3)}}{(s_{123}-q^2)(s_{1234}-q^2)},\nonumber\\
    &v_{5,(\mathbf{1},\mathbf{1})}[1,5,3,4,2]=-\frac{\tau_{31}\tau_{42}\tau_{51}}{(s_{123}-q^2)(s_{1234}-q^2)}, \qquad  v_{5,(\mathbf{1},\mathbf{1})}[1,4,3,5,2]=\frac{\tau_{32}\tau_{41}\tau_{52}}{(s_{123}-q^2)(s_{1234}-q^2)}, \nonumber \\
    & v_{5,(\mathbf{1},\mathbf{1})}[1,5,4,3,2]=\frac{\tau_{32}\tau_{51}}{s_{1234}-q^2}\Big(\frac{\tau_{4,(2+3)}}{s_{123}-q^2}+1\Big),\quad v_{5,(\mathbf{1},\mathbf{1})}[1,4,5,3,2]= \frac{\tau_{32}\tau_{41}\tau_{5,(2+3)}}{(s_{123}-q^2)(s_{1234}-q^2)} \,, \nonumber 
\end{align}
and
\begin{align}
    &v_{5,(\mathbf{1},\sigma_2)}[1,3,4,5,2]=-\frac{\tau_{31}\tau_{4,(1+3)}\tau_{52}}{(s_{123}-q^2)(s_{1235}-q^2)}, \  v_{5,(\mathbf{1},\sigma_2)}[1,3,5,4,2]= -\frac{\tau_{31}\tau_{42}}{s_{1235}-q^2}\Big(\frac{\tau_{5,(1+3)}}{s_{123}-q^2}+1\Big),\nonumber\\
    &v_{5,(\mathbf{1},\sigma_2)}[1,5,3,4,2]=\frac{\tau_{31}\tau_{41}\tau_{52}}{(s_{123}-q^2)(s_{1235}-q^2)}, \quad     v_{5,(\mathbf{1},\sigma_2)}[1,4,3,5,2]=-\frac{\tau_{32}\tau_{41}\tau_{52}}{(s_{123}-q^2)(s_{1235}-q^2)}, \nonumber \\
    & v_{5,(\mathbf{1},\sigma_2)}[1,5,4,3,2]=\frac{\tau_{32}\tau_{42}\tau_{51}}{(s_{123}-q^2)(s_{1235}-q^2)},\quad\  v_{5,(\mathbf{1},\sigma_2)}[1,4,5,3,2]= \frac{\tau_{32}\tau_{51}}{s_{1235}-q^2}\Big(\frac{\tau_{4,(2+3)}}{s_{123}-q^2}+1\Big) \,. \nonumber 
\end{align}
And now we can check term by term that the following permutational covariance holds (note that $\bar{\rho}=(4,5)$,  $\bar{\kappa}=\mathbf{1}$ and $(4,5)^2=\mathbf{1}$)
\begin{equation}\label{eq:permvvectors1}
   {v}_{5,(\mathbf{1},\sigma_2)}[\beta]= (45){v}_{5,(\mathbf{1},\mathbf{1})}[(45)\beta]\,,
\end{equation}
say
\begin{equation}
\begin{aligned}
   v_{5,(\mathbf{1},\sigma_2)}[1,5,3,4,2]=&\,\frac{\tau_{31}\tau_{41}\tau_{52}}{(s_{123}-q^2)(s_{1235}-q^2)}\\
   =v_{5,(\mathbf{1},\mathbf{1})}[1,4,3,5,2]\Big|_{p_4\leftrightarrow p_5}=&\,\frac{\tau_{32}\tau_{41}\tau_{52}}{(s_{123}-q^2)(s_{1234}-q^2)}\Big|_{p_4\leftrightarrow p_5}\,.
\end{aligned}
\end{equation}

Next we look at a more non-trivial example, where the inverse ($\bar{\kappa}^{-1}/\bar{\rho}^{-1}$) plays a role. Consider the following two vector elements
\begin{equation}
\begin{aligned}
	v_{6,(\mathbf{1},\mathbf{1})}[1,3,4,6,5,2]=& \,\frac{\tau_{31}\tau_{52}\tau_{6,(1+3+4)}}{(s_{1234}-q^2)(s_{12345}-q^2)}\Big(\frac{\tau_{4,(1+3)}}{s_{123}-q^2}+1\Big)\,,\\
	v_{6,(\mathbf{1},(4,5,6))}[1,3,6,5,4,2]=& \,\frac{\tau_{31}\tau_{62}\tau_{4,(1+3+5)}}{(s_{1235}-q^2)(s_{12356}-q^2)}\Big(\frac{\tau_{5,(1+3)}}{s_{123}-q^2}+1\Big)\,,\\
\end{aligned}
\end{equation}
showing that for the ordering
\begin{equation}
(4,5,6)^{-1}\{1,3,4,6,5,2\}=(6,5,4)\{1,3,4,6,5,2\}=\{1,3,6,5,4,2\}\,,
\end{equation} 
and the RHS of these two identities are consistent up to a permutation $p_4\rightarrow p_5\rightarrow p_6\rightarrow p_4$.

\subsection{The universality}\label{ssec:universalv}
In the previous subsection, we give the closed formula of the $\vec{v}$ vectors  for the $\operatorname{tr}(\phi^2)$ form factor. 
Remarkably, exactly the same expressions of $\vec{v}$ can lead to the hidden factorization relations and the matrix decomposition for other form factors, namely, form factors of different operators or different types of external states. In particular, $\vec{v}$ depends  only on the number of point $n$ and the ``spurious"-type pole $\QQ_m^2{-}q^2$. 
We refer to this as the universality of the $\vec{v}$ vectors. 

To be more precise, we use the following diagrammatic expression to represent the hidden factorization relation
\begin{equation}\label{eq:varrow1}
    \begin{aligned}
    \includegraphics[width=0.52\linewidth]{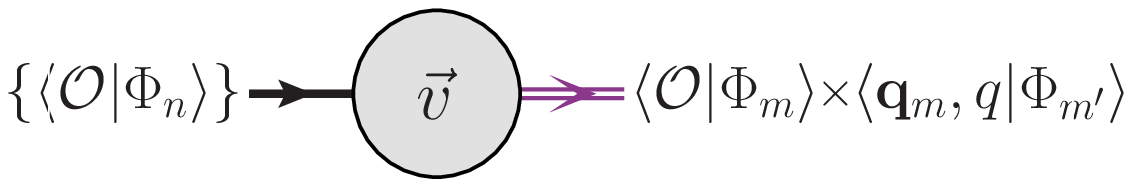}
    \end{aligned}\,,
\end{equation}
where on the LHS the input is a basis set of $n$-point form factors and on the RHS we have $m$-point form factors and $m'$-point amplitudes. 
Note that both the operator $\mathcal{O}$ and the asymptotic states $|\Phi_n\rangle =|\Phi_{m}\rangle \otimes |\Phi_{m'}\rangle$ can be arbitrary.

Similarly, we can discuss the matrix decomposition relations given as 
\begin{equation}\label{eq:varrow2}
    \begin{aligned}
        \includegraphics[width=0.56\linewidth ]{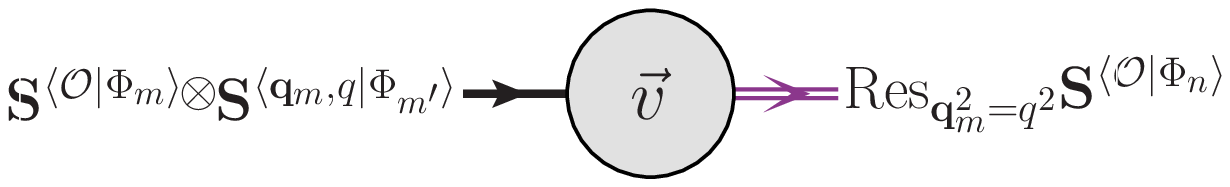}
    \end{aligned}\,.
\end{equation}
Here $\mathbf{S}^{\langle\star\rangle}$ are the KLT kernels defined for the form factor/amplitude ${\langle\star\rangle}$. Once again, \eqref{eq:varrow2}  is valid as long as the operators and the asymptotic states on both sides are consistent. 

Below we will decode the universality in \eqref{eq:varrow1} and  \eqref{eq:varrow2} by plugging in concrete form factors. 
To be more precise, we clarify the form factors involved in this subsection: 
\begin{itemize}
[topsep=3pt,itemsep=-1ex,partopsep=1ex,parsep=1ex]
\item 
the $\operatorname{tr}(\phi^2)$ form factors with $r$ $(r\geq 2)$ external scalars in the YMS+$\phi^3$ theory, as discussed in \cite{Lin:2022jrp};
\item the form factors of the high-length operators $\operatorname{tr}(\phi^k)$ with $k$ ($k>2$) external scalars in the YMS theory, which will be covered at length in Section~\ref{sec:highlength}. 
\end{itemize}
Note that an arbitrary number of external gluons is assumed. 
We discuss some further generalizations in Appendix~\ref{ap:scalartheory}.

\vskip 5pt

As mentioned above, $\vec{v}$ vectors will depend on the type of  the $s_{12\cdots m}{-}q^2$ poles. 
The first non-trivial case is the $\vec{v}$ vector defined for the $s_{123}{-}q^2$ pole.\footnote{The reason is that the only form factors that allow the $s_{12}{-}q^2$ spurious pole are the $\operatorname{tr}(\phi^2)$ form factors with two external scalars.}

\paragraph{The $s_{123}{-}q^2$ pole.}

When considering the  $s_{123}{-}q^2$ pole, \eqref{eq:varrow1} and \eqref{eq:varrow2} can be translated to the following statement: for the ${\rm tr}(\phi^2)$ form factors with two or three external scalars, and ${\rm tr}(\phi^3)$ form factors with  three external scalars, the same $\vec{v}$ vectors appear in the hidden factorization relations and the matrix decomposition. 
Explicitly, for the hidden factorization relation, we have (we use the red color to highlight some minor differences)
\begin{align}\label{eq:threephivFuniv}
    \sum_{\beta \in S_{n-2}} v[1,\beta,2] \mathcal{F}_{{\rm tr}({\color{red}\phi^2}),n}(1^{\phi},\beta\{3^{{\color{red} g}},4^{g},\ldots,& n^{g}\},2^{\phi})\big|_{s_{123 }=q^2}  \\
    =&\mathcal{F}_{{\rm tr}({\color{red}\phi^2}),3}(1^{\phi},3^{{\color{red} g}},2^{\phi})\mathcal{A}_{n-1}(\QQ_{3}^{S},4^{g},\ldots,n^{g},-q^{S})\,, \nonumber \\
     \sum_{\beta \in S_{n-2}} v[1,\beta,2] \mathcal{F}_{{\rm tr}({\color{red}\phi^2}),n}(1^{\phi},\beta\{3^{{\color{red} \phi}},4^{g},\ldots,& n^{g}\},2^{\phi})\big|_{s_{123 }=q^2}\nonumber\\
    =&\mathcal{F}_{{\rm tr}({\color{red}\phi^2}),3}(1^{\phi},3^{{\color{red} \phi}},2^{\phi})\mathcal{A}_{n-1}(\QQ_{3}^{S},4^g,\ldots,n^{g},-q^{S})\,, \nonumber\\
     \sum_{\beta \in S_{n-2}} v[1,\beta,2] \mathcal{F}_{{\rm tr}({\color{red}\phi^3}),n}(1^{\phi},\beta\{3^{{\color{red} \phi}},4^{g},\ldots,& n^{g}\},2^{\phi})\big|_{s_{123 }=q^2} \nonumber \\
    =&\mathcal{F}_{{\rm tr}({\color{red}\phi^3}),3}(1^{\phi},3^{{\color{red} \phi}},2^{\phi})\mathcal{A}_{n-1}(\QQ_{3}^{S},4^g,\ldots,n^{g},-q^{S})\,. \nonumber
\end{align}
The $n=4$ case can be found in (3.42) and (6.16) in \cite{Lin:2022jrp}, and  \eqref{eq:phi34ptrelations} in this paper.
Although these are equations for different form factors,  the $v[1,\beta,2]$ is universal for the three equations.
We give a few remarks here: (1)
in each one of them, the operator and asymptotic states on both sides are the same to make the equation consistent;
(2) all the elements of $v[1,\beta,2]$ are non-zero, due to the closed formulas discussed in the last subsection; and (3) the amplitude factor on the RHS is also universal. 

For completeness, we also give all the three matrix decomposition relations as 
\begin{equation}
\begin{aligned} 
   \text{Res}_{s_{123}=q^2}\big[\mathbf{S}_n^{\mathcal{F}_{{\rm tr}(\phi^2)},{\color{red}{\rm a}}}[\alpha_1|\alpha_2]\big]=\sum_{\bar{\rho}_{1,2}\in S_{n-4}}v_{(\mathbf{1},\bar{\rho}_1)}[\alpha_1]
    \big(\mathbf{S}^{\mathcal{F}_{{\rm tr}(\phi^2)},{\color{red}{\rm a}}}_3[\mathbf{1}|\mathbf{1}]\mathbf{S}^{\cal A}_{n-1}[\bar{\rho}_1|\bar{\rho}_2]\big)
    v_{(\mathbf{1},\bar{\rho}_2)}[\alpha_2]\,,\\
    \text{Res}_{s_{123}=q^2}\big[\mathbf{S}_n^{\mathcal{F}_{{\rm tr}(\phi^2)},{\color{red}{\rm b}}}[\alpha_1|\alpha_2]\big]=\sum_{\bar{\rho}_{1,2}\in S_{n-4}}v_{(\mathbf{1},\bar{\rho}_1)}[\alpha_1]
    \big(\mathbf{S}^{\mathcal{F}_{{\rm tr}(\phi^2)},{\color{red}{\rm b}}}_3[\mathbf{1}|\mathbf{1}]\mathbf{S}^{\cal A}_{n-1}[\bar{\rho}_1|\bar{\rho}_2]\big)
    v_{(\mathbf{1},\bar{\rho}_2)}[\alpha_2]\,,\\
    \text{Res}_{s_{123}=q^2}\big[\mathbf{S}_n^{\mathcal{F}_{{\rm tr}({\color{red}\phi^3})}}[\alpha_1|\alpha_2]\big]=\sum_{\bar{\rho}_{1,2}\in S_{n-4}}v_{(\mathbf{1},\bar{\rho}_1)}[\alpha_1]
    \big(\mathbf{S}^{\mathcal{F}_{{\rm tr}({\color{red}\phi^3})}}_3[\mathbf{1}|\mathbf{1}]\mathbf{S}^{\cal A}_{n-1}[\bar{\rho}_1|\bar{\rho}_2]\big)
    v_{(\mathbf{1},\bar{\rho}_2)}[\alpha_2]\,,
\end{aligned}
\end{equation}
where $\mathbf{S}^{\mathcal{F}_{{\rm tr}(\phi^2)},{\rm a}}$, $\mathbf{S}^{\mathcal{F}_{{\rm tr}(\phi^2)},{\rm b}}$ and $\mathbf{S}^{\mathcal{F}_{{\rm tr}(\phi^3)}}$ mean the KLT kernel for ${\rm tr}(\phi^2)$ form factors with two scalars, for ${\rm tr}(\phi^2)$ form factors with three scalars and for ${\rm tr}(\phi^3)$ form factors (with three scalars), respectively. 
Note that the expression  for the form factor  KLT kernel depends on the external asymptotic states, which is why we keep track of asymptotic states in \eqref{eq:varrow2}. 
Four-point examples are given in (3.44) and 
 (6.15) in \cite{Lin:2022jrp} and \eqref{eq:phi34ptrelations} in this paper.

\paragraph{The $s_{1234}{-}q^2$ pole and beyond.}

Although the above $s_{123}{-}q^2$ case has clarified \eqref{eq:varrow1} and \eqref{eq:varrow2} to a good extent, it is not the full story. 
There is a new feature in the $s_{12\cdots m}{-}q^2$ $(m\geq 4)$ cases, namely,
when considering the ${\rm tr}(\phi^4)$ or even higher-length operators, one encounters a counting mismatch: the $\vec{v}$ is originally defined with $(n-2)!$ vector elements for the ${\rm tr}(\phi^2)$ form factors, but the color-ordered form factors of ${\rm tr}(\phi^4)$, for instance, have $(n-2)!/2$ elements (which will be explained shortly). 

In order to deal with this, we take a closer look at the expression of the $\vec{v}$ vectors. Consider the  ${\rm tr}(\phi^2)$ form factors with two external scalars, the $\vec{v}$ vectors associated with the $\QQ_m^2{-}q^2$ pole (with $m\geq 4$) can have some zero matrix elements. 
For instance, we have the hidden factorization relation 
\begin{align}\label{eq:fact2}
    \sum_{\beta \in S_{n-2}/S_2} v[1,\beta,2] \mathcal{F}_{{\rm tr}(\phi^2),n}(1^{\phi},\beta\{3^{g},4^{g},\ldots,& n^{g}\},2^{\phi})\big|_{s_{1234 }=q^2}  \\
    =&\mathcal{F}_{{\rm tr}(\phi^2),4}(1^{\phi},3^{g},4^{g},2^{\phi})\mathcal{A}_{n-2}(\QQ_{4}^{S},5^g,\ldots,n^{g},-q^{S})\,. \nonumber 
\end{align}
Here $\beta\in S_{n-2}/S_2$ means that the permutation $\beta$ does not change the relative order of $\{3,4\}$. Other $\beta$'s that shift $\{3,4\}$ lead to zero $v[1,\beta,2]$ so they can be dropped from the sum. This solves the mismatch problem mentioned above. 

As discussed in Section~\ref{ssec:FFL4}, the number of form factor basis for the ${\rm tr}(\phi^4)$ operator is $(n-2)!/2$ which can be taken as $\mathcal{F}(1,\beta\{3,\ldots,n\},2)$ with $\beta\in S_{n-2}/S_2$. 
So it is reasonable to expect
\begin{align}\label{eq:fact3}
    \sum_{\beta \in S_{n-2}/S_2} v[1,\beta,2] \mathcal{F}_{{\rm tr}(\phi^4),n}(1^{\phi},\beta\{3^{\phi},4^{\phi},\ldots,& n^{g}\},2^{\phi})\big|_{s_{1234 }=q^2}  \\
    =&\mathcal{F}_{{\rm tr}(\phi^4),4}(1^{\phi},3^{\phi},4^{\phi},2^{\phi})\mathcal{A}_{n-2}(\QQ_{4}^{S},5^g,\ldots,n^{g},-q^{S})\,. \nonumber
\end{align}
Now we can observe the clear similarity between \eqref{eq:fact2} and \eqref{eq:fact3}. This also means the input process in \eqref{eq:varrow1} is to sum over $\beta\in S_{n-2}/S_2$.

Upon inspecting the KLT kernel matrix decomposition, we are facing the same problem. 
The $\mathbf{S}^{\cal F}$ for ${\rm tr}(\phi^{L})$ with $L=2,3$ and $\mathbf{S}^{\cal F}$ for ${\rm tr}(\phi^4)$ are not of the same size: the former is $(n-2)!$ by $(n-2)!$ and the latter is $(n-2)!/2$ by $(n-2)!/2$. 
The idea is that we take the upper-left $(n-2)!/2$ by $(n-2)!/2$ minor of $\mathbf{S}_n^{\mathcal{F}_{{\rm tr}(\phi^{2,3})}}$ as follows
\begin{equation}\label{eq:SFminor}
    \mathbf{S}_n^{\mathcal{F}_{{\rm tr}(\phi^{2,3})}}=\left(\begin{array}{c: c}
      {\color{red} \mathbf{S}^{\cal F}[\alpha_1|\alpha_2] } & \mathbf{S}^{\cal F}[\alpha_1|\widehat{\alpha}_2]\\\hdashline
       \mathbf{S}^{\cal F}[\widehat{\alpha}_1|\alpha_2] & \mathbf{S}^{\cal F}[\widehat{\alpha}_1|\widehat{\alpha}_2]\\
    \end{array}\right)
    \,,
\end{equation}
where $\alpha_{1,2}\in S_{n-2}/S_2$ permuting $\{3,\ldots,n\}$ but leave the relative order of $\{3,4\}$ invariant, while $\widehat{\alpha}_{1,2}\in S_{n-2}$ but gets $\{3,4\}$ swapped. 
And then the matrix decomposition formulas for the red minor look exactly the same as for $\mathbf{S}_n^{\mathcal{F}_{{\rm tr}(\phi^{4})}}$, which is 
\begin{equation}\label{eq:phi4FFmatdec}
    \text{Res}_{s_{1234}=q^2}\big[\mathbf{S}_n^{\mathcal{F}_{{\rm tr}(\phi^4)}}[\alpha_1|\alpha_2]\big]=\sum_{\bar{\rho}_{1,2}\in S_{n-5}}v_{(\mathbf{1},\bar{\rho}_1)}[\alpha_1]
    \big(\mathbf{S}^{\mathcal{F}_{{\rm tr}(\phi^4)}}_4[\mathbf{1}|\mathbf{1}]\mathbf{S}^{\cal A}_{n-2}[\bar{\rho}_1|\bar{\rho}_2]\big)
    v_{(\mathbf{1},\bar{\rho}_2)}[\alpha_2]\,,
\end{equation}
where $\alpha_{1,2}\in S_{n-2}/S_2$ and $v_{(\mathbf{1},\bar{\rho})}[\alpha_i]$ are exactly the non-zero vector elements of $\vec{v}$.%
\footnote{Based on the permutational covariance, $v_{(\mathbf{1},\bar{\rho})}[\alpha]$ is $\bar{\rho}v_{(\mathbf{1},\mathbf{1})}[\bar{\rho}^{-1}\alpha]$. Since $\bar{\rho}$ only act on the gluons $\{5,\ldots,n\}$ and $\alpha\in S_{n-2}/S_2$ permuting $\{3,4,5,\ldots,n\}$ without changing the relative position of $\{3,4\}$, we have $\bar{\rho}^{-1}\alpha$ also belongs to $S_{n-2}/S_2$. Therefore, $v_{(\mathbf{1},\bar{\rho})}[\alpha]$ is non-zero.}  
We see again that the ``correct" number of non-zero elements for certain $\vec{v}$ is the key.
It is interesting to notice that for the $\operatorname{tr}(\phi^{2,3})$ form factors, we have exactly the same form
\begin{equation}\label{eq:phi23FFmatdec}
    \text{Res}_{s_{1234}=q^2}\big[\mathbf{S}_n^{\mathcal{F}_{{\rm tr}(\phi^{2,3})}}[\alpha_1|\alpha_2]\big]=\sum_{\bar{\rho}_{1,2}\in S_{n-5}}v_{(\mathbf{1},\bar{\rho}_1)}[\alpha_1]
    \big(\mathbf{S}^{\mathcal{F}_{{\rm tr}(\phi^{2,3})}}_4[\mathbf{1}|\mathbf{1}]\mathbf{S}^{\cal A}_{n-2}[\bar{\rho}_1|\bar{\rho}_2]\big)
    v_{(\mathbf{1},\bar{\rho}_2)}[\alpha_2]\,,
\end{equation}
if we also confine ourselves to $\alpha_{1,2}\in S_{n-2}/S_2$. 
We give a comment here. 
From the original version of the matrix decomposition \eqref{eq:nptSFdecomposition}, one may expect the following equation for the ${\rm tr}(\phi^{2,3})$ case
\begin{align}\label{eq:phi23FFmatdec2}
    \text{Res}_{s_{1234}=q^2}\big[\mathbf{S}_n^{\mathcal{F}_{{\rm tr}(\phi^{2,3})}}&[\alpha_1|\alpha_2]\big]=\\
    &\sum_{\bar{\kappa}_{1,2}\in S_2}\sum_{\bar{\rho}_{1,2}\in S_{n-5}}v_{(\bar{\kappa}_1,\bar{\rho}_1)}[\alpha_1]
    \big(\mathbf{S}^{\mathcal{F}_{{\rm tr}(\phi^{2,3})}}_4[\bar{\kappa}_1|\bar{\kappa}_2]\mathbf{S}^{\cal A}_{n-2}[\bar{\rho}_1|\bar{\rho}_2]\big)
    v_{(\bar{\kappa}_2,\bar{\rho}_2)}[\alpha_2]\nonumber\,.
\end{align}
However, if we consider only $\alpha_{1,2}\in S_{n-2}/S_2$, then in the first sum, only $\bar{\kappa}_{1,2}=\mathbf{1}$ gives non-trivial contribution. If either one of $\bar{\kappa}_{1,2}$ is $\sigma_2$, which is the other element in $S_2$, then $v_{(\sigma_2,\bar{\rho}_{i})}[\alpha_{i}]$ is zero.\footnote{We can even go one step further. The matrix $\mathbf{S}_4^{\mathcal{F}_{{\rm tr}(\phi^{2,3})}}$ is a 2 by 2 matrix, which has 4 matrix elements. For each one of the matrix elements, we can define a $(n-2)!/2$ by $(n-2)!/2$ minor as in \eqref{eq:phi23FFmatdec} by replacing $\mathbf{S}_4^{\mathcal{F}_{{\rm tr}(\phi^{2,3})}}[\mathbf{1}|\mathbf{1}]$ with other matrix element of $\mathbf{S}_4^{\mathcal{F}_{{\rm tr}(\phi^{2,3})}}$ and change the corresponding $\vec{v}$ vector.
These four minors are exactly the four minors in \eqref{eq:SFminor}. 
}

\vskip 6pt

For general $s_{12\cdots m}{-}q^2$ cases, the diagrammatic equations \eqref{eq:varrow1} and \eqref{eq:varrow2} are always valid. 
Translating these diagrammatic equations into explicit expressions, we have 
\begin{equation}\label{eq:generalhidfac}
    \sum_{\beta \in S_{n-2}/S_{m-2}} v[\beta]  \mathcal{F}_n(1,\beta,2)\big|_{s_{12\cdots m}-q^2}=\mathcal{F}_m(1,3,\ldots,m,2) \mathcal{A}_{m'}(\QQ_m,(m{+}1),\ldots,n,-q), 
\end{equation}
and
\begin{equation}\label{eq:generalmatdec}
    \text{Res}_{\QQ_m^2=q^2}\big[\mathbf{S}_n^{\mathcal{F}}[\alpha_1|\alpha_2]\big]=\sum_{\bar{\rho}_{1,2}\in S_{m'-3}}v_{(\mathbf{1},\bar{\rho}_1)}[\alpha_1]
    \big(\mathbf{S}^{\mathcal{F}}_m[\mathbf{1}|\mathbf{1}]\mathbf{S}^{\cal A}_{m'}[\bar{\rho}_1|\bar{\rho}_2]\big)
    v_{(\mathbf{1},\bar{\rho}_2)}[\alpha_2]\,,
\end{equation}
where $\alpha_{1,2}\in S_{n-2}/S_{m-2}$, which are permutations acting on $\{3,\ldots,n\}$ but leave the relative ordering of $\{3,\ldots,m\}$ invariant. 
The form factor $\mathcal{F}$ can be any form factor that has the $s_{12\cdots m}{-}q^2$ pole after double copy, as long as we are consistently using the same class of form factors on both sides of the above equations. 

\vskip 5pt

In summary, the $\vec{v}$ vectors can be regarded as \textbf{universal} quantities that relate higher- and lower-point form factors, as indicated in \eqref{eq:varrow1} and \eqref{eq:varrow2}. They depend solely on $n$ (the number of particles) and the spurious pole $\QQ_m^2{-}q^2$. 
Such a universality may have a more profound interpretation, see discussions in Section~\ref{sec:discussion}.

\section{Towards the double copy of the ${\rm tr}(F^2)$ form factor}\label{sec:trF2}

In this section, we consider form factors of $\operatorname{tr}(F^2)$ in the pure Yang-Mills theory. 
As we will see, these form factors have very different structures and the double-copy construction requires a genuinely new prescription.

Let us start with the simple two-point minimal form factor
\begin{equation}
\mathcal{F}_{\operatorname{tr}(F^2)}(1,2) = 
({\rm f}_1)^{\mu}_{~\nu} ({\rm f}_2)^\nu_{~\mu} \,,
\end{equation}
where we recall the definition
${\rm f}_i^{\mu\nu}\equiv p_i^{\mu}\varepsilon_i^{\nu}{-}p_i^{\nu}\varepsilon_i^{\mu}$.
In this case, one can make the double copy by simply squaring the form factor as
\begin{equation}
    \mathcal{G}_2(1,2)=\left(\mathcal{F}_{\operatorname{tr}(F^2)}(1,2)\right)^2 \,, 
\end{equation}
which is well-defined and corresponds to the form factor of $R^2=R_{\mu\nu\sigma\rho}R^{\mu\nu\sigma\rho}$ operator. 
Since the form factors of $\operatorname{tr}(F^2)$ can be understood as Higgs plus gluon amplitudes in the Higgs EFT with an effective interaction vertex $H{\rm tr}(F_{\mu\nu}F^{\mu\nu})$(see Appendix~\ref{ap:scalartheory}), the double-copy quantities are expected to correspond to Higgs and graviton amplitudes with the interaction vertex $H R_{\mu\nu\sigma\rho}R^{\mu\nu\sigma\rho}$.

For higher-point non-minimal form factors, however,  the double-copy generalization is non-trivial. In  Section~\ref{ssec:3ptF2}, we will explain using the three-point example which can capture most of the salient features. In Section~\ref{ssec:4ptF2}, we discuss the more complicated four- and higher-point cases.

\subsection{The three-point case}\label{ssec:3ptF2}

We begin with the first non-minimal case: the three-point form factor. In this example, we will explicitly reveal the difficulties appearing in a naive application of the previous double-copy procedure and show how to tackle them. 
Before the concrete discussion, we would like to emphasize first that a physical double-copy construction must satisfy diffeomorphism invariance and have consistent factorization properties on all poles, which will be of central importance below.

\paragraph{Problem of an undesirable pole.}

\begin{figure}
    \centering
    \includegraphics[width=0.65\linewidth]{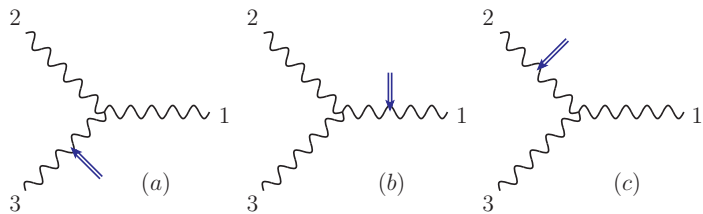}
    \caption{The cubic diagram for the three-point ${\rm tr}(F^2)$ form factor.}
    \label{fig:FsqDC3ptdiag}
\end{figure}

The three-gluon form factor of ${\rm tr}(F^2)$ in general involves three cubic diagrams as shown in Figure~\ref{fig:FsqDC3ptdiag} and has the following form
\begin{equation}\label{eq:F3trF2-1}
    \itbf{F}_{\operatorname{tr}(F^2),3}(1^{g},2^{g},3^{g})=\frac{C_a N_a}{s_{12}}+\frac{C_b N_b}{s_{13}}+\frac{C_c N_c}{s_{23}}\,.
\end{equation}
We emphasize 
that there are three cubic diagrams contributing to this form factor, which is different from the form factor ${\cal F}_{{\rm tr}(\phi^2),3}(1^\phi,2^\phi,3^g)$ involving only two diagrams.
The different numbers of diagrams will give different pole structures. 
Let us naively follow the previous double-copy procedure as in the ${\rm tr}(\phi^2)$ form factors. 
We impose the CK duality and require the numerators to satisfy
\begin{equation}\label{eq:trF23ptglobalCK}
    C_a=C_b=C_c\quad \Rightarrow \quad  N_a=N_b=N_c\,.
\end{equation}
The CK-dual numerators can be solved uniquely as
\begin{equation}
    N^{\rm CK}=N_a=N_b=N_c=\frac{s_{12}s_{13}s_{23}\mathcal{F}_{\operatorname{tr}(F^2),3}(1^{g},2^{g},3^{g})}{s_{12} s_{23}+s_{13} s_{23}+s_{12} s_{13}}\,.
\end{equation}
where $\mathcal{F}_{\operatorname{tr}(F^2),3}(1^{g},2^{g},3^{g})$ is the color-ordered form factor.
Then we make the naive double copy of \eqref{eq:F3trF2-1} as
\begin{equation}\label{eq:G3wrong}
    \mathcal{G}_{3}^{\prime}(1^{h},2^{h},3^{h})=\frac{N_a^2}{s_{12}}+\frac{N_b^2}{s_{13}}+\frac{N_c^2}{s_{23}}
    =\frac{s_{12}s_{23}s_{13}}{s_{12} s_{23}+s_{13} s_{23}+s_{12}s_{13}} \left(\mathcal{F}_{\operatorname{tr}(F^2),3}(1^{g},2^{g},3^{g})\right)^2 \,.
\end{equation}
Although this expression satisfies the diffeomorphism invariance, an obvious problem is that there is an ``unwanted" pole $(s_{12} s_{23}{+}s_{13} s_{23}{+}s_{12}s_{13})$.
This pole can no longer be understood as a massive Feynman propagator as in the ${\rm tr}(\phi^2)$ form factor.
One can also check that, after plugging in the explicit expression for $\mathcal{F}_3$ in terms of Lorentz product of $\varepsilon_i$ and $p_i$, there is no possibility to eliminate this unwanted pole.
Thus the double-copy quantity \eqref{eq:G3wrong} cannot be explained as a Higgs plus graviton amplitude or any other physical quantity with only local propagators.\footnote{One may not exclude the possibility that this new type of pole may have some higher-derivative kinetic terms or some non-local interpretation.}

\paragraph{Special MHV case.}

To understand the above problem better, we notice that 
for the MHV form factor in four dimensions, 
the form factors of ${\rm tr}(F^2)$ and ${\rm tr}(\phi^2)$ are proportional to each other:
\begin{equation}
	\mathcal{F}_{\operatorname{tr}(F^2),3}(1^{-},2^{-},3^{+})= \langle 12 \rangle^2 \mathcal{F}_{\operatorname{tr}(\phi^2),3}(1^{\phi},2^{\phi},3^{+})=\frac{\langle 12 \rangle^4}{\langle 12 \rangle \langle 23 \rangle \langle 31 \rangle}\,.
\end{equation}
In this case, one may simply define the double copy of the ${\rm tr}(F^2)$ form factor as
\begin{equation}\label{eq:R2G3MHV}
	\mathcal{G}_{3}(1^{-},2^{-},3^{+}) = \langle 12 \rangle^4 \mathcal{G}_{{\rm tr}(\phi^2),3}(1^{\phi},2^{\phi},3^{h}) \,.
\end{equation}
One can check that the  double copy defined as such indeed satisfies the factorization property and corresponds to a Higgs and three gravitons (-,-,+) amplitude.

Diagrammatically, this comes from the fact that for the MHV form factor, the diagram (a) in Figure~\ref{fig:FsqDC3ptdiag} does not contribute, because the three-point vertex coupled to three minus helicity gluons vanishes. 
In particular, the two negative-helicity gluons play the roles of the two scalars in the ${\rm tr}(\phi^2)$ form factors, thus the cubic diagrams of the MHV ${\rm tr}(F^2)$ form factors are the same as the ${\rm tr}(\phi^2)$ form factors. 
Therefore, the same propagator matrices appear of which the inverse are free of the above unwanted type of poles.
This picture also applies to higher-point MHV form factors.

\paragraph{NMHV case and general solution.}

\begin{figure}
    \centering
    \includegraphics[width=0.65\linewidth]{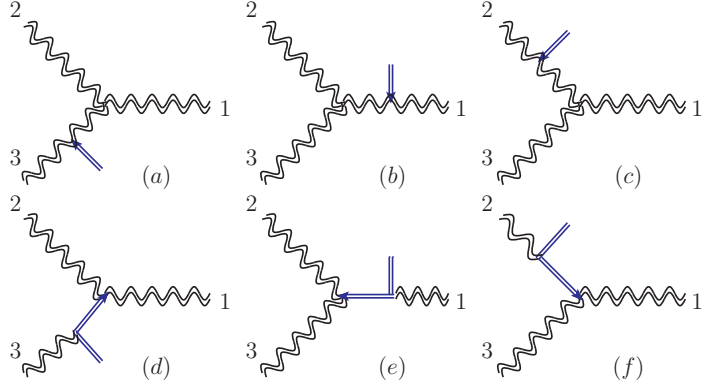}
    \caption{Feynman diagrams for the double copy of the three-point ${\rm tr}(F^2)$ form factor. }
    \label{fig:R2G3FD}
\end{figure}

Can the MHV story be generalized to the general non-MHV case?
While the structure of non-MHV form factors is certainly more complicated than the MHV cases,
the above MHV picture provides a clue. 
Let us consider the NMHV case with three minus-helicity gluons. 
Like the MHV form factors where the two minus-helicity gluons are mapped to the two external scalars (in the ${\rm tr}(\phi^2)$ form factors), is it possible to also map the NMHV form factor to the ${\rm tr}(\phi^2)$ form factors with three external scalars, of which the double copy is known?
This picture turns out to be true!

Specifically, the above idea implies that we should make connections between the ${\rm tr}(F^2)$ form factors and the ${\rm tr}(\phi^2)$ ones, in order to make use of the double copy of the latter. 
This connection is intuitively natural: one can ``pull out" part of the polarization vectors and expand the ${\rm tr}(F^2)$ form factors in terms of the ${\rm tr}(\phi^2)$ ones. For the three-point form factor, one has the following decomposition 
\begin{equation}
\begin{aligned}
\mathcal{F}_{{\rm tr}(F^2),3}(1^{-},2^{-},3^{+})&=\langle 12 \rangle^2 \mathcal{F}_{{\rm tr}(\phi^2),3}(1^{\phi},2^{\phi},3^{+}) \\
    \mathcal{F}_{{\rm tr}(F^2),3}(1^{-},2^{-},3^{-})&=\langle 12 \rangle^2\mathcal{F}_{{\rm tr}(\phi^2),3}(1^{\phi},2^{\phi},3^{-})+\langle 13 \rangle^2\mathcal{F}_{{\rm tr}(\phi^2),3}(1^{\phi},2^{-},3^{\phi})\\
    &\quad  +\langle 23 \rangle^2\mathcal{F}_{{\rm tr}(\phi^2),3}(1^{-},2^{\phi},3^{\phi})+ \langle 12 \rangle \langle 23 \rangle \langle 31 \rangle \mathcal{F}_{{\rm tr}(\phi^2),3}(1^{\phi},2^{\phi},3^{\phi}).
\end{aligned}
\end{equation}
These two equations  can be encoded in one unified $D$-dimensional form 
\begin{equation}\label{eq:trF2Decomp}
\begin{aligned}
	    \mathcal{F}_{{\rm tr}(F^2),3}(1^{g},2^{g},3^{g})=&{\rm tr}^{\rm f}(1,2)\mathcal{F}_{{\rm tr}(\phi^2),3}(1^{\phi},2^{\phi},3^{g})+{\rm tr}^{\rm f}(1,3)\mathcal{F}_{{\rm tr}(\phi^2),3}(1^{\phi},2^{g},3^{\phi})\,  \\
    &+ {\rm tr}^{\rm f}(2,3)\mathcal{F}_{{\rm tr}(\phi^2),3}(1^{g},2^{\phi},3^{\phi})+2{\rm tr}^{\rm f}(1,2,3)\mathcal{F}_{{\rm tr}(\phi^2),3}(1^{\phi},2^{\phi},3^{\phi}),
\end{aligned}
\end{equation}
where we introduce the notation ${\rm tr}^{\rm f}(i_1,..,i_k)$ as 
\begin{equation}
    {\rm tr}^{\rm f}(i_1,..,i_k)={\rm f}_{i_1,\mu_1}^{\mu_k}{\rm f}_{i_2,\mu_2}^{\mu_1}\cdots {\rm f}_{i_k,\mu_k}^{\mu_{k-1}}\,.
\end{equation}

Now let us see how to make use of such a connection. 
The double copy of all the scalar-Yang-Mills blocks in the RHS of \eqref{eq:trF2Decomp} are known from previous discussions. 
We use the special cubic-diagram form that manifests CK-duality for all the scalar form factor blocks, such as $\mathcal{F}_{{\rm tr}(\phi^2),3}(1^{\phi},2^{\phi},3^{g}) = N_1\times(1/s_{13}+1/s_{23})$.
As a result, we have the following \emph{pre-double-copy form}, which is a specific cubic diagram expansion with numerators composed of CK-dual blocks
\begin{align}
    \mathcal{F}_{{\rm tr}(F^2),3}(1^{g},2^{g},3^{g})=& \ {\rm tr}^{\rm f}(1,2)\left(\frac{N_1}{s_{13}}+\frac{N_1}{s_{23}}\right)+{\rm tr}^{\rm f}(1,3)\left(\frac{N_2}{s_{12}}+\frac{N_2}{s_{23}}\right)  \nonumber \\
    & \ + {\rm tr}^{\rm f}(2,3)\left(\frac{N_3}{s_{13}}+\frac{N_3}{s_{12}}\right)+2{\rm tr}^{\rm f}(1,2,3)\left(\frac{1}{s_{12}}+\frac{1}{s_{23}}+\frac{1}{s_{13}}\right)\nonumber\\
    =& \ 
    \frac{{\rm tr}^{\rm f}(1,2)N_1+{\rm tr}^{\rm f}(2,3)N_3+2{\rm tr}^{\rm f}(1,2,3)}{s_{13}}+
    {\rm cyc}(1,2,3) , \label{eq:F23ptCK}
\end{align}
where $N_{1,2,3}$ are the CK numerators given in \cite{Lin:2022jrp}
\begin{equation}
    N_{1}=\frac{2p_2\cdot {\rm f}_3\cdot p_1}{s_{12}-q^2}, \qquad N_{2}=\frac{2p_1\cdot {\rm f}_2\cdot p_3}{s_{13}-q^2}, \qquad N_{3}=\frac{2p_3\cdot {\rm f}_1\cdot p_2}{s_{23}-q^2}\,.
\end{equation}
And the full-color three-point form factor is simply
\begin{equation}\label{eq:F23ptCK2}
    \itbf{F}_{{\rm tr}(F^2),3}(1^{g},2^{g},3^{g})=f^{123}\mathcal{F}_{{\rm tr}(F^2),3}(1^{g},2^{g},3^{g})\,,
\end{equation}
since the three cubic diagrams all contribute to $\mathcal{F}_3$ and have identical color factors.

We need to point out that, given the numerator representation in \eqref{eq:F23ptCK}, \eqref{eq:F23ptCK2} does not satisfy the ``global" CK-duality like \eqref{eq:trF23ptglobalCK}. 
To satisfy that duality, the numerators of the three cubic diagrams with propagators $s_{12}$, $s_{13}$ and $s_{23}$ have to be identical. 
The three numerators in \eqref{eq:F23ptCK} (and thus in \eqref{eq:F23ptCK2}) are related by the cyclic permutation, so that the numerator $N_{s_{13}}={\rm tr}^{\rm f}(1,2)N_1+{\rm tr}^{\rm f}(2,3)N_3+2{\rm tr}^{\rm f}(1,2,3)$ would have to be cyclic invariant to make the CK-duality valid. 
By examining the explicit expression, however, it is not the case. 
Therefore, we would not have a meaningful double-copy result by simply squaring these numerators. 

To cure this problem, we inspect the representation \eqref{eq:F23ptCK} and notice the sub-blocks in the numerators are gauge invariant. 
This implies that we may relax the requirement of CK duality while we can still keep the double copy quantity diffeomorphism invariant.
Concretely, we note that there are three terms in the numerator $N_{s_{13}}$ of the $s_{13}$ propagator in \eqref{eq:F23ptCK}, denoted as $\tilde{n}_i$
\begin{equation}
\tilde{n}_1=N_{1} {\rm tr}^{\rm f}(1,2) \,, \qquad \tilde{n}_2=N_3 {\rm tr}^{\rm f}(2,3) \,, \qquad \tilde{n}_3= {\rm tr}^{\rm f}(1,2,3) \,.
\end{equation} 
Rather than squaring the full numerator $N_{s_{13}}$, we propose an ansatz of the double-copy numerator given in a form allowing quadratic mixing between different $\tilde{n}_i$'s:
\begin{equation}\label{eq:N3ansatztrF2}
    \mathcal{N}_{s_{13}}^{\rm ansatz}=\sum_{i,j=1}^{3}M_{ij}\tilde{n}_i\tilde{n}_j \,,
\end{equation}
where $M_{ij}$ are just rational numbers.
We will refer to $M$ as the mixing coefficient matrix. 
The double copy quantity can be obtained as
\begin{equation}\label{eq:R2G3ansatz}
    \mathcal{G}_{3}^{\rm ansatz}=\frac{\mathcal{N}_{s_{13}}^{\rm ansatz}}{s_{13}}+{\rm cyc}(1,2,3)\,.
\end{equation}
The above ansatz form can be viewed as a generalization of the usual double copy operation where one simply replaces color factors with dual kinematic numerators. Here the CK duality is only used at the level of scalar form factor blocks. As already emphasized above, since our ansatz is built out of gauge invariant blocks, the double-copy result is manifestly invariant under the linear diffeomorphism transformation. 

Given the ansatz, we still need to solve for the constant matrix $M_{ij}$ in \eqref{eq:N3ansatztrF2}. We achieve this by requiring \eqref{eq:R2G3ansatz} to have consistent factorization properties.
Also, we expect that the result of \eqref{eq:R2G3ansatz} should match the diagrams in Figure~\ref{fig:R2G3FD}. 
We need to mention that it is a priori not known whether the ansatz works \eqref{eq:R2G3ansatz}|this needs to be justified by explicit calculations. 
Fortunately, after inspecting all factorization relations (in four-dimensional kinematics), we find there exists indeed a solution; in this case, it is also unique.
The mixing coefficient matrix $M_{ij}$ is  
\begin{equation}
M = \begin{pmatrix}
1~ & 0~ & 2~ \\
0 & 1 & 2 \\
2 & 2 & 6 
\end{pmatrix} \,.
\end{equation}
We point out that the definition of the matrix $M_{ij}$ in \eqref{eq:N3ansatztrF2} is not unique, and we will choose the matrix to be symmetric (which is always possible) just for convenience. 
The double-copy result is given as
\begin{equation}\label{eq:R2G3resD}
\begin{aligned}
    \mathcal{G}_3(1^{h},2^{h},3^{h})=&\frac{1}{s_{13}}\big[({\rm tr}^{\rm f}(1,2)N_1)^2+({\rm tr}^{\rm f}(2,3)N_3)^2+4{\rm tr}^{\rm f}(1,2,3){\rm tr}^{\rm f}(2,3)N_3+\\
    &\qquad\quad  4{\rm tr}^{\rm f}(1,2,3){\rm tr}^{\rm f}(1,2)N_1 + 6 ({\rm tr}^{\rm f}(1,2,3))^2\big] +{\rm cyc}(1,2,3)\,.
\end{aligned}
\end{equation}
We can specify it to the MHV and NMHV helicity sectors, which are given respectively as\footnote{Note that ${\rm tr}^{\rm f}(-+)$ or ${\rm tr}^{\rm f}(--+)$ are zero, which explains why the MHV case is simple.} 
\begin{align}\label{eq:R2G3res}
    \mathcal{G}_3(1^{-},2^{-},3^{+})=&\frac{({\rm tr}^{\rm f}(1^{\scriptscriptstyle -},2^{\scriptscriptstyle -})N_1^{\scriptscriptstyle \phi\phi+})^2}{s_{13}}+\frac{({\rm tr}^{\rm f}(1^{\scriptscriptstyle -},2^{\scriptscriptstyle -})N_1^{\scriptscriptstyle \phi\phi+})^2}{s_{23}}\,,\\
    \mathcal{G}_3(1^{-},2^{-},3^{-})=&\frac{1}{s_{13}}\big[({\rm tr}^{\rm f}(1^{\scriptscriptstyle -},2^{\scriptscriptstyle -})N_1^{\scriptscriptstyle \phi\phi-})^2+({\rm tr}^{\rm f}(2^{\scriptscriptstyle -},3^{\scriptscriptstyle -})N_3^{\scriptscriptstyle -\phi\phi})^2+4{\rm tr}^{\rm f}(1^{\scriptscriptstyle -},2^{\scriptscriptstyle -},3^{\scriptscriptstyle -}){\rm tr}^{\rm f}(2^{\scriptscriptstyle -},3^{\scriptscriptstyle -})N_3^{\scriptscriptstyle -\phi\phi} \nonumber \\
    &\qquad + 4{\rm tr}^{\rm f}(1^{\scriptscriptstyle -},2^{\scriptscriptstyle -},3^{\scriptscriptstyle -}){\rm tr}^{\rm f}(1^{\scriptscriptstyle -},2^{\scriptscriptstyle -})N_1^{\scriptscriptstyle \phi\phi-} + 6 ({\rm tr}^{\rm f}(1^{\scriptscriptstyle -},2^{\scriptscriptstyle -},3^{\scriptscriptstyle -}))^2\big] +{\rm cyc}(1,2,3)\,. \nonumber
\end{align}
Here $N_i^{\scriptscriptstyle \phi\phi+}$ and $N_i^{\scriptscriptstyle \phi\phi-}$ are the four-dimensional CK-dual numerators with special helicities, \emph{e.g.}
\begin{equation}
N_1^{\scriptscriptstyle \phi\phi+} = 2 \times {\langle 12\rangle [13] [32] \over s_{12}-q^2} \,, \qquad N_1^{\scriptscriptstyle \phi\phi-} = 2 \times {\langle 32\rangle [12] \langle 13\rangle \over s_{12}-q^2} \,.
\end{equation}
The MHV expression is consistent with the previously discussed result \eqref{eq:R2G3MHV}.
Moreover, both the MHV and NMHV results given by $\mathcal{G}_3$ in \eqref{eq:R2G3resD} match the Feynman diagram results as shown in Figure~\ref{fig:R2G3FD}. 
In particular, it satisfies the factorization on the spurious-type poles like $s_{12}{-}q^2$ in the second line diagrams in Figure~\ref{fig:R2G3FD}.

\paragraph{Subtlety in $D$ dimensions.}
It may be tempting to conclude that this $\mathcal{G}_3(1^{g},2^{g},3^{g})$ in \eqref{eq:R2G3resD} is just the object that we are looking for. 
But one should still ask whether the four-dimensional valid result is true in $D$ dimensions. 
Indeed, it shares most of the desired properties. For example, the spurious-type poles like $s_{12}{-}q^2$ are simple poles and $\mathcal{G}_3$ has consistent factorization on those poles. 
However, when checking the massless poles like $s_{12}$, we find no solution from our ansatz \eqref{eq:R2G3ansatz} that matches the $D$-dimensional tree products. 

Since the solution \eqref{eq:R2G3resD} is already consistent in four dimensions, 
the missing terms can only be given in an expression that is non-zero in $D$ dimensional kinematics but vanishes in four dimensions. Such terms will be named as \emph{evanescent corrections}.\footnote{A systematic construction of gluonic evanescent operators in YM theory, as well as their relation to $D$-dimensional form factors, has been recently studied in \cite{Jin:2022ivc, Jin:2022qjc}.}
It turns out that an evanescent correction $\Delta_3$ is needed to make the $D$-dimensional factorization correct: 
\begin{equation}\label{eq:R2G3cut}
   \text{Res}_{s_{12}=0} \Big( \mathcal{G}_{3}+\Delta_{3} \Big)=\sum_{{\rm helicity} I}\mathcal{G}_2(3^{h},P^I(\varepsilon)) 
   \mathcal{M}_3(-P^I(\bar{\varepsilon}),1^{h},2^{h})\,,
\end{equation}
where
\begin{equation}
    \Delta_{3}=\left(\frac{1}{s_{12}}+\frac{1}{s_{23}}+\frac{1}{s_{13}}\right){\rm Gram}(p_1,p_2,p_3,\varepsilon_1,\varepsilon_2,\varepsilon_3)
\end{equation}
with the Gram determinant defined as 
\begin{equation}
    {\rm Gram}(u_1,\ldots,u_r)=\det\big((T_{ab})_{r\times r}\big)\,, \ \text{with } T_{ab}=u_a\cdot u_b,\ a,b=1,\ldots,r\,.
\end{equation}
Clearly, ${\rm Gram}(p_1,p_2,p_3,\varepsilon_1,\varepsilon_2,\varepsilon_3)$ (and thus $\Delta_{3}$) is zero in four dimensions.\footnote{${\rm Gram}(p_1,p_2,p_3,\varepsilon_1,\varepsilon_2,\varepsilon_3)$ can be viewed as the on-shell presentation for the gravitational operator 
$\delta_{\mu_1\mu_2\mu_3\mu_4\mu_5\mu_6}^{\nu_1\nu_2\nu_3\nu_4\nu_5\nu_6} R^{\mu_1\mu_2}_{\qquad\nu_1\nu_2}R^{\mu_3\mu_4}_{\qquad\nu_3\nu_4}R^{\mu_5\mu_6}_{\qquad\nu_5\nu_6}$, where $\delta_{\mu_1\mu_2\mu_3\mu_4\mu_5\mu_6}^{\nu_1\nu_2\nu_3\nu_4\nu_5\nu_6} = \epsilon_{\mu_1\mu_2\mu_3\mu_4\mu_5\mu_6} \epsilon^{\nu_1\nu_2\nu_3\nu_4\nu_5\nu_6}$ is the rank-6 Kronecker symbol, and $R^{\mu_1\mu_2}_{\qquad\nu_1\nu_2}$ corresponds to $(p^{\mu_1}\varepsilon^{\mu_2}-p^{\mu_2}\varepsilon^{\mu_1})(p_{\nu_1}\varepsilon_{\nu_2}-p_{\nu_2}\varepsilon_{\nu_1})$. The Kronecker symbol makes it clear that the Gram vanishes in four dimensions. It is straightforward to generalize to higher-point Gram with higher-rank Kornecker symbols.}

Finally, we comment that given the ansatz $\mathcal{G}_{3}^{\rm ansatz}$ in \eqref{eq:R2G3ansatz}, there is no way to get $\Delta_3$ by adjusting the $M_{ij}$ matrix therein. 
Therefore, the original ansatz can only produce a consistent double copy in four dimensions. It needs to be complemented by some evanescent terms to make it valid in $D$ dimensions.

\subsection{Higher-point generalizations}\label{ssec:4ptF2}

\begin{figure}
    \centering
    \includegraphics[width=0.8\linewidth]{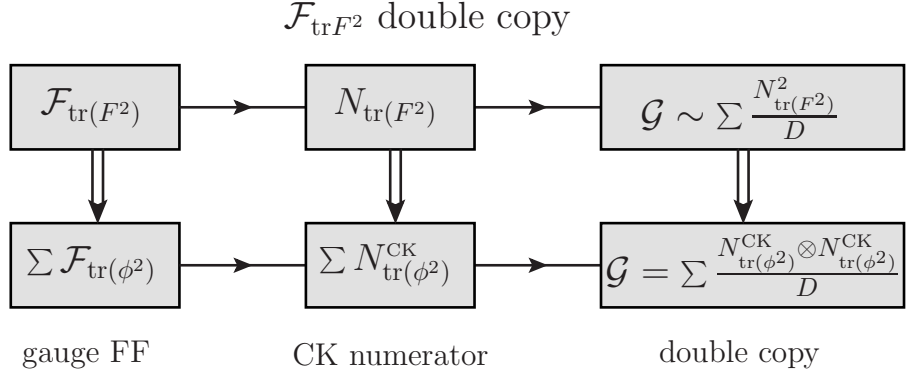}
    \caption{
    An illustrative figure for the main strategy of the double copy prescription for the ${\rm tr}(F^2)$ form factor.
    First we relate the ${\rm tr}(F^2)$ form factor to the ${\rm tr}(\phi^2)$ ones via a gauge invariant decomposition.
    Next we express the ${\rm tr}(F^2)$ numerators in terms of ${\rm tr}(\phi^2)$ CK numerators. 
    Then the double-copy numerators are written in the general quadratic form of those scalar numerators.
    }
    \label{fig:F2DCgeneral}
\end{figure}

We can generalize the above procedure to higher points. 
Let us recapitulate the main steps used in the previous three-point example: 
\begin{enumerate}[topsep=3pt,itemsep=-1ex,partopsep=1ex,parsep=1ex]
    \item First, we use a gauge-invariant decomposition to decompose the ${\rm tr}(F^2)$ form factor into the ${\rm tr}(\phi^2)$ ones, with coefficients encoding the polarization information given in terms of traces of field strength, as in \eqref{eq:trF2Decomp}.
    \item Next, we take the CK-dual numerators of the ${\rm tr}(\phi^2)$ ones and make a numerator ansatz in general quadratic form of those scalar numerators, as in \eqref{eq:N3ansatztrF2}.
    A (numeric) mixing matrix $M$ is introduced.
    \item Finally, we solve for the ansatz, in particular the numbers in the matrix $M$, by requiring the double copy ansatz to be consistent with all factorization conditions (at least in four-dimensional kinematics). 
\end{enumerate} 
The three-point example is fully symmetric and relatively simple.
Below we will consider in detail the four-point case which can provide more explanations needed for the higher-point cases. 
For the reader's convenience, we also summarize the above strategy for performing double copy in Figure~\ref{fig:F2DCgeneral}.

\paragraph{1) The gauge-invariant decomposition.}
To express the decomposition in different helicity sectors, we follow the same strategy and obtain: 
\begin{align}\label{eq:4pttrF21}
	{\cal F}_{\operatorname{tr}(F^2),4}(1^{-},2^{-},3^{+},4^{+})=&\langle 12\rangle^2 {\cal F}_{\operatorname{tr}(\phi^2),4}(1^{\phi},2^{\phi},3^{+},4^{+}),\nonumber \\
	{\cal F}_{\operatorname{tr}(F^2),4}(1^{-},2^{-},3^{-},4^{+})=&\langle 12\rangle^2 {\cal F}_{\operatorname{tr}(\phi^2),4}(1^{\phi},2^{\phi},3^{-},4^{+})+\langle 13\rangle^2 {\cal F}_{\operatorname{tr}(F^2),4}(1^{\phi},2^{-},3^{\phi},4^{+})\nonumber \\
	&+\langle 23\rangle^2 {\cal F}_{\operatorname{tr}(F^2),4}(1^{-},2^{\phi},3^{\phi},4^{+})+
	\langle 12 \rangle \langle 23 \rangle \langle 31\rangle {\cal F}_{\operatorname{tr}(\phi^2),4}(1^{\phi},2^{\phi},3^{\phi},4^{+}), \nonumber \\
	{\cal F}_{\operatorname{tr}(F^2),4}(1^{-},2^{-},3^{-},4^{-})=&\sum_{i<j}\langle i j \rangle^2 \mathcal{F}_{\operatorname{tr}(\phi^2),4}(\ldots,i^{\phi},\ldots,j^{\phi},\ldots)\\
	&+\langle 12 \rangle \langle 23 \rangle \langle 31\rangle {\cal F}_{\operatorname{tr}(\phi^2),4}(1^{\phi},2^{\phi},3^{\phi},4^{-})+\mathrm{cyc}(1,2,3,4)\nonumber\\
	&+\left(\langle 12 \rangle^2 \langle 34\rangle^2-\langle 13 \rangle^2 \langle 24\rangle^2\right)
 \mathcal{F}_{\rm a}^{\phi\phi\phi\phi}
 +\left(\langle 14 \rangle^2 \langle 23\rangle^2-\langle 13 \rangle^2 \langle 24\rangle^2\right) \mathcal{F}_{\rm b}^{\phi\phi\phi\phi}\,.
 \nonumber
\end{align}
Since we need to consider CK duality later, we would like to  present the cubic diagram representation for these  ${\rm tr}(\phi^2)$ form factor blocks as  
\begin{equation}\label{eq:4ptYMSFFblocks}
\begin{aligned}
     \includegraphics[width=0.8\linewidth]{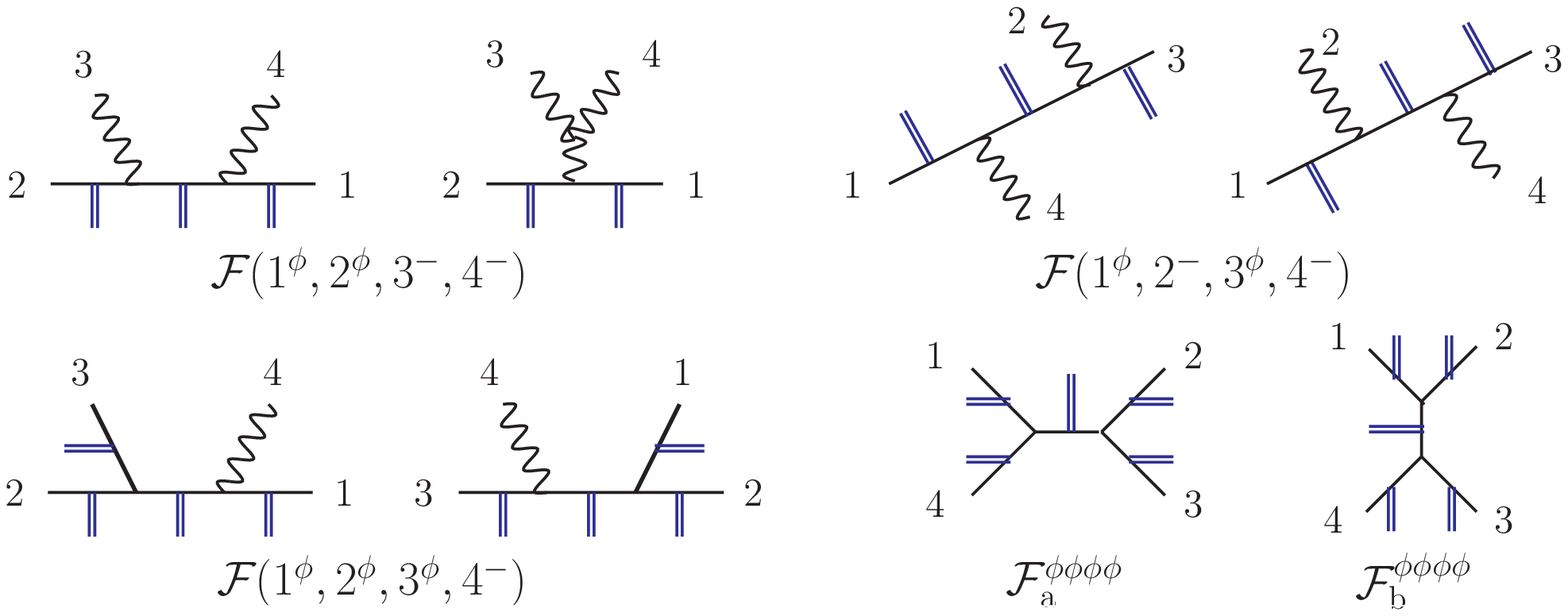}
\end{aligned}
\end{equation}
in which we have used the diagrammatic convention that:
a single diagram with multiple double-line $q$-legs is actually a sum of diagrams containing a single $q$-leg in all positions.
For example,
\begin{equation}
\begin{aligned}
        \includegraphics[width=0.89\linewidth]{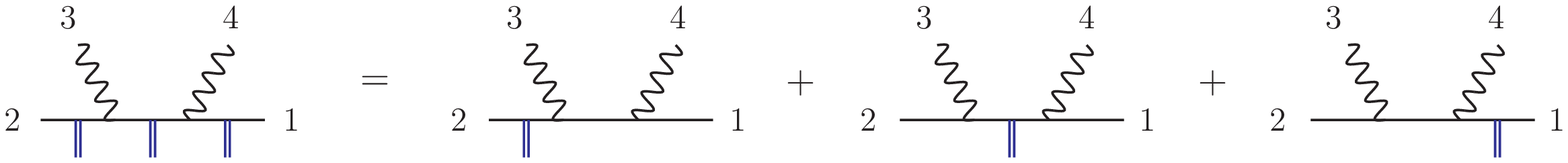}
	\end{aligned} .
\end{equation}
Note that the last two $\mathcal{F}_{\rm a,b}^{\phi\phi\phi\phi}$ in \eqref{eq:4ptYMSFFblocks} are two four-scalar blocks, for example,
\begin{equation}
\mathcal{F}_{\rm a}^{\phi\phi\phi\phi} =  {1\over s_{23}s_{123}} + {1\over s_{23}s_{234}} +  {1\over s_{14}s_{234}} +  {1\over s_{14}s_{412}} +  {1\over s_{14}s_{23}} \,.
\end{equation}

Moreover, the $D$-dimensional version encoding all the helicity configuration in \eqref{eq:4pttrF21} is%
\begin{align}\label{eq:4pttrF22}
	{\cal F}_{\operatorname{tr}(F^2),4}&(1^g,2^g,3^g,4^g)=\Big(\sum_{i<j}\operatorname{tr}^{\rm f}(i,j) \mathcal{F}_{\operatorname{tr}(\phi^2),4}(\ldots,i^{\phi},\ldots,j^{\phi},\ldots)\Big)\\
	&+2\Big(\operatorname{tr}^{\rm f}(1,2,3)  {\cal F}_{\operatorname{tr}(\phi^2),4}(1^{\phi},2^{\phi},3^{\phi},4^{g})+\mathrm{cyc}(1,2,3,4)\Big) \nonumber\\
	&+ 2\Big(\big(\operatorname{tr}^{\rm f}(1,2,3,4) -\operatorname{tr}^{\rm f}(4,2,3,1)\big)
  \mathcal{F}_{\rm a}^{\phi\phi\phi\phi}
 +\big(\operatorname{tr}^{\rm f}(1,2,3,4) -\operatorname{tr}^{\rm f}(2,1,3,4)\big)
 \mathcal{F}_{\rm b}^{\phi\phi\phi\phi}
 \Big)\,, \nonumber
\end{align}
which goes back to \eqref{eq:4pttrF21} by specifying to appropriate helicity sectors.

We mention that the gauge invariant decomposition can be generalized to higher points conveniently if we rewrite it in terms of double-ordered form factors in the Yang-Mills-Scalar+bi-adjoint-$\phi^3$ theory. The expansion is (see \cite{Dong:2022bta} for a systematic study)
\begin{align}
     {\cal F}_{\operatorname{tr}(F^2),4}(1^{g},2^{g},3^{g},4^{g})&
     =
      \sum_{i_1<i_2}\operatorname{tr}^{\rm f}(i_1,i_2) F_4^{\operatorname{tr}(\phi^2)}(i_1,i_2|1,2,3,4) \nonumber\\
     &+2\sum_{i_1<i_2<i_3}\operatorname{tr}^{\rm f}(i_1,i_2,i_3) F_4^{\operatorname{tr}(\phi^2)}(i_1,i_2,i_3|1,2,3,4) \nonumber\\
     &+2\sum_{\sigma\in S_{4}/(\mathbb{Z}_4\times S_2)}\operatorname{tr}^{\rm f}(\sigma(1,2,3,4)) F_4^{\operatorname{tr}(\phi^2)}(\sigma(1,2,3,4)|1,2,3,4) \nonumber \,,
\end{align}
where the $F(\alpha|\beta)$ are the double ordered form factors. 
For the purpose of the present paper, however, we still prefer the diagrammatic representation in \eqref{eq:4ptYMSFFblocks}.

\paragraph{2) CK-dual numerators for the ${\rm tr}(\phi^2)$ form factors.}

With the gauge invariant decomposition, our next target is to write the gauge-theory form factors in a form similar to \eqref{eq:F23ptCK}, where the numerators of cubic diagrams are composed of gauge-invariant CK-dual blocks. 
We first directly spell out such blocks and then explain where they are from. 
The three types of blocks are 
\begin{equation}
\begin{aligned}\label{eq:numstrF24pt}
    \tilde{n}&={\rm tr}^{\rm f}(a,b)N_1(a^{\phi},c^{g},d^{g},b^{\phi}) \,, \\  
    \tilde{n}^{\prime}&={\rm tr}^{\rm f}(a,b,c)N_2(a^{\phi},d^{g},b^{\phi},c^{\phi}) \,,\\  
    \tilde{n}^{\prime\prime }&={\rm tr}^{\rm f}(a,b,c,d)-{\rm tr}^{\rm f}(b,a,c,d) \,,
\end{aligned}
\end{equation}
with $a,b,c,d$ should be properly chosen  as a permutation of $1,2,3,4$, and 
\begin{align}\label{eq:4pttrF2CKnum1}
	N_1(a^{\phi},c^{g},d^{g},b^{\phi})=&-\frac{2\left({\rm f}_{c}^{\mu\nu}{\rm f}_{d,\nu\rho}p_{a,\mu}p_{b}^{\rho}\right)}{(s_{ab}-q^2)}+ 
	\frac{4\left({\rm f}_{c}^{\mu\nu}p_{a,\mu}p_{b,\nu}\right)\left({\rm f}_{d}^{\mu\nu}p_{b,\mu}q_{\nu}\right)}{(s_{ab}-q^2)(s_{abc}-q^2)}+\frac{4\left({\rm f}_{d}^{\mu\nu}p_{a,\mu}p_{b,\nu}\right)\left({\rm f}_{c}^{\mu\nu}p_{a,\mu}q_{\nu}\right)}{(s_{ab}-q^2)(s_{abd}-q^2)}\,,\nonumber \\
	N_2(a^{\phi},d^{g},b^{\phi},c^{\phi})=&-\frac{2 {\rm f}_{d,\mu\nu}p_{a}^{\mu}(p_{b}+p_{c})^{\nu}}{(s_{abc}-q^2)}\,.
\end{align}

These numerator blocks emerge from the CK-dual representation of the form factors appearing in the gauge-invariant decomposition \eqref{eq:4pttrF21}.
Let us look at some examples. 
For the following $\phi^2$ form factors, one can take the manifestly CK-dual representation as 
\begin{align}\label{eq:trF24ptblockexp}
	&{\cal F}_{\operatorname{tr}(\phi^2),4}(1^{\phi},2^{\phi},3^{g},4^{g})=\\
	&\quad N_1(2^{\phi},3^{g},4^{g},1^{\phi})
  P\bigg(\hskip -3pt 
 \begin{aligned}
        \includegraphics[width=0.155\linewidth]{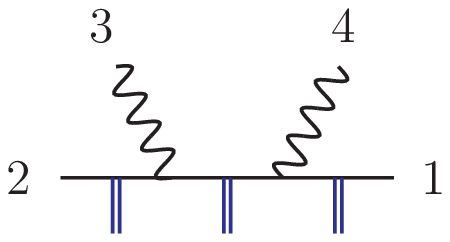}
	\end{aligned}
 \hskip -3pt 
 \bigg)
 \hskip -2pt +\big (N_1(2^{\phi},3^{g},4^{g},1^{\phi}) -N_1(2^{\phi},4^{g},3^{g},1^{\phi})\big)P\bigg(\hskip -3pt\begin{aligned}
        \includegraphics[width=0.13\linewidth]{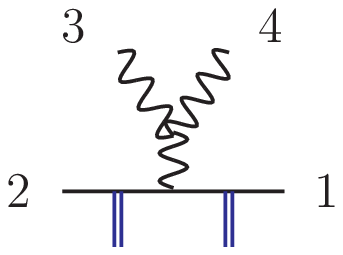}
	\end{aligned}
 \hskip -3pt
 \bigg)
 \nonumber \\
 &{\cal F}_{\operatorname{tr}(\phi^2),4}(1^{\phi},2^{\phi},3^{g},4^{g})=\nonumber \\
	&\quad N_1(2^{\phi},3^{g},4^{g},1^{\phi}) P\bigg(\hskip -3pt\begin{aligned}
        \includegraphics[width=0.15\linewidth]{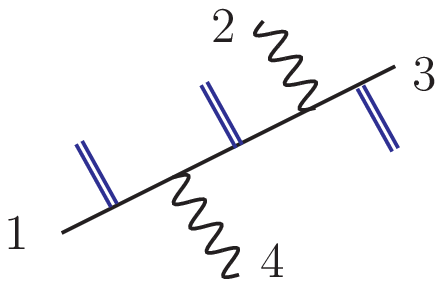}
	\end{aligned}
 \hskip -2pt
 \bigg)
 \hskip -2pt +\big (N_1(2^{\phi},3^{g},4^{g},1^{\phi}) -N_1(2^{\phi},4^{g},3^{g},1^{\phi})\big) P\bigg(\hskip -3pt\begin{aligned}
        \includegraphics[width=0.15\linewidth]{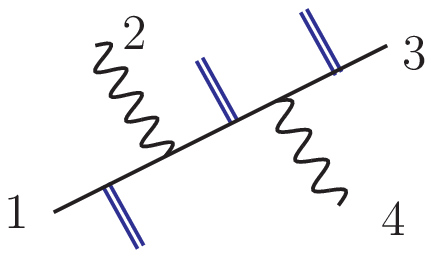}
	\end{aligned}\hskip -3pt \bigg) \nonumber \\
&{\cal F}_{\operatorname{tr}(\phi^2),4}(1^{\phi},2^{\phi},3^{\phi},4^{g})=N_2(1^{\phi},2^{\phi},4^{g},3^{\phi})
 P\bigg(\hskip -3pt 
 \begin{aligned}
        \includegraphics[width=0.15\linewidth]{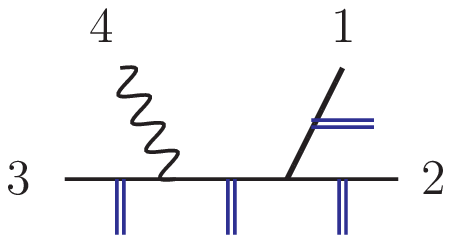}
	\end{aligned}
 \hskip -4pt \bigg)
 \hskip -2pt +N_2(2^{\phi},3^{\phi},4^{g},1^{\phi})
 P\bigg(\hskip -3pt 
 \begin{aligned}
        \includegraphics[width=0.155\linewidth]{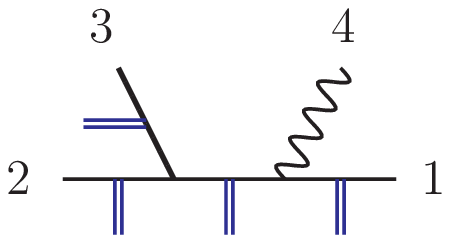}
	\end{aligned} 
 \hskip -4pt\bigg) \nonumber \\
	&{\cal F}_{\operatorname{tr}(\phi^2),4{\rm a}}^{\phi\phi\phi\phi}=
P\bigg(\hskip -3pt\begin{gathered}
        \includegraphics[width=0.15\linewidth]{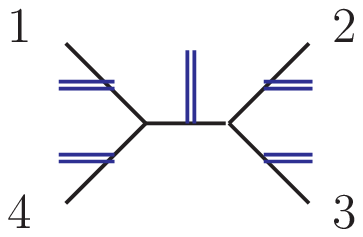}
\end{gathered}\hskip -3pt \bigg),
\qquad
{\cal F}_{\operatorname{tr}(\phi^2),4{\rm b}}^{\phi\phi\phi\phi}=
P\bigg(\hskip -3pt \begin{gathered}
        \includegraphics[width=0.12\linewidth]{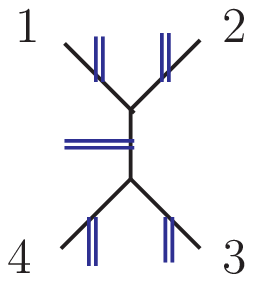}
\end{gathered}\hskip -3pt \bigg).  \nonumber
\end{align}
where $P(\Gamma)$ means the propagator corresponding to the cubic diagram $\Gamma$. 
For the detailed derivations, see \cite{Lin:2022jrp}. 

Substituting these form factors back in the expansion \eqref{eq:4pttrF22}, we see that first we also get a cubic diagram form for the ${\rm tr}(F^2)$ form factors, and then the corresponding ${\rm tr}(F^2)$ numerators now include the ${\rm tr}^{\rm f}$ factor and the  ${\rm tr}(F^2)$ CK-dual numerators $N_{1,2}$.
This means that the first two rows of \eqref{eq:numstrF24pt} are recovered. 
As for the third one in \eqref{eq:numstrF24pt}, it naturally comes from the ${\cal F}^{\phi\phi\phi\phi}$ terms in the expansion \eqref{eq:4pttrF22}. 
Interestingly, although the numerators of pure scalar form factors are trivial, the corresponding ${\rm tr}^{\rm f}$ factors do bear the CK structures.
Concretely, the  ${\rm tr}^{\rm f}$  factor before $\mathcal{F}_{\rm a}^{\phi\phi\phi\phi}$ is $\operatorname{tr}^{\rm f}(1,2,3,4) -\operatorname{tr}^{\rm f}(4,2,3,1)$, the ${\rm tr}^{\rm f}$  factor before $\mathcal{F}_{\rm b}^{\phi\phi\phi\phi}$ is $\operatorname{tr}^{\rm f}(1,2,3,4) -\operatorname{tr}^{\rm f}(4,2,3,1)$, and we can exchange $2\leftrightarrow 1$ in $\mathcal{F}_{\rm a}^{\phi\phi\phi\phi}$ and get a new diagram with the following  $\operatorname{tr}^{\rm f}$ factor $\operatorname{tr}^{\rm f}(1,3,4,2) -\operatorname{tr}^{\rm f}(1,3,2,4)$. 
Therefore, we have the following three diagrams
\begin{equation}
    \quad  \begin{aligned}
        \includegraphics[width=0.17\linewidth]{figure/s4q.eps}
    \end{aligned} \qquad \qquad \qquad \quad
    \begin{aligned}
        \includegraphics[width=0.15\linewidth]{figure/t4q.eps}
    \end{aligned} \qquad \qquad \qquad \qquad
    \begin{aligned}
        \includegraphics[width=0.17\linewidth]{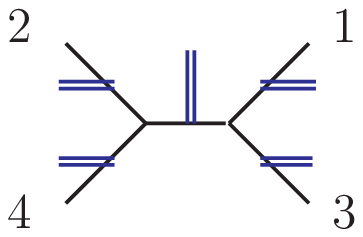}
    \end{aligned}
\end{equation}
with the associated ${\rm tr}^{\rm f}$ factor satisfying 
\begin{align}
    &\big[\operatorname{tr}^{\rm f}(1,2,3,4) -\operatorname{tr}^{\rm f}(4,2,3,1)\big] 
    - \big[\operatorname{tr}^{\rm f}(1,2,3,4) -\operatorname{tr}^{\rm f}(2,1,3,4)\big]
    =\big[\operatorname{tr}^{\rm f}(1,3,4,2) -\operatorname{tr}^{\rm f}(1,3,2,4)\big]\nonumber
\end{align}
which looks exactly the same as the dual Jacobi relation for the three $s,t,u$-channel diagrams in 4-point amplitudes. 

In the end, we see conclusively that the ${\rm tr}(F^2)$ numerators are indeed composed of the blocks in \eqref{eq:numstrF24pt}. 
For example, for the diagram$\begin{aligned}
    \includegraphics[width=0.12\linewidth]{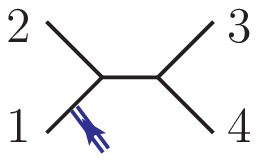}
\end{aligned}$, the numerator is 
\begin{align}\label{eq:trF24ptPDC1}
	&{\rm tr}^{\rm f}(1,4)N_1(1^{\phi},2^{g},3^{g},4^{\phi})-{\rm tr}^{\rm f}(1,3)N_1(1^{\phi},2^{g},4^{g},3^{\phi})+\nonumber \\
	&{\rm tr}^{\rm f}(1,2)\big[ N_1(1^{\phi},3^{g},4^{g},2^{\phi})-N_1(1^{\phi},4^{g},3^{g},2^{\phi}) \big] +{\rm tr}^{\rm f}(1,2,3)N_2(1^{\phi},2^{\phi},3^{g},4^{\phi})+\\
	&{\rm tr}^{\rm f}(1,2,3)N_2(4^{\phi},3^{\phi},2^{g},1^{\phi})-{\rm tr}^{\rm f}(1,2,3)N_2(1^{\phi},2^{\phi},4^{g},3^{\phi})+\big[\operatorname{tr}^{\rm f}(1,2,3,4) -\operatorname{tr}^{\rm f}(2,1,3,4)\big] \nonumber
\end{align}
with $3+3+1=7$ blocks in total. Another type of cubic diagram is like $\begin{aligned}
    \includegraphics[width=0.12\linewidth]{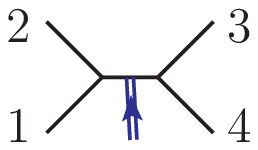}
\end{aligned}$, and there are $4+4+1=9$ blocks in the numerator.

\paragraph{3) Performing double copy.}

The discussion above is actually reaching the  so-called pre-double-copy form like \eqref{eq:F23ptCK}, where the ${\rm tr}(F^2)$ form factor is expressed via a cubic diagram expansion of which the numerators are like \eqref{eq:trF24ptPDC1}.
Starting from this pre-double-copy form, we perform the double copy by introducing a mixing matrix $M_{ij}$ similar to \eqref{eq:R2G3ansatz}. The only difference is that we have two $M_{ij}$s here, defined for the diagrams $\begin{aligned}
    \includegraphics[width=0.12\linewidth]{figure/s4pt.eps}
\end{aligned}$ and $\begin{aligned}
    \includegraphics[width=0.12\linewidth]{figure/t4pt2.eps}
\end{aligned}$ respectively. 
Therefore, we can write the ansatz as
\begin{align}\label{eq:R2G4ansatz}
    \mathcal{G}_{4}^{\rm ansatz}(1^{h},2^{h},3^{h},4^{h})=&\bigg(\frac{\sum_{i,j=1}^{9}M^{(1)}_{ij}\tilde{n}_{1}^{i}\tilde{n}_{1}^{j}}{s_{34}s_{234}}+{\rm perm}(1,2,3,4)/{\rm perm}(3,4)\bigg)+
    \\
    &\bigg(\frac{\sum_{i,j=1}^{11}M^{(2)}_{ij}\tilde{n}_{2}^{i}\tilde{n}_{2}^{j}}{s_{12}s_{34}}+{\rm perm}(1,2,3,4)/({\rm perm}(3,4)\times {\rm perm}(1,2)) \bigg)\nonumber \,,
\end{align}
where $\tilde{n}_{m}^{i}$ take the form of the building blocks defined in \eqref{eq:numstrF24pt}, \emph{e.g.}~for the diagram 
$\begin{gathered}
\includegraphics[width=0.12\linewidth]{figure/s4pt.eps}
\end{gathered}$, 
the blocks in \eqref{eq:trF24ptblockexp} are denoted as those  $\tilde{n}_{1}^{i}$.\footnote{We need a more detailed  clarification about $\tilde{n}_{m}^{i}$ here. Technically, $\tilde{n}_{1}^{i}$ does not have a term by term correspondence with \eqref{eq:trF24ptPDC1}. $\tilde{n}_{1}^{1}$ to $\tilde{n}_{1}^{6}$ are indeed the first six terms in \eqref{eq:trF24ptPDC1} (the two particle and three particle parts). But we also have $\tilde{n}_{1}^{7}={\rm tr}^{\rm f}(1,2,3,4)$, $\tilde{n}_{1}^{8}={\rm tr}^{\rm f}(1,3,2,4)$ and $\tilde{n}_{1}^{9}={\rm tr}^{\rm f}(1,2,4,3)$, each is gauge invariant. Thus in total we choose nine $\tilde{n}_{1}^{i}$ as in the sum in \eqref{eq:R2G4ansatz}. This is also the case with $\tilde{n}_2$, where we have $9+2=11$ terms.} 
The total unfixed numbers from the two matrices $M_{ij}$ is 111.

To solve for the ansatz, we first impose some general conditions, which should be valid regardless of dimensions or the specific helicity configurations
\begin{enumerate}[topsep=3pt,itemsep=-1ex,partopsep=1ex,parsep=1ex]
    \item Diagrammatic symmetry: The numerator has the same symmetry as the cubic diagram. For example, for the diagram $\begin{aligned}
    \includegraphics[width=0.12\linewidth]{figure/s4pt.eps}
\end{aligned}$, we demand its numerator $\sum_{i,j}M^{(1)}_{ij}\tilde{n}_{1}^{i}\tilde{n}_{1}^{j}$ to be invariant under the $(3,4)$ permutation. 
    
    \item 
    Crossing symmetry: 
    \begin{equation}
        \mathcal{G}_{4}^{\rm ansatz}(1^{h},2^{h},3^{h},4^{h})=\mathcal{G}_{4}^{\rm ansatz}(\sigma(1^{h},2^{h},3^{h},4^{h}))\,, \ \forall \sigma\in S_4\,.
    \end{equation}
    Note that the first two symmetry requirements have also been used in the three-point ansatz \eqref{eq:R2G3ansatz} construction. 
    \item
No double pole: \begin{equation*}
   \lim_{\delta\rightarrow 0} \left( \delta \times \mathcal{G}_{4}^{\rm ansatz}|_{s_{12}-q^2=\delta }\right)= \text{finite},\qquad  \lim_{\delta\rightarrow 0} \left( \delta \times \mathcal{G}_{4}^{\rm ansatz}|_{s_{123}-q^2=\delta }\right)= \text{finite}\,.
\end{equation*}
\end{enumerate}
These conditions put strong constraints on $M_{ij}$, and there are only 32 free parameters remaining.

Next, we consider the factorization conditions. 
We will temporarily consider the result in four-dimensional kinematics.
There are four channels  $s_{12}{=}0,s_{123}{=}0$ and $s_{12}{=}q^2,s_{123}{=}q^2$:
\begin{equation}\label{eq:G4F2fact}
\begin{aligned}
    &\text{Res}_{s_{12}=0} \mathcal{G}_4^{\rm ansatz}(1,2,3,4)= \sum_{{\rm helicity} I} \mathcal{G}_3(P_I(\varepsilon),3^h,4^h)\,
    \mathcal{M}_{3}(-P_I(\bar{\varepsilon}),1^h,2^h)\,,\\
    &\text{Res}_{s_{123}=0} \mathcal{G}_4^{\rm ansatz}(1,2,3,4)= \sum_{{\rm helicity} I}\mathcal{G}_2(P_I(\varepsilon),4^h)\, 
    \mathcal{M}_{4}(-P_I(\bar{\varepsilon}),1^h,2^h,3^h)\,,\\
    &\text{Res}_{s_{12}=q^2} \mathcal{G}_4^{\rm ansatz}(1,2,3,4)= \mathcal{G}_2(1^h,2^h)\,\mathcal{M}_{4}(\QQ_2^{\phi},3^h,4^h,-q^{\phi})\,,\\
    &\text{Res}_{s_{123}=q^2} \mathcal{G}_4^{\rm ansatz}(1,2,3,4)= \mathcal{G}_3(1^h,2^h,3^h) \,\mathcal{M}_{3}(\QQ_3^{\phi},4^h,-q^{\phi})\,.
\end{aligned}
\end{equation}
In the first two equations,   $\mathcal{M}$ represents pure graviton amplitudes (which explains the helicity sum), while in the last two equations the $\mathcal{M}$ is amplitudes containing two scalars. 
Practically, in performing each concrete $4$-dimensional factorization, helicity configurations for external gluons must be specified. 
We are looking for proper values of the $M_{ij}$ so that our ansatz $\mathcal{G}_{4}^{\rm ansatz}$ has desired factorization properties for all helicity configurations. This means each of the equations in \eqref{eq:G4F2fact} represents three equations. 
Via all these conditions, the mixing matrices $M_{ij}^{(1),(2)}$ are both completely determined, as in  the three-point case. One can find the result, which still holds a nice structure, in Appendix~\ref{ap:Fsqsols}.

Since the complete result is a little bit lengthy, here we only present a simplified version of the four-point solution, specified to the NMHV form factor, to give a taste. This expression shows the consistency between higher- and lower-point double-copy solutions.
When considering the NMHV form factors (say $1^{-},2^{-},3^{-},4^{+}$), in which all the ${\rm tr}^{\rm f}(a,b,c,d)$ factors are actually zero, we have 
\begin{align}
    \sum_{i,j=1}^{9}&M^{(1)}_{ij}\tilde{n}_{1,i}\tilde{n}_{1,j}= \big[ {\rm tr}^{\rm f}(1^-,3^-)N_1(1^{\phi},2^{-},4^{+},3^{\phi}) \big]^2\\
    &+ \big[ {\rm tr}^{\rm f}(1^-,2^-)(N_1(1^{\phi},3^{-},4^{+},2^{\phi})-N_1(1^{\phi},4^{+},3^{-},2^{\phi})) \big]^2\,\nonumber\\
    &+ 4{\rm tr}^{\rm f}(1^-,2^-,3^-)N_2(2^{\phi},1^{\phi},4^{-},3^{\phi}){\rm tr}^{\rm f}(1^-,3^-)N_1(1^{\phi},2^{-},4^{+},3^{\phi})\,\nonumber\\
    &+ 4{\rm tr}^{\rm f}(1^-,2^-,3^-)N_2(1^{\phi},2^{\phi},4^{-},3^{\phi})\big[ {\rm tr}^{\rm f}(1^-,2^-)(N_1(1^{\phi},3^{-},4^{+},2^{\phi})-N_1(1^{\phi},4^{+},3^{-},2^{\phi})) \big]\,\nonumber\\
    &+ 6 \big[ {\rm tr}^{\rm f}(1^-,2^-,3^-)N_2(2^{\phi},1^{\phi},4^{-},3^{\phi})\big]^2\nonumber\,.
\end{align}
We see that the numerical coefficients are simple integers $\{1,4,6\}$, which are the matrix elements of $M_{ij}$, and are the same as the previous three-point result \eqref{eq:R2G3res}. 
In fact, if taking the soft limit of $4^{+}$, one can see this has to be true. 

Finally, we want to comment on the generalizations to $D$ dimensions.
When trying to make \eqref{eq:G4F2fact} valid in $D$ dimensions, we examine the four equations, and the only one that can be trivially satisfied is the last one.
Like in the three-point discussion, we need to add an evanescent contribution $\Delta_4$.
To make $(\mathcal{G}_4{+}\Delta_4)$ satisfy all the factorization channels in $D$ dimensions, we need $\Delta_4$ to have $s_{ij},\, s_{ij}{-}q^2,\,s_{ijk}{-}q^2$ poles.
To construct $\Delta_4$, one straightforward attempt is to make use of the known $\Delta_3$ in certain factorization channel: for example, one can construct an ansatz for  $\Delta_4$ including the factors like  $\text{Gram}(p_1,p_2,p_3,\varepsilon_1,\varepsilon_2,\varepsilon_3) (p_1 \cdot \text{f}_4 \cdot p_2) (p_2 \cdot \text{f}_4 \cdot p_3)$. However, our calculation shows that such a naive ansatz fails to give a $\Delta_4$ consistent with all factorization channels. This suggests that more complicated structures like $\text{Gram}(p_1,p_2,p_3,p_4,\varepsilon_1,\varepsilon_2,\varepsilon_3,\varepsilon_4)$ or $\text{Gram}(p_1,p_2,p_3,\varepsilon_1,\varepsilon_2) \times \text{Gram}(p_2,p_3,p_4,\varepsilon_3,\varepsilon_4)$ are required. To construct $\Delta_4$ systematically, one would need to understand the complete basis for the evanescent space, which is an interesting technical problem. We leave further analysis of these evanescent contributions to future study.

\vskip 5pt

In summary, we have proposed a prescription of performing double copy for $\mathcal{F}_{\operatorname{tr}(F^2)}$ in pure YM theory. 
The first crucial step is a gauge-invariant decomposition of  $\mathcal{F}_{\operatorname{tr}(F^2)}$, which has an $n$-point closed formula in \cite{Dong:2022bta}.
This is to connect the $\mathcal{F}_{\operatorname{tr}(F^2)}$ double copy to the $\mathcal{F}_{\operatorname{tr}(\phi^2)}$ ones, since the double-copy of the latter is known for all-multiplicity, which is discussed in the previous paper \cite{Lin:2022jrp} (also see  discussions in Section~\ref{ssec:universalnum}).
By allowing mixings between different terms in the decomposition, we can get a double-copy solution that is valid for any helicity configurations for four-dimensional kinematics. We have introduced some numerical matrix $M_{ij}$s to describe the mixing, and some consistency between higher- and lower-point mixing matrices has been shown.  
Checks are performed up to 5 points verifying that the strategy works, which is non-trivial considering the complexity if one tries to evaluate Feynman diagrams of 5 gravitons and one scalar. 
We comment that the pattern for the number matrix $M_{ij}$ in general higher-point cases deserves study, and it would be interesting to understand the existence of evanescent corrections in general $D$ dimensions. 

\section{Discussion}\label{sec:discussion}

In this paper, and in the previous one \cite{Lin:2022jrp}, we discuss the double-copy construction for form factors in detail, a topic first reported in the letter \cite{Lin:2021pne}. 
A graphic summary of the structure of these two papers can also be found in Figure~1 in \cite{Lin:2022jrp}. 
We summarize the main new results as follows.
\begin{itemize}
\item 
A key new observation is that new poles appear when performing the double copy for form factors.
These poles, referred to as \emph{spurious-type poles}, are not singularities of gauge-theory form factors, but they become exposed as physical poles in gravity during the double-copy procedure. 

\item 
The origin of the spurious-type poles can be attributed to the new feature of form factors: the operator insertion introduces new color relations (determined by the operator's color factor), and requiring the double-copy quantity to satisfy diffeomorphism invariance introduces new numerator relations, named as \emph{operator-induced relations}, which are dual to these color relations. 

\item
The spurious-type poles have a clear physical interpretation in the gravity theory:
they take the form of massive Feynman propagators with the mass square equivalent to $q^2$. 
The factorization properties are nicely maintained when considering the residues on such poles. 

\item
Examining these factorizations in more detail turns out to be very fruitful. 
On one hand, the factorization can be understood at the level of matrix decomposition, which is related to the factorization of the KLT kernel. 
On the other hand, it reveals a hidden factorization property of gauge-theory form factors when taking the spurious poles to be zero. 

\item
Within the factorization relations, an important quantity is the so-named \emph{${\vec v}$ vector}.  
The ${\vec v}$ vectors have been observed to be insensitive to the types of operators and external states.
Moreover, the all-multiplicity construction rules are given for the expression of ${\vec v}$ vectors. 
Although these statements still lack rigorous mathematical proof, numerous non-trivial checks have been carried out.

\item
Regarding the CK-dual numerators, one distinct feature that differs from usual amplitude cases is that the form-factor numerators are uniquely fixed; in other words, there is no so-called ``generalized gauge transformation" for the form-factor CK-dual numerators. At the practical level, this is because the form-factor propagator matrices are invertible. Consequently, the CK-dual numerators can be related to the color-ordered form factors in an unambiguous way.

\item
Since the numerators are uniquely defined, we also managed to obtain a conjectured compact closed form of CK-dual numerators which has passed various checks.
Another type of universality also appears for the expressions of numerators, essentially stating that, neglecting the flavor factors, 
CK-dual numerators do not depend on
the operator (at least in the context of operators composed of scalar fields).

\item
While the most fundamental examples considered involve the ${\rm tr}(\phi^2)$ operator in the Yang-Mills-scalar theory, we also show that the above properties apply for form factors with more general operators, including high-length operators or operators with different matter fields. 
For the double copy of these form factors, an important novelty is that new types of color relations involving four or more color factors appear, but the CK duality and  double copy work perfectly.

\item
The double copy for form factors of ${\rm tr}(F^2)$ with purely gluon external states is also presented. In this case, a non-trivial new prescription is required, which involves a decomposition for the form factor and a mixture of different BCJ numerators appearing in the expansion. 

\end{itemize}

\vskip.1cm

As a concluding remark, it should be instructive to briefly recap the fundamental reason behind all these new features.
In essence, \emph{color-singlet} is the key.
Our new approach depicts the double copy of form factors, which are intrinsically defined to have a color-singlet operator insertion, as well as the amplitudes with {color-singlet} particles such as amplitudes of Higgs plus quarks and gluons, in both the standard model or the Higgs EFT.
For normal (gauge-theory) amplitudes with colored external particles, each cubic diagram has its unique color factor. 
In contrast, in the case of form factors, the insertion of the color-singlet operator can generate different diagrams (with various propagator structures) that share the same color factor. A similar complication exists for amplitudes involving an external color-singlet particle. 
Therefore, imposing CK duality for the form factor diagrams is significantly more non-trivial. 
Interestingly, despite such complications, the propagator matrices exhibit elegant structures for a large class of form factors, including the simplicity of their determinants and factorization properties. This is not always the case though, the diagrams for the ${\rm tr}(F^2)$ form factor in pure YM theory are too intricate to impose CK duality directly, and we have to introduce an additional decomposition and apply CK duality only for the simpler sub-blocks.


\vskip.5cm
To give some outlook, we also list open questions deserving further exploration based on the previous results.
\begin{itemize}

\item 
First, it would be important to understand the various factorization properties better, such as the hidden factorization relation for the gauge form factors and the associated $\vec{v}$ vectors. As discussed in \eqref{eq:nptbcjN}-\eqref{eq:BCJampexample}, the hidden factorization relations in some channels are analogous to the fundamental BCJ relations for amplitudes. It may be possible to understand these relations along the lines of proof for amplitudes using recursion relations \cite{Feng:2010my} or as string theory relations in the low energy limit \cite{Bjerrum-Bohr:2009ulz, Stieberger:2009hq}.
This may also help to understand the universality property of the $\vec{v}$ vectors and provide proof for their construction rules.

\item 
Understanding the algebraic structure for duality-satisfying numerators is another problem to investigate.
An explicit derivation for the proposed closed formula for CK-dual form factor numerators should make a concrete step. 
One promising approach is to adopt the Hopf-algebra-related method \cite{Chen:2019ywi,Chen:2021chy,Brandhuber:2021bsf,Brandhuber:2021kpo,Chen:2022nei,Brandhuber:2022enp,Ben-Shahar:2022ixa}. 
In particular, equivalence between some of the numerator results in this paper and in \cite{Chen:2022nei} have been observed, 
and proof of these findings would be welcome.
It would be interesting to also discuss the kinematic algebra from a more physical perspective, such as those given in \cite{Monteiro:2011pc,Cheung:2016prv,Ben-Shahar:2021zww}.

\item 
While we have presented the form factor double copy mainly in the YMS or YMS+$\phi^3$ theories, it should be straightforward to generalize to theories with more general matter fields. We have discussed operators with fermions like ${\rm tr}(\bar\psi\psi)$ in Higgs+quarks+gluons context in \cite{Lin:2022jrp}. 
As mentioned before, the gauge-invariant operator in the form factor can be identified as a color-single scalar, thus 
it would be interesting to apply our prescription to EFT amplitudes that contain a color-single scalar particle.

\item 

For the form factors of ${\rm tr}(F^2)$ in pure YM theory, we have constructed the double-copy using an ansatz and found 
concrete solutions up to five external gluons.
It would be interesting to have a systematic extension to higher-point cases and understand the general pattern of double-copy results.
Ideally, one would have some explicit rules for the solutions without recourse to an ansatz. 
Additionally, a more comprehensive understanding of the difference, existing in the current constructions, between four-dimensional and $D$-dimensional kinematics would be desirable. 

We have only considered the length-2 ${\rm tr}(F^2)$ operator and one may try to extend this to form factors of more general operators such as ${\rm tr}(F^m)$ (see \cite{Broedel:2012rc} and also more recent papers \cite{Carrasco:2019yyn,Carrasco:2021ptp} for the study of such operators in the context of EFT amplitudes).
By examining the procedure in this paper, it appears that such a generalization may be feasible, in the sense that: (1) the gauge invariant decomposition that relates the ${\rm tr}(F^2)$ form factors to the ${\rm tr}(\phi^2)$ ones appears applicable for the ${\rm tr}(F^m)$ case, and (2) the ${\rm tr}(\phi^m)$ form factors, serving as building blocks in the decomposition, have also had their double copy as discussed in this paper.

\item
For amplitudes, the double copy possesses a natural extension to string theory, as evidenced by the initial discovery of the KLT relation within string theory \cite{Kawai:1985xq}.
One may wonder if it is possible to achieve these in the context of form factors. 
A modest starting point may be to include some $\alpha'$ corrections into the discussion about the double copy of form factors, like the Z-theory discussion in amplitudes \cite{Broedel:2013tta, Carrasco:2016ldy}. More ambitiously, it is worth investigating whether a full string theory extension could exist for form factors, possibly in the context of string field theories. In particular, it would be important to explore whether the hidden factorization relations for form factors could be interpreted as certain string monodromy relations  \cite{Bjerrum-Bohr:2009ulz, Stieberger:2009hq}.

\item 
Furthermore, the CHY formula~\cite{Cachazo:2013hca, Cachazo:2014xea}
is a string-inspired method describing a large class of massless amplitudes in general spacetime dimensions. 
Given the abundant form factor double copy results presented in this and previous papers, it is natural to ask whether a CHY formalism exists.
Several attempts have been made to address some specific form factors in four dimensions \cite{He:2016jdg,Brandhuber:2016xue,He:2016dol}, but it would be valuable to pursue a consistent description of form factors in general dimensions.

\item
In the context of EFTs, the KLT bootstrap has been developed in \cite{Chi:2021mio,Bonnefoy:2021qgu,Chen:2022shl}.
One may test these within the framework of the form factor double copy. Conversely, the form factor double copy may provide new insights for these methods. 
For example, in the KLT-bootstrap method, certain analytic properties of the KLT kernel are assumed, however, allowing the new spurious-type poles, as appearing in the form factors, may lead to novel possible solutions.

\item 
So far all the discussions about the form factor double copy are at tree level.
It would be important to have the loop-level generalization. 
We would like to briefly comment on the difference between the new double copy prescription for form factors and the previous CK-dual loop constructions \cite{Boels:2012ew,Yang:2016ear,Lin:2021kht,Lin:2021qol, Lin:2021lqo,Lin:2020dyj,Li:2022tir}.
In the previous loop constructions, operator-induced relations are not considered and the CK duality (for the Jacobi relations) was primarily used as an ansatz input to simplify loop integrand constructions. 
In contrast, to fulfill the double-copy construction for form factors, operator-induced relations are necessitated by diffeomorphism invariance.
A full loop-level double-copy construction should take into account these operator-induced relations as well. The challenge would be to find the CK-dual solutions that are consistent with physical requirements like unitarity cuts.

Apart from the inherent significance of loop constructions, such a generalization could potentially offer further clarity on the physical interpretation. At the tree level, one can perceive the operator insertion as a massive particle, which makes its coupling to gravity more or less natural after double copy. A possible alternative picture to interpret the operator after double copy is through a semi-classical graviton dressed operator, similar to those discussed in worldline theory \cite{Alawadhi:2021uie, Bastianelli:2021rbt, Shi:2021qsb}.
A concrete loop-level construction should help to understand the picture better.

\item 
CK duality and double copy have been explored for operator-associated quantities in other contexts: including but not limited to the double copy structure between the gluon and gravity boundary correlators in curved space~\cite{Li:2018wkt,Farrow:2018yni,Lipstein:2019mpu,Fazio:2019iit,Albayrak:2020fyp,Armstrong:2020woi,Zhou:2021gnu,Jain:2021qcl,Diwakar:2021juk,Sivaramakrishnan:2021srm,Cheung:2022pdk,Alday:2022lkk,Herderschee:2022ntr,Drummond:2022dxd, Armstrong:2023phb, Mei:2023jkb}.
It could be interesting to explore potential connections to the form factors here. 
It would also be highly interesting to extend our discussion to generalized form factors that involve multiple operator insertions.

\end{itemize}

\acknowledgments

We thank Gang Chen, Jin Dong, Song He, and Congkao Wen for discussions. 
This work is supported in part by the National Natural Science Foundation of China (Grants No.~11935013, 12175291, 12047503) and by the CAS under Grants No.~YSBR-101.
We also thank the support of the HPC Cluster of ITP-CAS.

\appendix

\section{Some further remarks on the $\vec{v}$ vectors}\label{ap:vremarks}

In this appendix, we give some more remarks on the $\vec{v}$ vectors. 

\paragraph{The $\vec{v}$ vectors as null vectors.}
In this first part, we show that the $\vec{v}$ vectors are null vectors of the propagator matrix, if evaluated on the spurious pole. 

Let us consider first the simple but illustrative four-point example.
Concretely, we need the fact that the propagator matrix $\Theta^{\cal F}$ is the inverse of $\mathbf{S}^{\cal F}$, implying that the $\vec{v}$ vectors involved in the decomposition of $\mathbf{S}^{\cal F}$ must lie in the null space of $\Theta^{\cal F}$. 
Let us see how it happens for the four-point case where $\delta=(q^2- s_{12})\shortrightarrow 0 $
\begin{equation}
	\Theta_{4}^{\cal F}\cdot \mathbf{S}_{4}^{\cal F}=\mathbf{1}\quad  \Rightarrow  \quad \lim_{\delta\shortrightarrow 0 }\left(\Theta_{4}^{\cal F}\cdot \mathbf{S}_{4}^{\cal F}\right)=\mathbf{1}\,.
\end{equation}
We know that $\Theta_{4}^{\cal F}$ is finite when $\delta\shortrightarrow 0 $ and  $ \mathbf{S}_{4}^{\cal F}$ is divergent. Then we have a Laurent series expansion:
\begin{equation}
	\Big(\Theta_{4}^{\mathcal{F}}\big|_{q^2=s_{12}}+ \cdots\Big)\cdot \Big(\frac{1}{\delta}\text{Res}\left[\mathbf{S}_{4}^{\mathcal F}\right]_{q^2=s_{12}}+\mathbf{S}_{4}^{\mathcal{F}}\big|_{q^2=s_{12}} + \cdots\Big)=\mathbf{1}\,,
\end{equation}
where the RHS is finite and irrelevant to $\delta$, so that 
\begin{equation}
\begin{aligned}
& \Theta_{4}^{\mathcal{F}}\big|_{q^2=s_{12}} \cdot \text{Res}\left[\mathbf{S}_{4}^{\mathcal F}\right]_{q^2=s_{12}} =0\,.
\end{aligned}
\end{equation}
We also have a similar equation by commuting $\Theta$ and $\mathbf{S}$. 

Then we plug in the matrix decomposition for the four-point kernel 
\begin{equation}
\begin{aligned}
    \mathrm{Res}_{s_{12}=q^2}\left[{\bf S}^{\mathcal{F}}_{4}\right]&= \begin{pmatrix}
        \frac{\tau_{31}\tau_{42}}{\tau_{3\QQ_2}} \\   \frac{ \tau_{32}\tau_{41}}{\tau_{3\QQ_2}}
    \end{pmatrix}
\cdot \left(1\times {\tau_{13}\tau_{34} \over s_{14}}\right)\cdot 
    \left( \frac{\tau_{31}\tau_{42}}{\tau_{3\QQ_2}} \ \   \frac{ \tau_{32}\tau_{41}}{\tau_{3\QQ_2}} \right)
    \\
    &=({\vec v}^{A}_4)^{\rm \scriptscriptstyle T}\cdot \left({\bf S}^{\mathcal{F}}_{2}\otimes{\bf S}^{\mathcal{A}}_{4}\right) \cdot  {\vec v}^{A}_4\,,
\end{aligned}
\end{equation}
and get 
\begin{equation}
	\Theta_{4}^{\mathcal{F}}\big|_{q^2=s_{12}} \cdot \Big(({\vec v}^{A}_4)^{\rm \scriptscriptstyle T} \cdot \left({\bf S}^{\mathcal{F}}_{2}\otimes{\bf S}^{\mathcal{A}}_{4}\right) \cdot {\vec v}^{A}_4 \Big)=0=\Big(({\vec v}^{A}_4)^{\rm \scriptscriptstyle T} \cdot \left({\bf S}^{\mathcal{F}}_{2}\otimes{\bf S}^{\mathcal{A}}_{4}\right) \cdot {\vec v}^{A}_4\Big) \cdot \Theta_{4}^{\mathcal{F}}\big|_{s_{12}=q^2}\,.
\end{equation}
Since the $\mathbf{S}$ kernels are full-ranked matrices, the following relation must be true
\begin{equation}
	\Theta_{4}^{\mathcal{F}}\big|_{q^2=s_{12}} \cdot ({\vec v}^{A}_4)^{\rm \scriptscriptstyle T}=({\vec v}^{A}_4)\cdot \Theta_{4}^{\mathcal{F}}\big|_{q^2=s_{12}} = 0\,.
\end{equation}
stating that the $\vec{v}$ is the null vector of $\Theta_{4}^{\mathcal{F}}\big|_{q^2=s_{12}}$.

Apparently, this conclusion can be generalized to higher points as 
\begin{equation}\label{eq:nptnullvector}
   \Theta_{n}^{\mathcal{F}}\big|_{\QQ_m^2=q^2}\cdot \vec{v}^{\scriptscriptstyle \rm T}_{(\bar{\kappa},\bar{\rho})}=\vec{v}_{(\bar{\kappa},\bar{\rho})}\cdot \Theta_{n}^{\mathcal{F}}\big|_{\QQ_m^2=q^2}=0 \,,
\end{equation}
where $\Theta_{n}^{\cal F}\big|_{\QQ_m^2=q^2}$ is the propagator matrix $\Theta_{n}^{\cal F}$ evaluated on the special kinematics $\QQ_m^2=q^2$.
We comment that the null vector condition  \eqref{eq:nptnullvector} is a particularly useful condition to determine the  $\vec{v}$ vectors.

\paragraph{Comment on the alternating list in the $\vec{v}$ vector closed formula.} In this part we will provide some observation (and the proof) of the $\vec{v}$ vector closed formula in Section~\ref{ssec:closedv}.

We will first review the formula, especially the \textbf{Step 3}.
Given an ordering $\beta$ for which we want to express $v[\beta]$, we need to split the $\beta$ ordering into two parts, based on their different contribution to the vector element $v[\beta]$. 

The first part consists of particles $j_a$ such that either $\{j_a-1,j_a,j_a+1\}$ or $\{j_a+1,j_a,j_a-1\}$ is a subordering of $\beta$. 
Although this first part will not be the target of the discussion here, it is helpful to emphasize its definition for comparison.
On the other hand, the second part, denoted as $\overline{\beta}$, is just simply defined as the complement of $\{j_1,j_2,\ldots,j_s\}$ mentioned above. 
Note that if we write $\beta$ as $\{1,{\rm x}_1,\ldots,{\rm x}_{-1},2\}$, both ${\rm x}_1$ and ${\rm x}_{-1}$ must belong to $\overline{\beta}$. 

As mentioned in Section~\ref{ssec:closedv} (the \textbf{Step.3}), given an element ${\rm x}_b\in \overline{\beta}$, its contribution to the vector element $v[\beta]$ takes the form of either $\tau_{{\rm x}_b,\Xi_{L}({\rm x}_b,\beta)}$ or $\tau_{{\rm x}_b,\Xi_{R}({\rm x}_b,\beta)}$.
Previously, we have observed that ${\rm x}_1$ gives $\tau_{{\rm x}_1,\Xi_{L}({\rm x}_1,\beta)}$; now we further assert that ${\rm x}_{-1}$ gives $\tau_{{\rm x}_{-1},\Xi_{R}({\rm x}_{-1},\beta)}$. 
This is a natural expectation based on the reflection of $\beta$. 
The rest of this part of the appendix is to prove this point, in which we get to better understand the structure of $\beta$ and $\overline{\beta}$. 

Let us recall the alternating list construction in Section~\ref{ssec:closedv} (the \textbf{Step.3}). 
First, we sort $\overline{\beta}$ and get $\overline{\beta}'$. 
Then we find ${\rm x}_1$ in $\overline{\beta}'$, assign $\tau_{\Xi_R}$ to the first element on ${\rm x}_1$'s right, assign $\tau_{\Xi_L}$ to the second element on ${\rm x}_1$'s right and so on. 
Therefore, to have $\tau_{{\rm x}_{-1},\Xi_{R}({\rm x}_{-1},\beta)}$, there must be even number of elements between ${\rm x}_1$ and ${\rm x}_{-1}$
\begin{equation}
    \overline{\beta}'=\{..,{\rm x}_1\underbrace{,...,}_{\text{even number}}{\rm x}_{-1},..\}
\end{equation}
(suppose ${\rm x}_1$ is on the left side of ${\rm x}_{-1}$; it is similar if ${\rm x}_1$ and ${\rm x}_{-1}$ are switched).
We first examine this via some examples.

{\small
\begin{enumerate}[topsep=3pt,itemsep=-1ex,partopsep=1ex,parsep=1ex]
    \item $\beta=\{1,3,4,5,6,7,8,9,2\}$. 

    This is the standard ordering in which $\overline{\beta}=\{3,9\}$. We sort $\overline{\beta}$ and get a $\overline{\beta}^{\prime}=\{3,9\}$. We see importantly that there is an even number of elements (which are zero in this case) between ${\rm x}_1=3$ and ${\rm x}_{-1}=9$ in $\overline{\beta}^{\prime}$.

    \item $\beta=\{1,3,4,7,6,5,8,9,2\}$. 

    In this example, we swap $5,7$ compared with the first standard ordering above. 
    Then $\overline{\beta}=\{3,7,5,9\}$, and we see again that there is an even number of elements (now there are two) in $\overline{\beta}$ between ${\rm x}_1=3$ and  ${\rm x}_{-1}=9$.

    A heuristic argument on why this is true is that when exchanging a pair of particles not in $\overline{\beta}$ (for instance 5,7 in this case) will make this pair show up in $\overline{\beta}$, because the exchange will spoil the three-element subordering condition. 
    
    \item $\beta=\{1,3,5,4,8,7,6,9,2\}$.

    This is a more disordered list and $ \overline{\beta}$ now is $\{3,5,4,8,6,9\}$, and again even number of elements are between ${\rm x}_1=3$ and ${\rm x}_{-1}=9$. 

    \item $\beta=\{1,4,3,5,6,7,9,8,2\}$. 
    Now we turn to more non-trivial examples. In this case, $\overline{\beta}=\{4,3,9,8\}$ and $\overline{\beta}'=\{3,4,8,9\}$ so that there is zero (which is an even number) element between $\mathrm{x}_1=4$ and $\mathrm{x}_{-1}=8$.

    \item $\beta=\{1,4,3,7,6,5,9,8,2\}$.
    This is another example following the above example 4. 
    In this case, we have $\overline{\beta}=\{4,3,7,5,9,8\}$ and $\overline{\beta}^{\prime}=\{3,4,5,7,8,9\}$, so that there are two element between ${\rm x}_1=4$ and   ${\rm x}_{-1}=8$.

    To understand, one can also try to  consider a subordering of $\beta$ consists of ${\rm y}_i \in \beta$ satisfying ${\rm x}_1<{\rm y}_i<{\rm x}_{-1}$ (or ${\rm x}_1>{\rm y}_i>{\rm x}_{-1}$ if ${\rm x}_1>{\rm x}_{-1}$). 
    Adapting the approach, we first delete 3,9 (and 1,2 of course) in $\beta$ and get a reduced ordering $\{4,7,6,5,8\}$. Then we can use the argument in the second example that exchanging two particles is equivalent to adding two particles in $\overline{\beta}$. 
    The deleting process does not change the number of elements between $4$ and $8$ in the sorted $\overline{\beta}'$, because $3$ and $9$ are not in the range $ {\rm x}_1< {\rm y}_i<{\rm x}_{-1}$. 

\end{enumerate}
}

The proof is given below, following the same argument outlined in the fifth example. 
First, we delete the elements in $\beta$ which are not in the range $({\rm x}_1,{\rm x}_{-1})$ (for simplicity we assume ${\rm x}_1<{\rm x}_{-1}$).
Then we have a list taking the form $\tilde{\beta}=\{\mathrm{x}_1,\mathrm{x}_2,..,{\rm x}_{-1}\}$. 
And as what we have done on $\beta$, splitting $\beta$ into parts is required, and defining $\overline{\tilde{\beta}}$ and getting it sorted to get $\overline{\tilde{\beta}}'$ is possible. 
The point is, if we define $\overline{\tilde{\beta}}'$ and calculate the number of elements between $\mathrm{x}_1$ and ${\rm x}_{-1}$, it must be the same as the number in $\overline{{\beta}}'$. 
Then we just need to make sure such a number is even. 

If ${\rm x}_1<{\rm x}_2<\cdots<{\rm x}_{-1}$, then $\overline{\tilde{\beta}}'=\{{\rm x}_1,{\rm x}_{-1}\}$, which means there are zero elements between ${\rm x}_1$ and ${\rm x}_{-1}$. 
If the elements in $\tilde{\beta}$ do not follow the standard ordering, we need to be more careful. 
There exists a unique permutation bringing $\tilde{\beta}$ back to the standard ordering, and since every permutation can be formed by a series of two-element swaps, it is sufficient to consider only two-element swaps.
The claim is that each two-element-swap can only possibly change the number of elements in $\overline{\tilde{\beta}}$ by 0 or 2. 
Verifying this requires a careful inspection of all possibilities.
Here we just give one case. 
Suppose $\tilde{\beta}$ is $\tilde{\beta}=\{{\rm x}_1,..,{\rm x}_i{-}1,{\rm x}_i,{\rm x}_i{+}1,..,{\rm x}_j{-}1,{\rm x}_j,{\rm x}_j{+}1,..,{\rm x}_{-1}\}$. 
Then we see ${\rm x}_i$ and ${\rm x}_j$ are not in $\overline{\tilde{\beta}}$.\footnote{Remember that the difference between $\overline{\tilde{\beta}}$ and $\tilde{\beta}$ are the elements with the corresponding three-element ordered subset in $\tilde{\beta}$.}
Then, if we swap ${\rm x}_i$ and ${\rm x}_j$, both of them will be in $\overline{\tilde{\beta}}$. 
All possible cases include all possible relative positions of ${\rm x}_{i,j}$ ${\rm x}_{i,}\pm 1$, which have been checked thoroughly.

\section{The Lagrangians and generalizations}\label{ap:scalartheory}

Here we give the Lagrangians of the gauge and gravity theories mentioned in the main text. In particular, we further generalize the discussions in Section~\ref{sec:highlength} and give the form factor double copy with more general high-length operators and external states.

\subsection{The Lagrangians}
The most basic one is the Yang-Mills-scalar theory, of which the Lagrangian is
 \begin{align}\label{eq:L}
    \mathcal{L}^{\rm YMS}=&\frac{1}{2} \operatorname{tr}_{\mathrm{C}}\left(D_{\mu} \phi D^{\mu} \phi\right) -\frac{1}{4}{\rm tr}_{\rm C} (F^{\mu\nu}F_{\mu\nu})\,,
\end{align}
where  $\phi$ is the scalar field carrying color index, $F^{\mu\nu}$ is the Yang-Mills field strength, $g$ is the Yang-Mills coupling and the ${\rm tr}_{\rm C}$ means the color trace. 
The form factors in Section~\ref{sec:highlength}, except for the ${\rm tr}(\phi^2)$ numerator review in Section~3.3.1, are defined in this theory.  

The first one is the most commonly used theory in this paper, which is the Yang-Mills-scalar+bi-adjoint $\phi^3$ 
 scalar theory. 
 \begin{align}\label{eq:L2}
    \mathcal{L}^{\text{YMS}+\phi^3}=&\frac{1}{2} \operatorname{tr}_{\mathrm{C}}\left(D_{\mu} \phi^{I} D^{\mu} \phi^{I}\right) -\frac{1}{4}{\rm tr}_{\rm C} (F^{\mu\nu}F_{\mu\nu}) -\frac{g^{2}}{4} \operatorname{tr}_{\mathrm{C}}\left(\left[\phi^{I}, \phi^{J}\right]^{2}\right) \nonumber \\
    &-\frac{g'}{3!}{\tilde f}^{IJK}f^{abc}\phi^{I,a}\phi^{J,b}\phi^{K,c}\,,
\end{align}
where the bi-adjoint scalar $\phi$ carries a scalar index $I$ and a color index $a$. 
The form factors in Section~\ref{sec:vvec} and 3.3.1 are defined specifically via  this Lagrangian, and the operator of these form factors is ${\rm tr}(\phi^2)=\sum_{a,I}\phi^{a,I}\phi^{a,I}$. 

The following one is used to define the propagator matrix 
\begin{equation}
\begin{aligned}
    \mathcal{L}^{\{\phi,\Phi\}}=&\frac{1}{2} \operatorname{tr}_{\mathrm{C}}\left(D_{\mu} \phi^{I} D^{\mu} \phi^{I}\right)
    +\frac{1}{2} \operatorname{tr}_{\mathrm{C}}\left(D_{\mu} \Phi^{I} D^{\mu} \Phi^{I}\right)
    -\frac{\lambda_3}{3 !} \tilde{f}^{I J K} f^{abc} \phi^{I, a} \phi^{J, b} \Phi^{K, c} \\
    &-\frac{\lambda_1}{3 !} \tilde{f}^{I J K} f^{abc} \phi^{I, a} \phi^{J, b} \phi^{K, c}-\frac{\lambda_2}{3 !} \tilde{f}^{I J K} f^{abc} \Phi^{I, a} \Phi^{J, b} \Phi^{K, c}
    \,, 
\end{aligned}
\end{equation}
where we use $\{I, J, K\}$ to denote the flavor (FL) index and $\{a, b, c\}$ to denote the color (C) index. 
The propagator matrices are considered in details in Section~\ref{ssec:FFL3} and Section~\ref{ssec:FFL4}; see also Section~4.2 and 6 in \cite{Lin:2022jrp}. 

In Section~\ref{sec:trF2}, the gluonic form factors can be regarded as the Higgs plus gluons amplitudes using the Higgs effective Lagrangian.
Although Higgs has no direct interaction with gluons (but through a quark loop) in the Standard Model when the top mass $m_{\rm t}$ is much larger than Higgs mass $m_{\rm H}$, the Higgs-gluon amplitudes can be greatly simplified using an effective theory where the top quark is integrated out, leaving a dimension-5 Higgs-gluon  interaction vertex in the effective Lagrangian as \cite{Wilczek:1977zn, Shifman:1979eb, Kniehl:1995tn}:
\begin{equation}
    \mathcal{L}^{\textrm{Higgs-EFT}}=\mathcal{L}^{\rm YM} + \lambda  H {\rm tr} (F^{\mu\nu}F_{\mu\nu})\,.
\end{equation}
In this case, the amplitude with a Higgs plus $n$ gluons is equivalent to the $n$-point ${\rm tr}(F^2)$ form factors considered in Section~\ref{sec:trF2}.

\subsection{Further generalizations of the high-length double copy} The target of the following discussion is to generalize the high-length double-copy discussions in Section~\ref{sec:highlength}, where only the YMS theory is considered and we require that the number of external scalars is the same as the number in the operator (called scalar-minimal).
The easier step is to reproduce the scalar-minimal results for the new YMS+$\phi^3$ Lagrangian. 
To achieve this, we realize first that the scalar field now bears both a scalar and a flavor indices. 
A natural generalization of the 
 ${\rm tr}(\phi^m)$ operator defined in Section~\ref{ssec:FFL4} would be
 \begin{equation}\label{eq:ophim}
     \widetilde{\mathcal{O}}_{\phi^m}=\sum_{a_k,I_k} V^{a_1\cdots a_m}\tilde{V}^{I_1\cdots I_m}\prod_{k=1}^m\phi^{a_k I_k}
\end{equation}
where $V$ can be any tensor with $m$ indices. 
We only require the tensor $V^{a_1\cdots a_m}$ and $\tilde{V}^{I_1\cdots I_m}$, for the color and flavor groups respectively, are the same.\footnote{For example, they can be both fully symmetric, antisymmetric, symmetric between the first two elements and so on.}
Such a definition  makes it reminiscent of the standard bi-adjoint scalar amplitudes bearing the color-kinematics duality structure.

After clarifying the operator, we give a simple argument on the reason why the minimal-scalar case for the new form factors with operator \eqref{eq:ophim} in the theory \eqref{eq:L2} is equivalent to our discussions in Section~\ref{sec:highlength}. 
Considering each Feynman diagram contributing to the form factor, we see that only the operator vertex has a flavor structure $\tilde{V}^{I_1\cdots I_m}$, and depriving this flavor factor will give back to the Feynman diagram in the YMS theory in Section~\ref{sec:highlength}. 
Dressing a universal $\tilde{V}^{I_1\cdots I_m}$ does not affect the color-kinematics duality and the double copy can be easily confirmed. 

The more non-trivial scenarios are to include more external scalars. 
We would like to justify two points in the following discussion: (1) these form factors can be double copied; (2) the numerators can be related to the ${\rm tr}(\phi^2)$ numerators given in Section~\ref{ssec:universalnum}.
In particular, the flavor structure will play a more non-trivial role here, and we discuss the form factor without gluon external lines to begin with. 
Concretely, we have 
\begin{equation}\label{eq:F5symmetric}
\begin{aligned}
\itbf{F}_{\widetilde{\mathcal{O}}_{\phi^4}}(1^{\phi},..,5^{\phi})&=\sum_{\sigma\in S_{5}/(S_3\times S_2)}\begin{aligned}
        \includegraphics[width=0.18\linewidth]{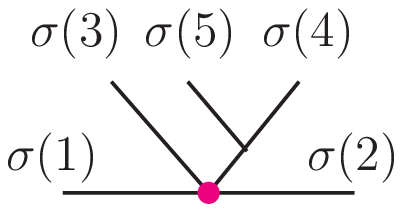}
    \end{aligned}\\
    &=\sum_{\sigma\in S_{5}/(S_3\times S_2)} \frac{\left(f^{\sigma(4)\sigma(3)\text{x}}V^{\sigma(1)\sigma(2)\sigma(3)\text{x}}\right)\left(\tilde{f}^{\sigma(4)\sigma(3)\text{x}}\tilde{V}^{\sigma(1)\sigma(2)\sigma(3)\text{x}}\right)}{s_{\sigma(4)\sigma(5)}}\,, 
\end{aligned} 
\end{equation}
where $\sigma\in S_5/(S_3\times S_2)$ to avoid over counting.
The number of diagrams is 10 in total. 
We see that such color and flavor factors are definitely satisfying the same algebraic relations, and the double copy of such a form factor is simply itself. 
This is a behavior similar to the bi-adjoint $\phi^3$ amplitudes.\footnote{It is also similar to the form factor propagator matrix defined in \emph{e.g.} \eqref{eq:deftheta4}. However, the difference is that we intentionally defined two kinds of scalar fields in \eqref{eq:deftheta4} so that the operator can only couple to one of them. In \eqref{eq:F5symmetric}, there is only one kind of scalar, making the number of diagrams greater compared with \eqref{eq:deftheta4} because the operator can now couple to any lines.}

Afterward, we can dress gluons on the pure scalar form factor. 
For the one-gluon form factor, there are three types of diagrams 
\begin{equation}\label{eq:F5symgluon}
    \begin{aligned}
        \includegraphics[width=0.9\linewidth]{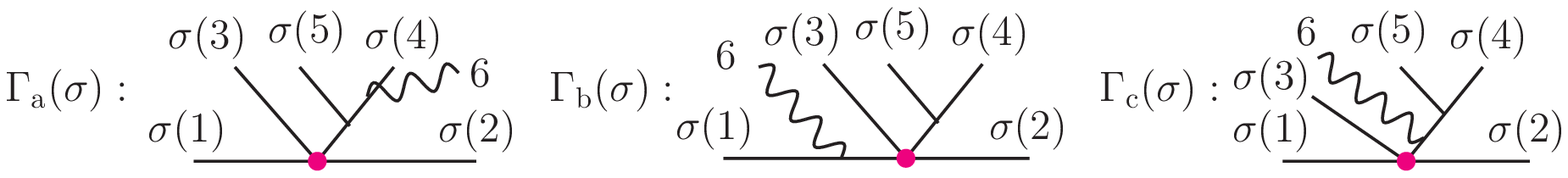}
    \end{aligned}
\end{equation}
A direct calculation shows that the double copy can also be achieved, of which the calculational detail will be given later.
The conclusion is, that in the double-copied quantity, the following diagrams show up 
\begin{equation}
    \begin{aligned}
        \includegraphics[width=0.9\linewidth]{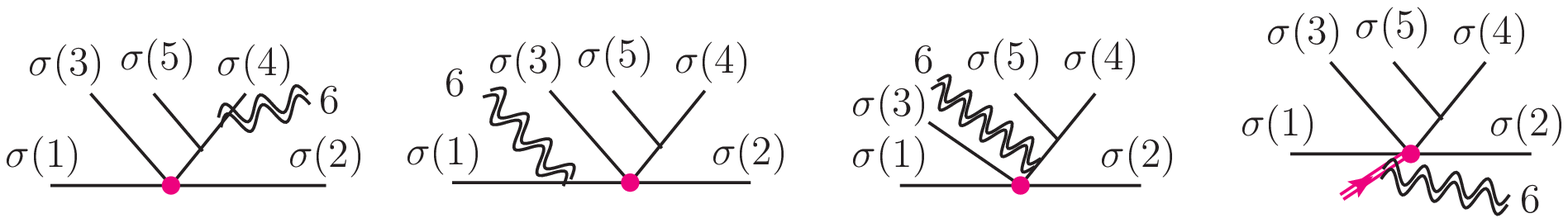}
    \end{aligned}
\end{equation}
with appropriate permutations $\sigma$.
The last set of diagrams having $s_{12345}{-}q^2$ is as expected. 

Concretely, we specify the CK numerators in the previous example. 
The numerators corresponding to the three diagrams in \eqref{eq:F5symgluon} are respectively 
\begin{equation}
\begin{aligned}
    N_{\rm a}(\sigma\{1,2,3,4,5\},6)\equiv N(\Gamma_{\rm a}(\sigma))=-\frac{2 p_{\sigma(1)\sigma(2)\sigma(3)\sigma(5)}\cdot  {\rm f}_6\cdot p_{\sigma(4)}}{s_{12345}-q^2}
    \tilde{f}^{\sigma(5)\sigma(4)\text{x}}\tilde{V}^{\sigma(1)\sigma(2)\sigma(3)\text{x}}\,,
    \\
    N_{\rm b}(\sigma\{1,2,3,4,5\},6)\equiv N(\Gamma_{\rm b}(\sigma))=-\frac{2 p_{\sigma(1)}\cdot  {\rm f}_6\cdot p_{\sigma(2)\sigma(3)\sigma(4)\sigma(5)}}{s_{12345}-q^2}\tilde{f}^{\sigma(5)\sigma(4)\text{x}}\tilde{V}^{\sigma(1)\sigma(2)\sigma(3)\text{x}}\,,\\
    N_{\rm c}(\sigma\{1,2,3,4,5\},6)\equiv N(\Gamma_{\rm c}(\sigma))=-\frac{2 p_{\sigma(1)\sigma(2)\sigma(3)}\cdot {\rm f}_6\cdot p_{\sigma(4)\sigma(5)}}{s_{12345}-q^2}\tilde{f}^{\sigma(5)\sigma(4)\text{x}}\tilde{V}^{\sigma(1)\sigma(2)\sigma(3)\text{x}}\,.
\end{aligned}
\end{equation}
Note that the numerators have flavor structures easily read out from the diagram. 
One can easily show that, with the help of these flavor structures, the following dual numerator relations hold
\begin{enumerate}[topsep=3pt,itemsep=-1ex,partopsep=1ex,parsep=1ex]
    \item One operator induced dual relation similar to \eqref{eq:phi24scalarCK2}
    \begin{equation}
    N_{\rm b}(1,2,3,4,5,6)+N_{\rm b}(2,1,3,4,5,6)+N_{\rm b}(3,1,2,4,5,6)+N_{\rm c}(1,2,3,4,5,6)=0\,.
\end{equation}
    \item Another operator induced dual relation similar to \eqref{eq:phi24scalarCK2} but with gluons replaced with scalars, which relies in particular on the flavor structure
    \begin{equation}
    N_{\rm b}(1,2,3,4,5,6)+N_{\rm b}(1,3,4,2,5,6)+N_{\rm b}(1,4,2,3,5,6)+N_{\rm a}(4,2,3,1,5,6)=0\,.
    \end{equation}
    \item Dual Jacobi relation 
    \begin{equation}
        N_{\rm a}(1,2,3,4,5,6)+N_{\rm a}(1,2,3,5,4,6)=N_{\rm c}(1,2,3,4,5,6)\,.
    \end{equation} 
\end{enumerate}

Another important point is that we recognize that after stripping off the flavor factor, the numerators look like the results given in Section~\ref{ssec:universalnum} in the sense that 
\begin{equation}
    N^{\mathcal{F}_{\widetilde{O}_{\phi^4}}}|_{\text{flavor factor}\rightarrow 1}=N^{\mathcal{F}_{{\rm tr}(\phi^2)}}|_{\text{flavor factor}\rightarrow 1}\,.
\end{equation}
To make this precise, we need to specify what are the diagrams to which the numerators correspond. 
We give only one example here
\begin{equation}
    N^{\mathcal{F}_{\widetilde{O}_{\phi^4}}} \left(\begin{aligned}
        \includegraphics[width=0.18\linewidth]{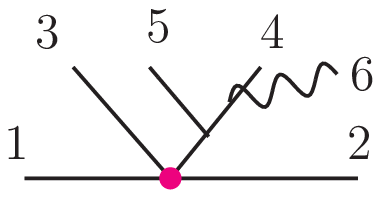}
    \end{aligned}\right)_{\text{flavor factor}\rightarrow 1}
    = N^{\mathcal{F}_{{\rm tr}(\phi^2)}}
    \left(\begin{aligned}
        \includegraphics[width=0.18\linewidth]{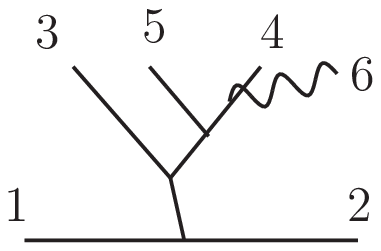}
    \end{aligned}\right)_{\text{flavor factor}\rightarrow 1}
\end{equation}
and the reader should be able to work out the general case.
This is another evidence of the universality of the numerators stated in Section~\ref{ssec:universalnum}. 

In general, we can add even more gluons and the double copy will still work. We have the spurious poles with the form $\QQ_m^2-q^2$ (with $m\geq \#$ of external scalars) which become real poles after double copy. 
Besides, the CK-dual numerators satisfying all the dual relations can be obtained by dressing the kinematical parts of the ${\rm tr}(\phi^2)$ numerator with appropriate flavor factors.
Furthermore, one can also further generalize the theory to introduce different types of scalars and more general operators, with the requirement that the color and flavor structure of the operator are the same.
One can easily verify that the form factors of these operators with solely scalar external lines can be double-copied. 
Although non-trivial to give proof, we believe that as long as the double copy makes sense for pure scalar form factors (without gluons), it works for the gauged case (with an arbitrary number of gluons). 
This belief is based on the universal properties mentioned throughout this paper and the general principles, such as gauge invariance and the CK duality, behind them.

\section{The four-point ${\rm tr}(F^2)$ double-copy solution}\label{ap:Fsqsols}

The four-point ${\rm tr}(F^2)$ double-copy solution is as follows. 

\begin{align}
    \sum_{i,j=1}^{9}&M^{(1)}_{ij}\tilde{n}_{1,i}\tilde{n}_{1,j}=({\rm tr}^{\rm f}(1,3)N_1(1^{\phi},2^{g},4^{g},3^{\phi}))^2+({\rm tr}^{\rm f}(1,4)N_1(1^{\phi},2^{g},3^{g},4^{\phi}))^2+\\
    &({\rm tr}^{\rm f}(1,2)(N_1(1^{\phi},3^{g},4^{g},2^{\phi})-N_1(1^{\phi},4^{g},3^{g},2^{\phi})))^2\,+\nonumber\\
    &4{\rm tr}^{\rm f}(1,2,3)N_2(2^{\phi},1^{\phi},4^{g},3^{\phi}){\rm tr}^{\rm f}(1,3)N_1(1^{\phi},2^{g},4^{g},3^{\phi})\,+\nonumber\\
    &4{\rm tr}^{\rm f}(1,2,3)N_2(1^{\phi},2^{\phi},4^{g},3^{\phi})({\rm tr}^{\rm f}(1,2)(N_1(1^{\phi},3^{g},4^{g},2^{\phi})-N_1(1^{\phi},4^{g},3^{g},2^{\phi})))\,+\nonumber\\
    &4{\rm tr}^{\rm f}(1,3,4)N_2(3^{\phi},4^{\phi},2^{g},1^{\phi}){\rm tr}^{\rm f}(1,3)N_1(1^{\phi},2^{g},4^{g},3^{\phi})\,-\nonumber\\
    &4{\rm tr}^{\rm f}(1,3,4)N_2(3^{\phi},4^{\phi},2^{g},1^{\phi}){\rm tr}^{\rm f}(1,4)N_1(1^{\phi},2^{g},3^{g},4^{\phi})\,+\nonumber\\
    &4{\rm tr}^{\rm f}(1,2,4)N_2(1^{\phi},2^{\phi},3^{g},4^{\phi}){\rm tr}^{\rm f}(1,4)N_1(1^{\phi},2^{g},3^{g},4^{\phi})\,-\nonumber\\
    &4{\rm tr}^{\rm f}(1,2,4)N_2(1^{\phi},2^{\phi},3^{g},4^{\phi})({\rm tr}^{\rm f}(1,2)(N_1(1^{\phi},3^{-},4^{+},2^{\phi})-N_1(1^{\phi},4^{+},3^{-},2^{\phi})))\,+\nonumber\\
    &8{\rm tr}^{\rm f}(1,2,3){\rm tr}^{\rm f}(1,3,4)N_2(2^{\phi},1^{\phi},4^g,3^{\phi})N_2(3^{\phi},4^{\phi},2^g,1^{\phi})\,-\nonumber\\
    &8 {\rm tr}^{\rm f}(1,2,4){\rm tr}^{\rm f}(1,3,4)N_2(1^{\phi},2^{\phi},3^g,4^{\phi})N_2(3^{\phi},4^{\phi},2^g,1^{\phi})\,-\nonumber\\
    &8 {\rm tr}^{\rm f}(1,2,3){\rm tr}^{\rm f}(1,2,4)N_2(1^{\phi},2^{\phi},3^g,4^{\phi})N_2(2^{\phi},1^{\phi},3^g,4^{\phi})\,+\nonumber\\
    &6 ({\rm tr}^{\rm f}(1,2,3)N_2(2^{\phi},1^{\phi},4^{g},3^{\phi}))^2+6 ({\rm tr}^{\rm f}(1,2,4)N_2(1^{\phi},2^{\phi},3^{g},4^{\phi}))^2+6 ({\rm tr}^{\rm f}(1,3,4)N_2(3^{\phi},4^{\phi},2^{g},1^{\phi}))^2+\nonumber\\
    & 4 {\rm tr}^{\rm f}(1,2)({\rm tr}^{\rm f}(1,2,3,4)-{\rm tr}^{\rm f}(2,1,3,4))(N_1(1^{\phi},3^{g},4^{g},2^{\phi})-N_1(1^{\phi},4^{g},3^{g},2^{\phi}))\,+\nonumber\\
    & 4 {\rm tr}^{\rm f}(1,3)({\rm tr}^{\rm f}(1,2,3,4)-{\rm tr}^{\rm f}(2,1,3,4))N_1(1^{\phi},2^{g},4^{g},3^{\phi})\,-\nonumber\\
    & 4 {\rm tr}^{\rm f}(1,4)({\rm tr}^{\rm f}(1,2,3,4)-{\rm tr}^{\rm f}(2,1,3,4))N_1(1^{\phi},2^{g},3^{g},4^{\phi})\,+\nonumber\\
    &12 {\rm tr}^{\rm f}(1,3,4)({\rm tr}^{\rm f}(1,2,3,4)-{\rm tr}^{\rm f}(2,1,3,4))N_2(3^{\phi},4^{\phi},2^{g},1^{\phi})\,+\nonumber\\
    &12 {\rm tr}^{\rm f}(1,2,3)({\rm tr}^{\rm f}(1,2,3,4)-{\rm tr}^{\rm f}(2,1,3,4))N_2(2^{\phi},1^{\phi},4^{g},3^{\phi})\,-\nonumber\\
    &12 {\rm tr}^{\rm f}(1,2,4)({\rm tr}^{\rm f}(1,2,3,4)-{\rm tr}^{\rm f}(2,1,3,4))N_2(1^{\phi},2^{\phi},3^{g},4^{\phi})\,+\nonumber\\
    &6 ({\rm tr}^{\rm f}(1,2,3,4)-{\rm tr}^{\rm f}(2,1,3,4))^2\,+\nonumber\\
    &\frac{1}{3}({\rm tr}^{\rm f}(1,2,3,4){\rm tr}^{\rm f}(1,3,2,4)-2{\rm tr}^{\rm f}(1,2,3,4){\rm tr}^{\rm f}(2,1,3,4)+{\rm tr}^{\rm f}(1,3,2,3){\rm tr}^{\rm f}(2,1,3,4))\,.\nonumber 
\end{align}

\begin{align}
    \sum_{i,j=1}^{11}&M^{(2)}_{ij}\tilde{n}_{2,i}\tilde{n}_{2,j}=({\rm tr}^{\rm f}(1,3)N_1(1^{\phi},2^{g},4^{g},3^{\phi}))^2+({\rm tr}^{\rm f}(1,4)N_1(1^{\phi},2^{g},3^{g},4^{\phi}))^2+\\
    &({\rm tr}^{\rm f}(2,3)N_1(2^{\phi},1^{g},4^{g},3^{\phi}))^2+({\rm tr}^{\rm f}(2,4)N_1(2^{\phi},1^{g},3^{g},4^{\phi}))^2\,+\nonumber\\
    &4{\rm tr}^{\rm f}(1,2,3)N_2(2^{\phi},1^{\phi},4^{g},3^{\phi}){\rm tr}^{\rm f}(1,3)N_1(1^{\phi},2^{g},4^{g},3^{\phi})\,+\nonumber\\
    &4{\rm tr}^{\rm f}(1,2,3)N_2(2^{\phi},1^{\phi},4^{g},3^{\phi}){\rm tr}^{\rm f}(2,3)N_1(2^{\phi},1^{g},4^{g},3^{\phi})\,+\nonumber\\
    &4{\rm tr}^{\rm f}(1,3,4)N_2(3^{\phi},4^{\phi},2^{g},1^{\phi}){\rm tr}^{\rm f}(1,3)N_1(1^{\phi},2^{g},4^{g},3^{\phi})\,-\nonumber\\
    &4{\rm tr}^{\rm f}(1,3,4)N_2(3^{\phi},4^{\phi},2^{g},1^{\phi}){\rm tr}^{\rm f}(1,4)N_1(1^{\phi},2^{g},3^{g},4^{\phi})\,+\nonumber\\
    &4{\rm tr}^{\rm f}(2,3,4)N_2(3^{\phi},4^{\phi},1^{g},2^{\phi}){\rm tr}^{\rm f}(2,3)N_1(2^{\phi},1^{g},4^{g},3^{\phi})\,+\nonumber\\
    &4{\rm tr}^{\rm f}(2,3,4)N_2(3^{\phi},4^{\phi},1^{g},2^{\phi}){\rm tr}^{\rm f}(2,4)N_1(2^{\phi},1^{g},3^{g},4^{\phi})\,+\nonumber\\
    &4{\rm tr}^{\rm f}(1,2,4)N_2(1^{\phi},2^{\phi},3^{g},4^{\phi}){\rm tr}^{\rm f}(1,4)N_1(1^{\phi},2^{g},3^{g},4^{\phi})\,-\nonumber\\
    &4{\rm tr}^{\rm f}(1,2,4)N_2(1^{\phi},2^{\phi},3^{g},4^{\phi}){\rm tr}^{\rm f}(2,4)N_1(2^{\phi},1^{g},3^{g},4^{\phi})\,+\nonumber\\
    &8{\rm tr}^{\rm f}(1,2,3){\rm tr}^{\rm f}(1,3,4)N_2(2^{\phi},1^{\phi},4^g,3^{\phi})N_2(3^{\phi},4^{\phi},2^g,1^{\phi})\,+\nonumber\\
     &8{\rm tr}^{\rm f}(1,2,3){\rm tr}^{\rm f}(1,3,4)N_2(2^{\phi},1^{\phi},4^g,3^{\phi})N_2(3^{\phi},4^{\phi},2^g,1^{\phi})\,-\nonumber\\
    &8 {\rm tr}^{\rm f}(1,2,3){\rm tr}^{\rm f}(2,3,4)N_2(2^{\phi},1^{\phi},4^g,3^{\phi})N_2(3^{\phi},4^{\phi},1^g,2^{\phi})\,-\nonumber\\
    &8 {\rm tr}^{\rm f}(1,2,4){\rm tr}^{\rm f}(2,3,4)N_2(1^{\phi},2^{\phi},3^g,4^{\phi})N_2(3^{\phi},4^{\phi},1^g,2^{\phi})\,+\nonumber\\
    &6 ({\rm tr}^{\rm f}(1,2,3)N_2(2^{\phi},1^{\phi},4^{g},3^{\phi}))^2+6 ({\rm tr}^{\rm f}(1,2,4)N_2(1^{\phi},2^{\phi},3^{g},4^{\phi}))^2+\\
    &6 ({\rm tr}^{\rm f}(1,3,4)N_2(3^{\phi},4^{\phi},2^{g},1^{\phi}))^2+6 ({\rm tr}^{\rm f}(2,3,4)N_2(3^{\phi},4^{\phi},1^{g},2^{\phi}))^2+\nonumber\\
    & 4 {\rm tr}^{\rm f}(1,3)({\rm tr}^{\rm f}(1,2,3,4)-{\rm tr}^{\rm f}(2,1,3,4))N_1(1^{\phi},2^{g},4^{g},3^{\phi})\,-\nonumber\\
    & 4 {\rm tr}^{\rm f}(1,4)({\rm tr}^{\rm f}(1,2,3,4)-{\rm tr}^{\rm f}(2,1,3,4))N_1(1^{\phi},2^{g},3^{g},4^{\phi})\,-\nonumber\\
    & 4 {\rm tr}^{\rm f}(2,3)({\rm tr}^{\rm f}(1,2,3,4)-{\rm tr}^{\rm f}(2,1,3,4))N_1(2^{\phi},1^{g},4^{g},3^{\phi})\,+\nonumber\\
    & 4 {\rm tr}^{\rm f}(2,4)({\rm tr}^{\rm f}(1,2,3,4)-{\rm tr}^{\rm f}(2,1,3,4))N_1(2^{\phi},1^{g},3^{g},4^{\phi})\,+\nonumber\\
    & 12 {\rm tr}^{\rm f}(1,3,4)({\rm tr}^{\rm f}(1,2,3,4)-{\rm tr}^{\rm f}(2,1,3,4))N_2(3^{\phi},4^{\phi},2^{g},1^{\phi})\,+\nonumber\\
    & 12 {\rm tr}^{\rm f}(1,2,3)({\rm tr}^{\rm f}(1,2,3,4)-{\rm tr}^{\rm f}(2,1,3,4))N_2(2^{\phi},1^{\phi},4^{g},3^{\phi})\,-\nonumber\\
    &12 {\rm tr}^{\rm f}(1,2,4)({\rm tr}^{\rm f}(1,2,3,4)-{\rm tr}^{\rm f}(2,1,3,4))N_2(1^{\phi},2^{\phi},3^{g},4^{\phi})\,-\nonumber\\
    &12 {\rm tr}^{\rm f}(2,3,4)({\rm tr}^{\rm f}(1,2,3,4)-{\rm tr}^{\rm f}(2,1,3,4))N_2(3^{\phi},4^{\phi},1^{g},2^{\phi})\,+\nonumber\\
    &6 ({\rm tr}^{\rm f}(1,2,3,4)-{\rm tr}^{\rm f}(2,1,3,4))^2\,+\nonumber\\
    &({\rm tr}^{\rm f}(1,2,3,4){\rm tr}^{\rm f}(1,3,2,4)-2{\rm tr}^{\rm f}(1,2,3,4){\rm tr}^{\rm f}(2,1,3,4)+{\rm tr}^{\rm f}(1,3,2,3){\rm tr}^{\rm f}(2,1,3,4))\,.\nonumber 
\end{align}



\providecommand{\href}[2]{#2}\begingroup\raggedright\endgroup

\end{document}